\documentclass[%
aps,
reprint,
nofootinbib,
 amsmath,amssymb,+--
 prl
]{revtex4-2}

\usepackage{graphicx}
\usepackage{dcolumn}
\usepackage{bm}
\usepackage{graphicx}
\usepackage{epstopdf, epsfig}
\usepackage{float}
\usepackage{subfig}
\usepackage{enumitem, amsmath,bm}
\usepackage{bbold}
\usepackage{dsfont}
\usepackage{comment}
\usepackage[dvipsnames]{xcolor}
\usepackage{mathtools}
\usepackage{float}
\usepackage{subfig}
\usepackage{amsmath,amssymb}
\usepackage{comment}
\usepackage[dvipsnames]{xcolor}
\usepackage{adjustbox}
\usepackage{multirow}
\usepackage{cases}
\usepackage{scalerel,stackengine}
\usepackage{appendix}
\usepackage{mathrsfs}
\usepackage{enumerate}

\stackMath

\newcommand\reallywidehat[1]{%
\savestack{\tmpbox}{\stretchto{%
  \scaleto{%
    \scalerel*[\widthof{\ensuremath{#1}}]{\kern.1pt\mathchar"0362\kern.1pt}%
    {\rule{0ex}{\textheight}}
  }{\textheight}%
}{2.4ex}}%
\stackon[-6.9pt]{#1}{\tmpbox}%
}

\usepackage{etoolbox}   

\makeatletter
\let\orig@addcontentsline\addcontentsline

\newcommand{\SuppressMainTOC}{%
  \renewcommand{\addcontentsline}[3]{}%
}

\newcommand{\RestoreTOC}{%
  \let\addcontentsline\orig@addcontentsline
}
\makeatother



\makeatletter

\newlength{\TOCPageWidth}
\def\@tocline#1#2#3#4#5#6#7{%
  \vskip #2%
  {\leftskip #3\relax \rightskip \@pnumwidth plus 1fil%
    \parfillskip -\TOCPageWidth%
    \parindent #3%
    \@afterindenttrue
    \interlinepenalty\@M
    \leavevmode
    #6\nobreak\leaders\hbox{$\mkern1mu.\mkern1mu$}\hfill
    \makebox[\TOCPageWidth][r]{#7}\par}%
}

\renewcommand{\@pnumwidth}{2.2em}

\makeatother

\newcommand{\sgn}{{\rm sgn}}
\newcommand{\be}{\mathbf{e}}
\newcommand{\bu}{\boldsymbol{u}}

\newcommand{\bv}{\boldsymbol{v}}
\newcommand{\bU}{\boldsymbol{U}}
\newcommand{\bforc}{\boldsymbol{f}}

\newcommand{\bx}{\boldsymbol{x}}

\newcommand{\bomega}{\boldsymbol{\omega}}
\newcommand{\bk}{\boldsymbol{k}}
\newcommand{\bq}{\boldsymbol{q}}
\newcommand{\bp}{\boldsymbol{p}}

\newcommand{\bh}{\boldsymbol{h}}

\newcommand{\uv}{\langle uv \rangle}

\newcommand{\Roe}{Ro_{\epsilon}}

\newcommand{\Vp}{V_{\bp\bk\bq}}

\newcommand{\pyf}{p_y^f}
\newcommand{\KyO}{K_y}
\newcommand{\KO}{K}

\newcommand{\Up}{{U'}}
\newcommand{\Vol}{\mathcal{V}}

\newcommand{\vv}{\langle vv \rangle}
\newcommand{\etav}{\langle \eta v \rangle}
\newcommand{\etaeta}{\langle \eta \eta \rangle}

\newcommand{\Tthree}{T_{\rm 3D-2D}}
\newcommand{\Ttwotwo}{T_{\rm 2D-2D}}

\newcommand{\dd}{\,\mathrm d}

\newcommand\SG[1]{{\color{black}#1}}
\newcommand\AF[1]{{\color{black}#1}}

\begin{document}
\preprint{APS/123-QED}

\SuppressMainTOC

\author{S\'ebastien Gom\'e
}
\author{Anna Frishman}

\affiliation{
Department of Physics, Technion Israel Institute of Technology, 32000 Haifa, Israel
}

\date{\today}
\title{
Helicity controls the direction of fluxes in rotating turbulence
}

\begin{abstract}
Turbulence sustains out-of-equilibrium energy fluxes shaped by conservation laws. 
Three-dimensional flows conserve energy and sign-indefinite helicity, both being transferred 
to small scales.
Yet in 3D rotating turbulence, energy is observed to flow simultaneously toward large-scale two-dimensional structures and small-scale three-dimensional 
motions. We identify a mechanism for
this dual behavior by considering inertial waves interacting with a large-scale 2D flow. 
%
For such interactions, sufficiently fast waves cannot exchange energy with waves of opposite helicity, so wave helicity is conserved separately for each sign.
This drives an inverse transfer of energy from 3D waves to 2D motions. 
For slower 3D modes, by contrast,  this conservation is broken: modes of opposite helicity exchange energy, producing a forward flux to small scales as in non-rotating turbulence.
Using a mean-wave kinetic theory, we derive analytical expressions for these competing bi-directional transfers, predicting the rotation- and Reynolds-number dependence of the large-scale 2D flow in fully nonlinear simulations, and unifying the picture from small to infinite rotation.
\end{abstract}

\maketitle


Inviscid invariants  are expected to control out-of-equilibrium fluxes in turbulent systems.
%
In two dimensions, an inviscid fluid conserves energy and enstrophy (squared vorticity). These
two sign-definite scale-related quantities
are constrained to flow in opposite directions: energy to large scales and enstrophy to small scales \citep{fjortoft1953changes, kraichnan1967inertial}. This leads to self-organization of turbulence into large-scale structures.
In contrast, three-dimensional turbulence
conserves energy and sign-indefinite helicity (the scalar product between velocity and vorticity)
\citep{kraichnan1973helical}.
%
This leaves the spectral fluxes unconstrained, with a simultaneous transfer of both energy and helicity to small scales \citep{chen2003joint, DitlevsenCascades, alexakis2017helically,knorr1990self}.


Defying this classification, 3D fluids in the presence of
external constraints, such as geometry, magnetic field or rotation,
can exhibit a dual organization of energy, with simultaneous transfers to large and small scales \citep{alexakis2018cascades,boffetta2011flux, smith1999transfer, falkovich2017vortices,  moffatt2019self, rubio2014upscale,  de2024pattern, benavides2017critical, van2019condensates,shaltiel2024direct,shavit2025turbulent}.
Rotation, a primary ingredient in geophysical and astrophysical flows, with a crucial role in climate and weather modeling \citep{vallis1993generation, ghil2020physics},
provides a striking example:
%
under rotation,
three-dimensional fluctuations take the form of waves and self-organize into large-scale 2D structures \citep{waleffe1993inertial,smith1999transfer,godeferd2015structure,buzzicotti2018energy,kolokolov2020structure,shaltiel2024direct, favier2014inverse, guervilly2014large, rubio2014upscale}, 
although the invariants are identical to those in 3D turbulence.
At the same time, when rotation is not too fast, 3D modes can also extract energy from large-scale 2D structures \cite{seshasayanan2018condensates, clark2020phase}, expected from, e.g., spontaneous generation of instabilities \citep{sipp2000three, billant2021taylor, seshasayanan2020onset}.
Flows in the Earth's oceans 
\citep{balwada2022direct} and atmosphere \citep{lindborg2005effect, czaja2019simulating} often lie in this regime, and support coexisting direct and inverse energy transfers \citep{geophysical_pouquet_2013}.
%
The bi-directional energy pathways in rotating 3D turbulence lack a general understanding: why and when do upscale and downscale energy fluxes coexist?
What sets the energy partition between large and small scales? 

Previous works have indicated that helicity should play a role~\cite{waleffe1992nature,waleffe1993inertial,biferale2012inverse,buzzicotti2018energy,clark2020phase}, but it has not been clear how. 
We have recently shown that wave-helicity is conserved separately by sign in interactions between fast inertial waves and a large-scale 2D flow, leading to an inverse energy transfer to the latter \citep{gome2026waves}.  
Here we reveal that, at moderate rotation rates, an additional forward energy flux emerges due to slow-frequency modes 
breaking this conservation.
We derive analytically the 
coexisting down- and upscale energy fluxes via a novel mean-wave kinetic theory, consistent with numerical simulations across rotation regimes. This is the first analytical framework capturing bi-directional fluxes and predicting their evolution with external constraints.
%
Our work reveals a general mechanism for the generation of competing fluxes in wave-dominated systems \citep{alexakis2024large, labarre2026distinguished}.


%

We consider the 3D Navier-Stokes equations (3DNSE) describing an incompressible fluid in a rotating frame, i.e.\ subject to a Coriolis force $-2 \Omega \be_z \times \bv$, in a domain of dimensions
$(L_x,L_y,L_z)$.
The flow is forced by a random three-dimensional isotropic forcing $\bforc$
where $\langle f_{\bp,i} (t) f_{\bq,j} (t') \rangle = \epsilon (\delta_{i,j}- p_i p_j /p^2) \chi_{\bp} \delta(\bp+\bq) \delta(t-t')$, 
$\chi_{\bp}$ is restricted to a shell around $k_f=2\pi/l_f$ and energy is injected at rate $\epsilon$. The Reynolds number, $Re= \epsilon^{1/3} k_f^{-4/3} /\nu$, where $\nu$ is the kinematic viscosity,
and
the Rossby number, $Ro = \epsilon^{1/3} k_f^{2/3} /(2\Omega)$, jointly determine the resulting state. In the following we will also use the combined parameter $\Roe=Ro\times \sqrt{Re}$. We use GHOST \citep{mininni2011hybrid} to solve this system numerically, taking $l_f = L_y/10$ and $L_x=L_z=L_y/2=\pi$, 
including hyperviscosity (see SM). 

\begin{figure}[t]
\includegraphics[width=\columnwidth]{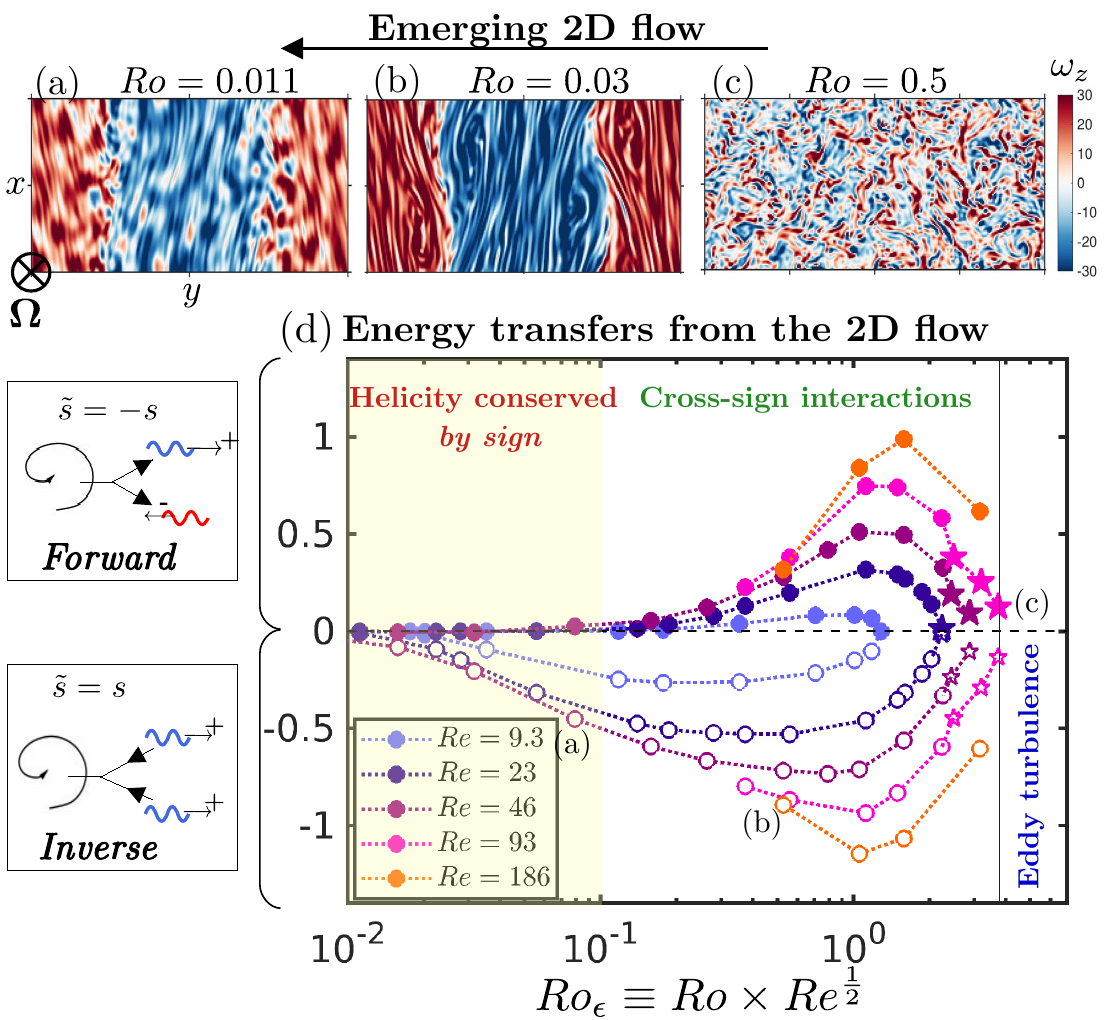} 
\caption{
(a-c) Flow visualizations of vertical vorticity in rotating turbulence.
(d) Energy transfers from the 2D condensate 
in DNS,
due to waves of same ($\tilde{s}=s$, hollow circles) or opposite helicity signs ($\tilde{s}=-s$, colored circles), as a function of $\Roe \equiv Ro \times Re^{\frac{1}{2}}$.
%
%
At large rotation, waves conserve their helicity by sign and energize the 2D flow.
At low rotation, modes with opposite helicities extract energy from the 2D flow.
%
%
}
\label{fig:cond_rescaled}
\end{figure}

For low enough Ro, a large-scale 2D mean flow, termed a condensate, spontaneously emerges 
when domain-scale dissipation is small. Here we focus on the statistical steady state after this condensate has formed. 
%
For our choice of domain geometry, the condensate takes the form of jets, i.e.\ a mean flow of the form
$\bU= \langle \bm{v}\rangle= U(y) \be_x$, see Fig.~\ref{fig:cond_rescaled}(a)-(b).
Typically, most of its energy is contained in the lowest mode $k_y= 2\pi /L_y$, 
implying $U(y) \approx \Up \sqrt{2}/k_y ~  \sin( k_y y)$, 
where $\Up \equiv\sqrt{ \int (\partial_y U)^2 (dy/L_y)}$ denotes the rms shear rate.

Under rotation fluctuations take the form of 3D inertial waves~\cite{greenspan1969theory}.
%
These waves can exchange energy with the condensate~\cite{greenspan1969theory,smith1999transfer}, with the total space-averaged transfer given by $\Tthree\equiv\int \langle uv\rangle \partial_yU \text{d}y/L_y$, where $\bm{u}=\bm{v}-U$, $u=\bm{u}_x, v=\bm{u}_y$ and $\langle uv\rangle$ is the Reynolds stress. 
Our goal in this Letter is to determine the direction and magnitude of this energy transfer as a function of Ro, Re and box geometry. The condensate amplitude $U'$, and thus the energy contained in the 2D component, is then set by the mean-flow energy balance:
\begin{align}
    \nu \Up^2 = \int \langle uv\rangle \partial_yU \frac{dy}{L_y} = \Tthree. \label{eq:mf1}
\end{align}

\paragraph*{Quasi-Linear Kinetic Theory \textemdash}
To obtain $\Tthree$, we 
employ the Quasi-Linear (QL) approximation~\cite{marston2023recent}, whereby wave-wave interactions are neglected, keeping only mean-wave coupling. 
This approximation holds as long as the mean-shear rate is sufficiently large compared to the fluctuation interaction time ($\bu\cdot \nabla\bu \ll \bU\cdot \nabla  \bu  \Leftrightarrow  1/U' \ll \tau_{nl} $).
In practice, this  limitation translates to a condition on the fluctuations wavenumber: $|\bp|\ll k_U$, where $k_U$ is a regime-dependent cutoff which will be discussed below.
We will also assume scale separation $l_f/L_y \ll 1$ in the following.

To exploit the wave character of the fluctuations, we write them in terms of inertial wave amplitudes $\bu (\bx) = \sum_{\bp, s} a_{\bp}^s (t) \bh_{\bp}^s e^{i \omega_{\bp}^s t + i \bp \cdot \bx }$,
as customary in wave-turbulence theory. Here, for each 3D mode $\bp$ ($p_z\neq0$) there are two wave polarizations $\bh_{\bp}^s$, such that $i \bp \times \bh_{\bp}^s = s p \bh_{\bp}^s $, with chiralities $s=\pm1$ and dispersion relation $\omega_{\bp}^s = s 2 \Omega p_z/p$. Each wave bears helicity of sign $s=\pm 1$. Its QL dynamics reads
%
\begin{align}
    \hspace{-1em}
 \partial_t a_{\bp}^s &=  \hspace{-1em}\sum_{\substack{\bq,\bk = \pm \frac{2\pi}{L_y}, \\ 
 \tilde{s}=\pm s}}
 \Vp^{s\tilde{s}}  U_{\bk}^* a_{\bq}^{\tilde{s}*}  (t) e^{-i (\omega_{\bp}^s + \omega_{-\bp-\bk}^{\tilde{s}} ) t}
 \delta_{\bk\bp\bq} +  f_{\bp}^s (t) e^{-i\omega_{\bp}^s t},
\label{eq:triad3D}
\end{align}
for $|\bp| \ll k_U$,
with $\delta_{\bk\bp\bq} = \delta_{\bk+\bp+\bq,0}$,
%
$\Vp^{s_1,s_2} =  - i p_x ~( \bh_{\bp}^{-s_1}  \cdot \bh_{\bq}^{-s_2})  + i k_y ~ (h_{\bp,x}^{-s_1} ~ h_{\bq,y}^{-s_2})$ the coupling coefficients with the 2D mode.

Rotation enters through the oscillating factor $e^{-i (\omega_{\bp}^s + \omega_{\bq}^{\tilde{s}} ) t}$, which determines the dynamical contribution of each triad.
Interactions are restricted to triads for which the 
frequency sum $ \omega_{\bp}^{s} + \omega_{-\bp-\bk}^{\tilde{s} } $ is sufficiently small, so that 
$e^{-i (\omega_{\bp}^s + \omega_{-\bp-\bk}^{\tilde{s}} ) t}\simeq 1$ for times $ t\ll 1/\Up$. This translates to
the near-resonance condition 
determining
allowed interactions:
\begin{align}
 |\omega_{\bp}^{s} + \omega_{-\bp-\bk}^{\tilde{s} } |< \Up, ~~~ \omega_{\bp}^s = s 2 \Omega \frac{p_z}{p},
 \label{eq:reso}
\end{align}
Otherwise, when condition \eqref{eq:reso} is not fulfilled (off-resonance), the triad decorrelates because $e^{-i (\omega_{\bp}^s + \omega_{-\bp-\bk}^{\tilde{s}} ) t}$ approximately averages out over times $t= o(1/\Up)$ \citep{bretherton1964resonant, smith1999transfer, zakharov2012kolmogorov,nazarenko2011wave, gome2026waves}.

There are two types of condensate-wave interactions: either involving waves of the same helicity sign $\tilde{s}= s$ (\emph{homochiral-wave interaction}), or of opposite helicity signs $\tilde{s}=- s$  (\emph{heterochiral-wave interaction}). 
Condition \eqref{eq:reso} is more easily satisfied by homochiral-wave interactions, for which it reads $|\omega_{\bp}^s+\omega_{-\bp-\bk}^s|\approx |k_y\partial_p\omega_{\bp}^s|< U'$, assuming scale separation $l_f/L_y\ll1$. 
%
%
For the theory developed here, we will assume that this resonance condition is always satisfied, corresponding to the regime $\Up/\Omega \gtrsim l_f/L_y \Leftrightarrow \Roe >l_f/(2L_y) $, see \cite{gome2026waves}.
Heterochiral-wave interactions, on the other hand, require that $\omega_{\bp}^{s} + \omega_{-\bp-\bk}^{-s} \approx 2 \omega_{\bp}^s < U'$. 
Modes $\bp$ which do not satisfy this condition have only homochiral-wave interactions.
%
This leads to the definition of two spectral sectors:
 \begin{align}
    \begin{cases}
    \text{Sector H: }  2\omega_{\bp}^s > \Up 
    \text{ (only homochiral interactions)}
    \\
     \text{Sector A: } 2\omega_{\bp}^s 
     \leq \Up
    \text{ (all interactions).} \end{cases}
\label{eq:sectors_main}
\end{align}
%
%
Modes in sector H are fast and constrained by rotation, while modes in sector A are slow and shear-dominated.

Using scale separation in the QL system \eqref{eq:triad3D}, and assuming $\Up/\Omega \gtrsim l_f/L_y$, we obtain the spectral balance for the wave energy $E_{\bp}^s\equiv \frac{1}{2} \langle |a_{\bp}^s|^2\rangle $ and helicity $H_{\bp}^s \equiv \frac{1}{2} s p \langle |a_{\bp}^s|^2\rangle $: 
\begin{align}
&\partial_t E_{\bp}^s 
+   \partial_{p_y} \Pi^{s}_{\bp} 
    = - \frac{\Up}{\sqrt{2}} \left(
    \uv_{\bp\bk}^{s,s}
    + \uv_{\bp\bk}^{s,-s} \mathbb{1}_{\bp\in A} \right)
       +  \epsilon \chi_{\bp}^s
  \label{eq:Econs_hetero}\\
&\partial_t  H_{\bp}^s + 
    \partial_{p_y} ( s p \Pi^{s}_{\bp}) 
    = 
    -  s p ~ \frac{\Up}{\sqrt{2}} \uv_{\bp\bk}^{s,-s} \mathbb{1}_{\bp\in A}
       +  s p \epsilon \chi_{\bp}^s, \label{eq:Hcons_hetero}
\end{align}
where $\Pi^s_{\bp}  = -\Up p_x (\langle a_{\bp}^s a_{-\bp + \bk}^s \rangle + \langle a_{\bp}^s a_{-\bp - \bk}^s \rangle) /(2\sqrt{2})$ is the spectral energy flux for waves of helicity sign $s$, and  $\uv_{\bp\bk}^{s,\tilde{s}} = \Re \left[h_{\bp,x}^s h_{-\bp, y}^{\tilde{s}}  \left(\langle a_{\bp}^s a_{-\bp+\bk}^{\tilde{s}} \rangle + \langle a_{\bp}^s a_{-\bp-\bk}^{\tilde{s}} \rangle \right) \right] $ is the Reynolds stress related to waves of chiralities $s$ and $\tilde{s}$. 
Note that the total helicity of the waves, $\int (H_{\bp}^s + H_{\bp}^{-s}) \text{d} \bp $, is conserved by this system (using $\uv_{\bp\bk}^{s,-s}=  \uv_{\bp\bk}^{-s,s}$ and \eqref{eq:Hcons_hetero}), since the helicity exchange between the waves and the condensate $(\propto l_f/L_y)$ is negligible in the scale separation limit $l_f/L_y \ll 1$.

The energy flux $\Pi_{\bp}^s$ in Eqs.~\eqref{eq:Econs_hetero}-\eqref{eq:Hcons_hetero} arises due to the shearing of the waves by the mean flow $U(y) \be_x$,
which moves their energy to increasing values of 
$\sigma p_y$ (with $\sigma \equiv \sgn(- \Up p_x)$) while leaving $p_x$ and $p_z$ unchanged. 
The flux to arbitrary small scales, $\Pi^s_{\sigma p_y\to\infty}$, determines the energy transfer to the condensate: integrating Eq.~\eqref{eq:Econs_hetero} over $p_y$,
\begin{align}
    \Tthree = \int \text{d}p_x \text{d} p_z
(\epsilon^s_{p_x,p_z}-\Pi^s_{\sigma p_y\to\infty}),
\label{eq:T32}
\end{align}
with $\epsilon_{p_x,p_z}^s$ the energy injected into modes with given ($p_x,p_z,s$).

For modes in sector H, the energy flux can be directly determined from Eq.~\eqref{eq:Hcons_hetero}.
For them, helicity is conserved \emph{by sign},
as $\uv_{\bp\bk}^{s,-s}=0$ and Eq.~\eqref{eq:Hcons_hetero} becomes a continuity equation for $H_{\bm{p}}^s$ \citep{gome2026waves}. 
Equivalently, wave action $\langle |a_{\bp}^s|^2 \rangle /\omega_{\bp}^s$ is spectrally conserved for each species $s$ \citep{bretherton1968wavetrains}.
In steady state, when forcing excites sector H directly, 
the energy flux decays with $p_y$, $\Pi_{\bp}^s= \epsilon^s_{p_x,p_z} k_f/p$ for $p>k_f$, resulting in an inverse energy transfer, from the waves to the condensate, as $\Pi_{\bp}^s < \epsilon_{p_x,p_z}^s$, see Eq.~\eqref{eq:T32}.
In the large rotation limit $U'/\Omega \ll 1$
all modes are in sector $H$, corresponding to the regime $\Roe \lesssim 0.1$ in Fig.~\ref{fig:cond_rescaled}(d), where only homochiral-wave interactions are present and the energy transfer is purely inverse \citep{gome2026waves}.

%

%

\begin{figure}
    \centering
    \includegraphics[width=\linewidth]{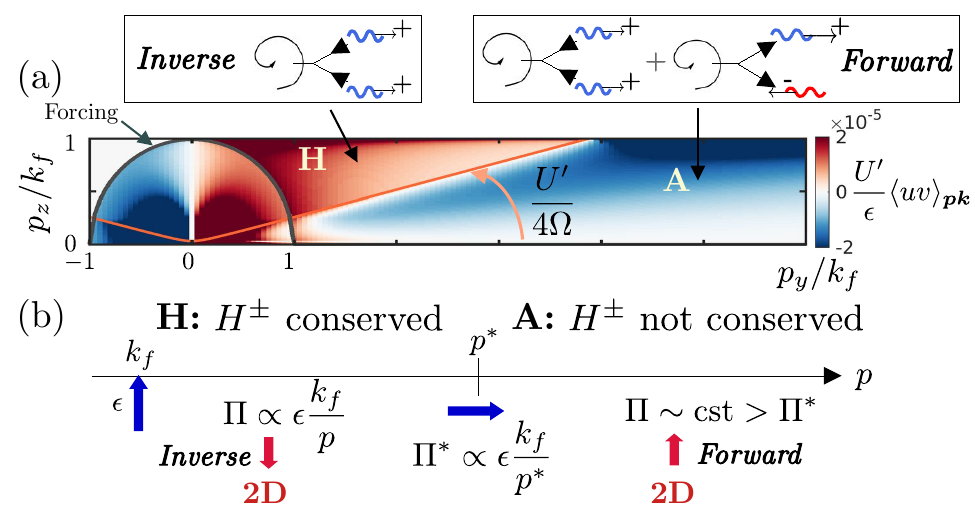}
    \vspace{-2em}
    \caption{(a) 
    Energy transfer to the large-scale 2D flow in spectral space $(p_y,p_z)$ from QL theory ($p_x=-8\pi/L_x, U'/\Omega=1$) and (b) schematic of the energy fluxes in the system.
    Orange line in (a) marks the boundary between
    Sectors H and A  \eqref{eq:sectors_main}, $p=p^*(p_z)$. 
    Waves in sector H conserve their sign-definite helicity $H_{\bp}^\pm$, constraining the forward energy flux and causing energy transfer to the 2D flow.
    In contrast, modes in sector A break this conservation and extract energy from the 2D flow, with resulting energy flux $\Pi_{\bp}> \Pi^*$ \eqref{eq:flux_hetero}, $\Pi^*$ being the energy input from sector H.
}
    \label{fig:theory}
\end{figure}

Decreasing  rotation progressively 
generates a coexisting forward energy transfer, from the condensate to small 3D scales, 
due to emerging heterochiral-wave interactions  when $\Roe \gtrsim 0.1$ --- see colored circles in Fig.~\ref{fig:cond_rescaled}(d).
%
%
%
Indeed, modes originally excited in sector H transfer energy to modes with decreasing frequency $\omega_{\bp}^s$, as $p_y$ increases due to shearing. Eventually, $2\omega_{\bp}^s < U'$ and the energy flux reaches sector A \eqref{eq:sectors_main}.
This occurs at $p=p^*(p_z) \equiv \frac{4 \Omega |p_z|}{\Up} $, where the energy flux is given by $\Pi^*=(\epsilon^+_{p_x,p_z} + \epsilon^-_{p_x,p_z} ) k_f/p^*$ 
--- see Fig.~\ref{fig:theory}.
In sector A, modes of opposite helicities interact 
($\uv_{\bp\bk}^{s,-s}\neq0$), single-sign helicity is no longer conserved, and energy is extracted from the condensate.

The small-scale energy flux in sector A can be determined quantitatively for a given mean shear rate $U'$.
The dynamics there is shear-dominated, effectively reducing to that of a non-rotating shear flow.
The natural variables are then cross-flow velocity and vorticity ($v_{\bp},\eta_{\bp}) \be_y$ instead of helical variables ($a_{\bp}^+, a_{\bp}^{-}$).
The system becomes analytically solvable in the scale-separation limit $l_f/L_y \ll 1$,
where the mean flow is effectively a linear shear $U' y \be_x$, see Eq.~\eqref{eq:corr_etav_main} and \citep{farrell1993optimal}.
%
%
This gives the steady-state energy flux to small scales:
\begin{align}
    &\Pi_{\bp} \underset{p \gg p^* }{=}
     \frac{\Pi^*}{2} \Bigg[ 
    1  +\frac{p^{*2}}{p_0^2} \frac{p_z^2}{p_x^2}
     \arcsin^2 \left( \frac{p_0}{p_y^*}\right)  \Bigg],      \label{eq:flux_hetero} \\
    & ~ \text{with }
    p^* = \frac{4 |p_z| \Omega}{\Up}, ~~~
    p_0 = \sqrt{p_x^2 + p_z^2}, ~~~
    p_y^* = \sqrt{p^{*2}- p_0^2}, \nonumber
\end{align}
for $p_y \gg p_y^*$, 
see End Matter \S A.
This small-scale flux is constant through scales (independent of $p_y$), implying the existence of an inertial range for the fluctuations \citep{frisch1995turbulence}. 
The second term in the RHS of \eqref{eq:flux_hetero}
corresponds to energy extraction from the mean flow due to
generation of vorticity $\eta_{\bp}$ through tilting of mean vorticity by $v_{\bp}$.
%
The total flux is dominated by contributions from
modes with the largest values of $p_z/p_x$ ($\sim L_x/l_f \gg 1$), which produce most tilting and interact with the shear for the longest time. For such $p_z/p_x \gg 1$ the leading order small-scale flux is $\Pi_{\bp} \sim  \frac{\Pi^*}{2} \frac{p_z^2}{p_x^2} $ when $U'/(4\Omega) \ll 1$. It is larger than the incoming flux $\Pi^*$, resulting 
in a net extraction of condensate energy by small scales.

The detailed spectral energy transfer to the 2D condensate, $U' \uv_{\bp\bk}$, can also be computed analytically. 
It is shown in Fig.~\ref{fig:theory}(a) for different $(p_z,p_y)$ and fixed $p_x$ and $U'/\Omega$, see Eq.~\eqref{eq:uv_py_2_main}. Modes with the largest $p_z/p_x$ are indeed seen to extract most of the energy. 
%
%
Note that some modes with small values of $p_z/p_x$ are directly excited in sector A.
They are treated differently from the description above and are included in Fig.~\ref{fig:theory}(a), see SM \S V.
Their contribution to the total flux is subleading.

\textit{Total energy transfer and closure ---} 
Using the leading- order flux to small scales in sector A, $\Pi_{\bp} \sim  \frac{\Pi^*}{2} \frac{p_z^2}{p_x^2} $, and the flux $\Pi^*$ entering this sector,  
the total energy transfer to the 2D flow \eqref{eq:T32}
is given by
\begin{align}
   \Tthree &\approx 
   \sum_{p_x,p_z} \epsilon_{p_x,p_z}
   \left( 1-
 \frac{ U' k_f }{8 \Omega |p_z|} \frac{p_z^2}{p_x^2} \right)
   =1  - \frac{\Up}{\Omega}  
\frac{\pi}{24}\frac{L_x}{l_f}
 \label{eq:uv_lowRo}
\end{align}
when $l_f/(4L_y) < U'/(4\Omega) \ll 1$,
with $\epsilon_{p_x,p_z} = \epsilon_{p_x,p_z}^+ + \epsilon_{p_x,p_z}^- $,
see SM \S V.A.
It reflects the combined effects of the inverse transfer in sector H and the forward transfer in sector A. 
Decreasing rotation (increasing $\Up/\Omega$ in Eq.~\eqref{eq:uv_lowRo}), 
the waves
extract more
energy from the 2D flow
as shear-tilting heterochiral interactions become more numerous.

Finally, to obtain the condensate amplitude, we insert Eq.~\eqref{eq:uv_lowRo} into the energy balance \eqref{eq:mf1}
and obtain the leading-order solution in the limit $l_f/L_x  \ll 1$:
\begin{align}
  \frac{\Up}{\Omega} & \simeq 2
     b \Roe^2 
     \Bigg(-1 + \sqrt{1 + \frac{1}{b^2 \Roe^2 } }
     \Bigg),  ~~~ b = \frac{\pi}{24} \frac{L_x}{l_f}.  \label{eq:sol_analytic}
\end{align}
This solution depends on the single parameter $\Roe= Ro \times Re^{1/2}$.

\begin{figure}[t]
\includegraphics[width=0.95\columnwidth]{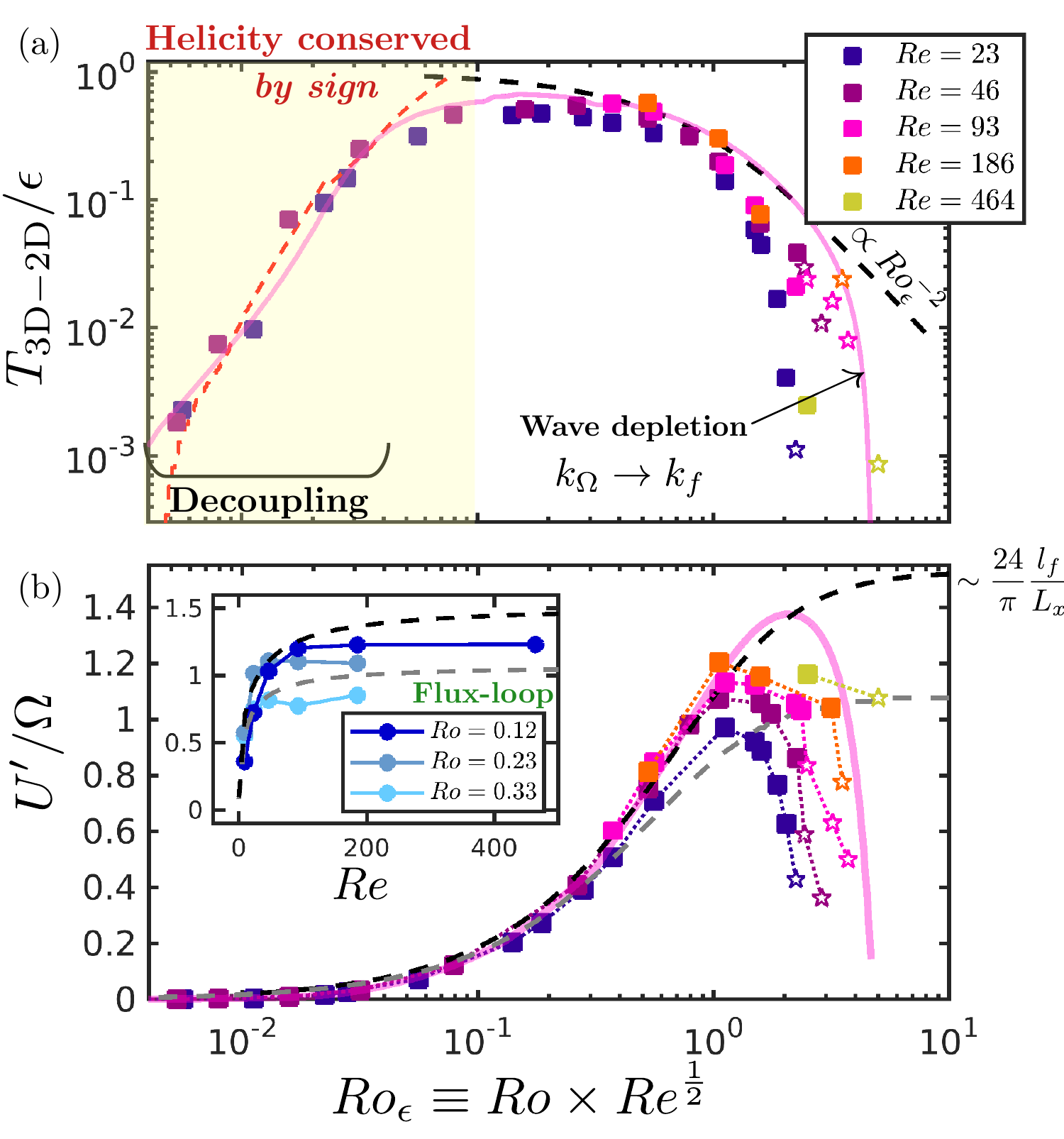} 
\vspace{-1em}
\caption{(a) Energy transfer from 3D waves to 2D condensate and (b) rescaled amplitude of the condensate, as measured from the DNS and predicted from the QL theory.
Dashed lines show the predictions in each regime: in black, Eq.~\eqref{eq:sol_analytic}; in red, the scalings derived in \cite{gome2026waves}; in grey, a numerical estimate including discretization effects.
Solid lines show a numerical solution of the QL system at $Re=93$, including a wave cutoff $k_\Omega$ and a smooth treatment of homochiral-wave near-resonances \eqref{eq:Fkf_Lorentzian_main}.
Inset in (b): rescaled amplitude of the 2D flow as a function of $Re$, showing the saturation to a flux-loop state.
}
\label{fig:recap_scalings}
\end{figure}

To verify this prediction, we measure the energy transfer from the waves to the 2D condensate (Fig.~\ref{fig:recap_scalings}(a)), and the amplitude of the latter $\Up/\Omega$ (Fig.~\ref{fig:recap_scalings}(b)) in DNS.
Equation \eqref{eq:sol_analytic} is shown as a black dashed line in Fig.~\ref{fig:recap_scalings}(b)
and the corresponding energy transfer, computed from Eqs.~\eqref{eq:uv_lowRo}-\eqref{eq:sol_analytic}, is shown as a black dashed line in Fig.~\ref{fig:recap_scalings}(a). 

For $\Roe\ll 1/b \sim 24 l_f/(\pi L_x)$, almost all interactions are purely homochiral and  Eqs.~\eqref{eq:uv_lowRo}-\eqref{eq:sol_analytic} give $\Tthree \simeq \epsilon$ and $U'\simeq \sqrt{\epsilon/\nu}$.
This is consistent with the DNS behavior in Fig.~\ref{fig:recap_scalings}(a) for $\Roe \sim 0.1$.
For $\Roe \lesssim l_f/(2L_y) \sim 0.05$, the transfer decreases with rotation due to the resonance condition \eqref{eq:reso} becoming restrictive for homochiral-wave interactions, see \cite{gome2026waves} and the red dashed line in Fig.~\ref{fig:recap_scalings}(a).
Note that the transition from the purely-homochiral to the mixed-interaction regime around $\Roe \simeq 0.1$ in the theory (red and black dashed-lines in Fig.~\ref{fig:recap_scalings}(a)) is smoothed out when using a smooth function of $\Up/\Omega$ \eqref{eq:Fkf_Lorentzian_main} instead of the sharp near-resonant condition \eqref{eq:reso} for $\tilde{s}=s$ (magenta solid line in Fig.~\ref{fig:recap_scalings}(a)). This improves the agreement with the DNS in this region.

At the other end, for $\Roe=\text{Ro}\times\text{Re}^{1/2}\gg 1/b$, Eq.~\eqref{eq:sol_analytic} predicts that $U'/\Omega \to 1/b= 24 l_f/(\pi L_x) $, a Reynolds-independent value. This asymptotic state corresponds to a vanishing energy transfer $\Tthree \to 0^+$ in Eq.~\eqref{eq:uv_lowRo}, i.e.\ to a \emph{flux-loop state}, where inverse and forward energy transfers are almost counterbalanced, see also \citep{boffetta2011flux, falkovich2017vortices, seshasayanan2018condensates, de2024pattern,clark2020phase}.
At fixed Ro, such a steady state with vanishing energy transfer and finite condensate amplitude $U'$ is obtained when $\rm{Re} \to \infty$, consistent with taking $\nu \to 0$ in the energy balance \eqref{eq:mf1}.
For fixed Ro and increasing Re in the DNS, $\Up/\Omega$ is indeed found to saturate to a constant value (see inset of Fig.~\ref{fig:recap_scalings}b), 
similar to the prediction
for low-enough $Ro$, and in accordance with previous observations
with 3D \citep{alexakis2015rotating} or 2D forcing \citep{seshasayanan2018condensates}.
%

 A similar steady state with ${\rm Re }\to \infty$ would be impossible if the system strictly conserved two sign-definite quantities, prohibiting a flux to small scales.
%
In our case, 
an exact balance between the inverse energy transfer from modes conserving $H_{\bp}^{\pm}$ (sector H), and the forward transfer due to modes breaking this conservation (sector A), 
can be achieved by tuning the condensate amplitude, and is realized as ${\rm Re }\to \infty$.
Note that here ${\rm Re }$ determines \emph{large-scale} dissipation, so  viscosity could be replaced by Ekman friction \citep{le2017inertial}.

\begin{figure}[t]
\includegraphics[width=0.9\columnwidth]{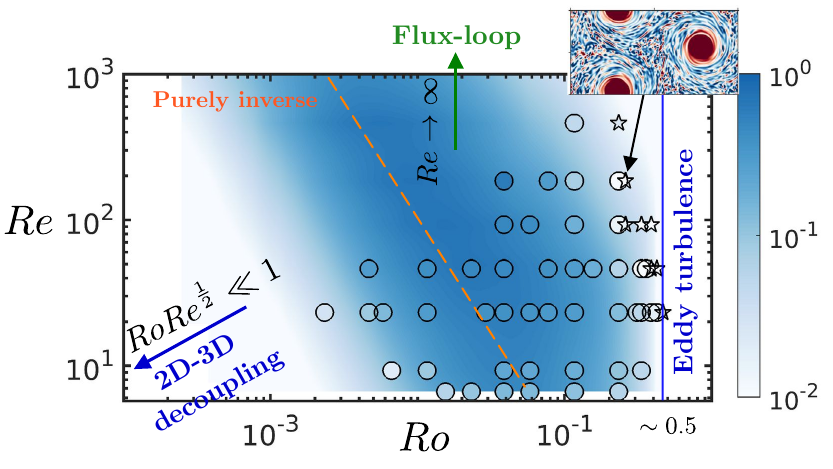}
\vspace{-1em}
\caption{Heatmap of the energy transfer to the 2D flow, $\Tthree$, in the $(Ro,Re)$ plane. 
The colored background corresponds to the prediction from the QL theory. 
Stars correspond to a scale-selected vortex lattice.
}
\label{fig:eps2D}
\end{figure}

\paragraph*{From 2D condensates to strong eddy turbulence \textemdash}
%
%
When, instead, fixing Re and increasing Ro, the scaling $U'\propto \Omega$ predicts that the large-scale 2D flow vanishes only in the zero-rotation limit ($\rm{Ro} \to \infty$). In DNS, however, condensates already give way to eddy turbulence at $Ro \simeq 0.5$; see Figs.~\ref{fig:cond_rescaled} and \ref{fig:eps2D}.
%
%
This is due to a depletion of waves at low rotation, causing the QL assumption to break in sector H, thus reducing the number of modes sustaining the condensate.
In this regime, the QL assumption can break due to eddy-eddy interactions, 
set by modes behaving as eddies rather than inertial waves, i.e\ such that $\tau_{nl}^{\rm eddy} < \tau_\Omega \equiv 2/ \omega_{\bp}$ (or, equivalently, $p > k_\Omega \equiv \Omega^{3/5} \epsilon^{-1/5} p_z^{3/5}$ \citep{zeman1994note}),
with $\tau_{nl}^{\rm eddy} = \epsilon^{-1/3} p^{-2/3}$ the Kolmogorov eddy turnover time.
In sector H, 
where $\tau_\Omega \lesssim 1/\Up$ \eqref{eq:sectors_main}, 
such eddy interactions are always faster than the interaction with the condensate, as for 
$p>k_\Omega$, $\tau_{nl}^{\rm eddy} < 1/\Up$.
Therefore, sector H, which is the source of energy for the condensate, is cutoff at $p = k_\Omega$, disappearing when $Ro=0.5$ for which
$k_\Omega( p_z=k_f) = k_f$. 
This explains the disappearance of 2D condensates for $Ro \gtrsim 0.5$ in Fig.~\ref{fig:eps2D}.
Note that the condensate could disappear earlier if wave-wave interactions dominate below $k_\Omega$, but it is not the case with our DNS parameters.

Incorporating the breaking of the QL assumption into our theory requires summing the spectral energy transfer only up to a maximum wavenumber. As discussed above, $k_\Omega$ is one such cutoff, but considering all competing time-scales of wave-wave or eddy-eddy interactions introduces additional cutoffs, affecting both sectors A and H, see SM.
Implementing this numerically reproduces the drop in 
energy transfer and condensate amplitude
observed in DNS as $\rm{Ro} \to 0.5$ due to $k_\Omega\to k_f$, see the magenta solid lines in Fig.~\ref{fig:recap_scalings}.
%
%
Note that corrections in $l_f/L_x$ and $l_f/L_z$ make $U'/\Omega$ lower than the leading-order prediction \eqref{eq:sol_analytic}, as shown numerically on a discrete set without cutoff, see the grey dashed line in Fig.~\ref{fig:recap_scalings}(b).
Introducing the cutoff to the latter increases the values of $U'/\Omega$ for $\Roe \lesssim 1$, making the magenta line accidentally close to the leading-order prediction (black dashed line).
This effect is due to sector A being cut-off in this regime, increasing the energy transfer to the condensate.


\paragraph*{Vortex lattice \textemdash}
While jets are found for most $Ro$ and $Re$ in our geometry, the condensate takes the form of cyclonic vortices at the boundary with 3D turbulence (stars in Figs.~\ref{fig:cond_rescaled} and \ref{fig:eps2D}), previously observed in \citep{marchetti2025spontaneous, seshasayanan2018condensates, clark2020phase}. Our theory roughly captures the condensate amplitude for these states, but cannot explain scale selection nor symmetry breaking.
The former may be due to time-scale competition between 2D modes of different scales, consistent with the metastability of such states \cite{estrada2025stability}.

\paragraph*{Conclusion \textemdash}

We have untangled the
organization of kinetic energy fluxes in rotating turbulence in the presence of a large-scale 2D flow, showing that it is controlled by helicity. Direct and inverse energy fluxes can coexist due to the presence of two sectors of modes which differ by their conservation laws. 
The inverse energy transfer to the large-scale 2D flow is induced by modes which are rotationally constrained to conserve helicity by sign. At the same time, energy is extracted from the 2D flow to small scales by modes which break this conservation. 
By determining both fluxes analytically, our theory explains why and how they coexist, and predicts how their balance changes with Ro and Re: from a regime in which the fluxes nearly cancel (flux-loop) to regimes dominated by one or the other, see Fig.~\ref{fig:eps2D}. 
Our results suggest a broader mechanism for bidirectional fluxes in systems containing waves: resonance conditions impose an additional sign-definite invariant on one sector of modes, while a coexisting sector, unconstrained by resonance, breaks it. This mechanism could be at play in system like chiral fluids \citep{de2024pattern} and
stratified flows
\citep{geophysical_pouquet_2013,resolving_marino_2015,chini2022exploiting,shah2024regimes,shavit2024sign, labarre2026distinguished}.
At the level of Navier–Stokes triads, our results suggest that the stability hypothesis \citep{waleffe1992nature, waleffe1993inertial, smith1999transfer} --- which links the direction of energy transfer to mode helicities and triad geometry --- may be linked to an additional conservation law for scale-separated triads.
Overall, the framework developed here suggests a route to determining how competing energy transfers and feedbacks govern self-organization in turbulent flows.




\acknowledgements

The authors would like to thank the Isaac
Newton Institute for Mathematical Sciences, Cambridge,
for support and hospitality during the programme ``Antidiffusive dynamics: from sub-cellular to astrophysical scales''. This work was supported by BSF grant No. 2022107 and ISF grant No. 486/23.

%
%

\newpage
\section*{End matter}

\subsection{Mean-wave Quasi-Linear Kinetic Theory}
\label{sec:theory}

We derive here the governing equations and solutions to the quasi-linear system describing the interactions between 3D inertial waves and a large-scale 2D condensate. 
We consider a sinusoidal mean flow 
of domain-scale mode $\bk=\pm 2\pi/L_y \hat{k}_y$, 
$U_{\bk} = \Up/\sqrt{2}\delta_{\bk- 2\pi/L_y \hat{k}_y}$, with $L_y k_f \gg 1$.
The fluctuation field $\bu (\bx)$ is projected onto the helical basis 
$\bh_{\bp}^s = (\bp \times [\bp \times \be_z]   - is p \bp \times \be_z)/(\sqrt{2} p p_\perp)
$, with $p=|\bp|$ and $p_\perp= \sqrt{p_x^2 + p_y^2}$.
%
%
Following wave kinetic theory, we obtain evolution equations for the wave correlators $\langle a_{\bp}^s a_{-\bp+\bk}^{s'} \rangle$, which we derive from the wave dynamical equation \eqref{eq:triad3D} in an expansion in $\Up/\Omega$ (see \citep{gome2026waves} and SM \S II), and result in the helicity balance \eqref{eq:Hcons_hetero}.

\paragraph*{Homochiral flux and transfer ---}
In the homochiral-only Sector H, Eq.~\eqref{eq:Hcons_hetero} in steady state is solved by
\begin{align}
 \Pi_{\bp}^s =
 \left\{
    \begin{aligned}
     &~ 0 ~~ \text{ if }  \sigma p_y  < -\pyf \\
&   \frac{\sigma \epsilon^s_{p_x,p_z}}{2}  \frac{k_f}{p }   
 ~~ \text{ if }  - \pyf <  \sigma p_y < \pyf \\
& \sigma \epsilon^s_{p_x,p_z}    \frac{k_f}{p }  
~~ \text{ if } \sigma  p_y  > \pyf,   \label{eq:flux_homo}
\end{aligned}
  \right.
\end{align}
with $\sigma = \sgn(- \Up p_x)$, $\pyf= (k_f^2 - p_x^2 - p_z^2)^{\frac{1}{2}}$ and $\epsilon^s_{p_x,p_z} = \frac{\epsilon }{4\pi \sqrt{k_f^2 - p_x^2 - p_z^2 } k_f \Vol}  $ the energy injected in each line $(p_x,p_z)$ and species $s$, and $\Vol = (L_x L_y L_z)/(2\pi)^3$. 
%
%
This flux decaying with $p_y$, waves lose energy to the 2D flow via the Reynolds-stress transfer:
\begin{align}
    \frac{\Up}{\sqrt{2}} \uv_{\bp\bk} &=
    - \frac{\Up}{2 \sqrt{2}} \frac{p_x p_y}{p^2} 
    \left(\langle a_{\bp}^s a_{-\bp + \bk}^s \rangle + \langle a_{\bp}^s a_{-\bp - \bk}^s \rangle \right) \nonumber\\
    &~~~~~~~~~~~~~~~~~~~~~~~~~~~~~~~~~~ + s\to -s
    \\
    &
    =
    \left\{
  \begin{aligned}
  & k_f \epsilon_{p_x,p_z} \frac{p_y}{2 p^3}, ~~~~~
   -\pyf < \sigma p_y < \pyf  ~~ \\
 &k_f \epsilon_{p_x,p_z} \frac{p_y}{p^3}, ~~~~~ 
  \sigma p_y > \pyf,  \label{eq:uv_H}  
    \end{aligned}
\right.   
    \end{align}
where $\epsilon_{p_x,p_z} = \epsilon^+_{p_x,p_z} + \epsilon^-_{p_x,p_z} $ denotes the energy injected in each line $(p_x,p_z)$.
The energy transfer to the large-scale 2D flow is therefore strictly positive when $\sigma p_y > \pyf$, provided that the homochiral interaction is resonant at $p=k_f$: $\omega_{\bp}^s + \omega_{-\bp-\bk}^s   < \Up $, see \citep{gome2026waves}.

\paragraph*{Heterochiral interactions ---}
As the 3D modes have low frequencies in sector A, $e^{i (\omega_{\bp}^s + \omega_{\bq}^{-s})t}\simeq 1$ from the near-resonant condition \eqref{eq:reso}, and rotation has no effect on the leading-order dynamics. 
Therefore, the fluctuations are effectively governed by the Rayleigh-Squire equations describing a non-rotating shear flow in terms of cross-flow velocity $v$ and vorticity $\eta \equiv (\nabla\times \bu)_{y}$:
\begin{align}
\partial_t \begin{bmatrix}
          \nabla^2 v \\
          \eta
           \end{bmatrix}
          =
          \begin{bmatrix}
           - (U \nabla^2 - \partial_y^2 U)\partial_x & 0 \\
           - \Up \partial_z  & - U \partial_x  
           \end{bmatrix}          
           \begin{bmatrix}
           v \\
          \eta
           \end{bmatrix}
           + 
           \begin{bmatrix}
			\nabla^2 f_y\\
			(\nabla\times \bforc)_y
           \end{bmatrix}. \label{eq:OS}
  \end{align}
With scale separation $l_f/L_y \ll 1$ (hence $\partial_y^2 U \ll U \nabla^2$), we obtain evolution equations for the Fourier-based correlators from Eq.~\eqref{eq:OS}:
%
\begin{subequations} \label{eq:corr_etav_main}
\begin{align}
    &\partial_t \frac{\langle |v_{\bp}|^2 \rangle}{2} - \frac{p_x U'}{\sqrt{2}} \frac{1}{p^4} \partial_{p_y} \Big(p^4 
     \Re ( \vv_{\bp\bk} )
     \Big)
    =  \epsilon \chi_{\bp,y} \label{eq:vv_main} \\
    & \partial_t \frac{\langle \eta_{\bp} v_{-\bp} \rangle}{2} - 
    \frac{1}{p^2}
    \frac{p_x U'}{\sqrt{2}} \partial_{p_y} \Big(p^2 
     \etav_{\bp\bk}  \Big)
    = - i \frac{p_z}{p_x} \vv_{\bp\bk}  \label{eq:etav_main} \\
    & \partial_t \frac{\langle |\eta_{\bp}|^2 \rangle}{2}  - \frac{p_x U'}{\sqrt{2}}\partial_{p_y} \Re (\etaeta_{\bp\bk} )
    =   - \frac{2 p_z}{p_x} \Im \etav_{\bp\bk}
    + \epsilon p^2 \chi_{\bp,y} \label{eq:etaeta_main} ,
\end{align}
\end{subequations}
with the notation $\langle a b \rangle_{\bp\bk} \equiv (\langle a_{\bp} b_{-\bp+\bk} \rangle + \langle a_{\bp} b_{-\bp-\bk} \rangle) /2 $ for two fields $a$ and $b$, and $ \chi_{\bp,y} \equiv (1- p_y^2/p^2) \chi_{\bp}/2$ the forcing correlation of the $y$ component.
System~\eqref{eq:corr_etav_main} can also be obtained directly from \eqref{eq:Hcons_hetero} by changing variables $(a_{\bp}^+, a_{\bp}^-)\to (\eta_{\bp}, v_{\bp})$.

Equation \eqref{eq:vv_main} results from the conservation of $\nabla^2 v$ in the mean-turbulent system. 
Cross-flow velocity $v$ then tilts the mean shear and generates cross-flow vorticity $\eta$, via the LHS of \eqref{eq:etav_main}.
This phenomenon generates enstrophy via Eq.~\eqref{eq:etaeta_main},  largest for modes with large $p_z/p_x$.
The associated energy flux can be computed through the balance of kinetic energy $E_{\bp} = p^2/(p_x^2+p_z^2)\langle |v_{\bp}|^2\rangle/2+1/(p_x^2+p_z^2)\langle |\eta_{\bp}|^2\rangle/2$, obtained from $p^2 \times \eqref{eq:vv_main} + \eqref{eq:etaeta_main}$:
\begin{align}
    \partial_t E_{\bp} + \partial_{p_y} \Pi_{\bp}
    = 
    \underbrace{
    \frac{U'}{\sqrt{2}}
    \left[  \frac{2p_x p_y }{p_0^2} \vv_{\bp\bk}  - \frac{2 p_z}{p_0^2}  \Im \etav_{\bp\bk}     \right]}_{ = - \frac{U'}{\sqrt{2}} \uv_{\bp\bk} }
   + \epsilon \chi_{\bp},
 \label{eq:energy_etav}
\end{align}
with  $p_0^2 = p_x^2 + p_z^2$, $\Pi_{\bp} = -\frac{p_x U' }{\sqrt{2}(p_x^2+ p_z^2)} \Big(p^2 \vv_{\bp\bk} + \etaeta_{\bp\bk} \Big) = \Pi^+_{\bp} + \Pi^-_{\bp} $ the total energy flux, and $\frac{U'}{\sqrt{2}} \uv_{\bp\bk} $ the Reynolds-stress transfer to the condensate.


Finding the energy flux $\Pi_{\bp}$ in sector A requires to solve Eqs.~\eqref{eq:vv_main},~\eqref{eq:etav_main} and~\eqref{eq:energy_etav} with spectral boundary conditions. 
For modes with large-enough $p_z$, boundary conditions are set by the values of correlators $\vv_{\bp\bk}$, $\etav_{\bp\bk}$ and flux $\Pi_{\bp}$ in sector H at $p=p^* \equiv 4 |p_z| \Omega/\Up$.
In sector H, 
mixed helicity correlators are zero,  implying that $\etav_{\bp\bk} =0$: indeed, projecting the $(v,\eta)$ representation onto helical modes,  $(v_{\bp},\eta_{\bp})=v_{\bm p}^+(1,ip)+v_{\bm p}^-(1,-ip) \equiv(v^+_{\bm{p}},\eta^+_{\bm{p}})+(v^-_{\bm{p}},\eta^-_{\bm{p}})$, gives 
$\langle \eta v\rangle_{\bp\bk}=\langle \eta^+ v^+\rangle_{\bp\bk}+\langle \eta^- v^- \rangle_{\bp\bk} \underset{k/p \ll 1}{\approx} i p \langle v^+v^+\rangle_{\bp\bk}-ip \langle v^-v^-\rangle_{\bp\bk}=0$, assuming isotropic forcing, i.e.\ a symmetry between
between positive and negative helicity modes.
To determine $\vv_{\bp\bk}$ in sector H, we notice that the energy transfer in sector H is given by
$-U' \uv_{\bp\bk}/\sqrt{2} = - 2 U' p_x p_y \vv_{\bp\bk} /(\sqrt{2} {p_0^2})$, see Eq.~\eqref{eq:energy_etav}. Finally, combining Eqs.~\eqref{eq:flux_homo} and \eqref{eq:uv_H}, we obtain at the boundary $p=p^*$: 
\begin{align}
    &\vv_{\bp\bk}(p^*) = -\frac{\Pi^* }{\sqrt{2} U' p_x} \frac{p_0^2}{ p^{*2}} , 
    ~~~~ \etav_{\bp\bk}(p^*) = 0, 
    \label{eq:BC1}
\end{align}
with the flux $\Pi^*= \Pi_{\bp}(p^*)=\sigma \epsilon_{p_x,p_z} \frac{k_f}{p^*}$.
The steady-state solution to Eqs.~\eqref{eq:vv_main}-\eqref{eq:etav_main}, using boundary conditions \eqref{eq:BC1}, reads:
\begin{align}
 \vv_{\bp\bk} 
&= -\frac{\Pi^*}{\sqrt{2} \Up p_x } \frac{p_0^2 p^{*2}}{p^4} \\
\etav_{\bp\bk}
&=
i \frac{p_z}{p_x} \frac{1}{p^2} \int_{p_y^*}^{p_y} q^2   \vv_{\bq\bk} ~ dq_y \nonumber \\
&= - i \frac{\Pi^* ~ p_0 p^{*2}}{ \sqrt{2} \Up p_x   } 
    \frac{p_z}{p_x} \frac{1}{p^{2}} 
\Big[ 
\arctan \Big( \frac{p_y}{p_0}\Big) -
\arctan \Big( \frac{p_y^*}{p_0}\Big) 
\Big],
\end{align}
%
%
with $p_y^* = (p^{*2} - p_0^2)^{\frac{1}{2}}$.
Finally, the steady-state energy flux solving Eq.~\eqref{eq:energy_etav} with boundary condition $\Pi_{\bp}(p=p^*)= \Pi^*$, is:
\begin{align}
    &\Pi_{\bp}
    = \Pi^*  + \frac{\sqrt{2} U'}{p_0^2}
    \int_{p_y^*}^{p_y}  \left(
    p_x p_y \vv_{\bp\bk}
     - p_z \Im \etav_{\bp\bk}  \right)\\
     &=\frac{ \Pi^*}{2} \Bigg[  1 +   \frac{p^{*2}}{p^2} +\frac{p^{*2}}{p_0^2} \frac{p_z^2}{p_x^2}  \Big( \arctan \Big( \frac{p_y}{p_0}\Big) -    \arctan \Big( \frac{p_y^*}{p_0}\Big) \Big)^2  \Bigg],
  \label{eq:flux_endmatter}
\end{align}
which leads to Eq.~\eqref{eq:flux_hetero} at small scales ($p \gg p^*$).
%
The energy transfer between the waves and the condensate for modes with $p> p^* > k_f$ then follows from 
\begin{align}
& \frac{\Up \uv_{\bp\bk}}{\sqrt{2}} = -\partial_{p_y} \Pi_{\bp},
\label{eq:uv_py_2_main}
\end{align}
with the energy flux $\Pi_{\bp}$ given in \eqref{eq:flux_endmatter}.

The above approach assumes that modes are excited in sector H and that energy flows to sector A at $p=p^*$.
However, modes can also be excited by the forcing in sector A when $p^* < k_f$. This case can be treated similarly, see SM \S E2. 
Note that the contribution of such modes is subleading,  as they have small $p_z/p_x$, hence 
weak shear tilting.
Figure \ref{fig:theory}(a) shows the analytical energy transfer \eqref{eq:uv_py_2_main} in Sector A (when $p^*> k_f$) and \eqref{eq:uv_H} in Sector H, as well as the expressions from the SM for the case $p^* < k_f$.

\subsection{Numerical closure with cutoff}
\label{sec:refinements}


To incorporate the breaking of the QL assumption at low-enough rotation, we sum the Reynolds stress up to relevant wavenumber cutoffs, above which fluctuations do not interact with the condensate. Specifically, 
wave-dominated behavior is lost above $k_\Omega= \Omega^{3/5} \epsilon^{-1/5} p_z^{3/5}$ (relevant for sector H), wave-wave interactions dominate above
$ k_U^{\rm wave} =     (\Omega p_z)^{1/3} {\Up}^{2/3} \epsilon^{-1/3}$, and eddy-eddy interactions dominate above $ k_U^{\rm eddy} =
{\Up}^{3/2} \epsilon^{-1/2}$ (relevant for sector A). 
The mean-flow energy balance then reads
\begin{align}
 &\frac{1}{4 \Roe^2 } \left(\frac{\Up}{\Omega} \right)^2  = \frac{\Tthree}{\epsilon} \Big[ \frac{\Up}{\Omega}\Big] =  \sum_{p_x,p_z} 
 \left( \epsilon_{p_x,p_z} - \Pi_{{\rm cutoff}} \right)
 F_{\rm Lor}, \label{eq:mf_bal}
\end{align}
with $p_z = 2\pi  m_z/L_z$ and $p_x= 2\pi m_x/L_x$ ($m_z,m_x \in \mathbb{Z}^*$),
which is solved numerically
by a root-finding procedure in variable $U'/\Omega$. Results are shown in Fig.~\ref{fig:recap_scalings} (magenta lines) for ${\rm Re}=93$. We discuss the effect of cutoffs on the Re-dependence of the QL solution in SM \S VI.2.

Note that for homochiral interactions $\tilde{s} = s$, we introduce a Lorentzian filter 
\begin{align}
     F_{\rm Lor}
     = \left( 1+ \left( \frac{\Omega}{\Up}  \frac{4\pi p_z \sqrt{k_f^2 - p_x^2 - p_z^2}}{L_y k_f^3} \right)^2 \right)^{-1}
    \label{eq:Fkf_Lorentzian_main}
\end{align}
in Eq.~\eqref{eq:mf_bal}, 
which smoothens out the resonance condition when $\omega_{\bp}^s + \omega_{\bq}^s \sim \Up$.
It mainly results in a smoothened transition between 
the homochiral and the heterochiral-interaction regimes around $\Roe \simeq 0.1$.
Furthermore, one can introduce a tuning parameter $\tilde{\omega}$ in the resonant condition \eqref{eq:reso} for heterochiral interactions, $\omega_{\bp}^+ + \omega_{\bq}^- < \tilde{\omega} \Up$. 
In Fig.~\ref{fig:recap_scalings}, we use $\tilde{\omega}=1$; when taking $\tilde{\omega}=1.2$, somewhat better agreement with DNS can be achieved, see SM \S VI.B.

It is noteworthy that the agreement between DNS and theory is qualitatively good, despite the fact that 
the theory for the fluctuations does not depend on the small-scale dissipation mechanism, while DNS has hyperviscosity acting at small scales.
This seems to suggest that using regular viscosity instead of hyperviscosity in more resolved DNS should give qualitatively similar results. Note that
quantitative disagreements with the theory can also be due to a limited inertial range in the DNS.

\bibliographystyle{apsrev4-2}
\bibliography{bib}

\onecolumngrid

\newpage
\appendix

\section*{Supplementary information}

\begingroup
\def\contentsname{}%
\setlength{\parindent}{0pt}
\InputIfFileExists{\jobname.app}{}{}
\endgroup

In this supplementary information, we detail the derivation of the mean-wave quasi-linear kinetic theory describing the interactions between a large-scale 2D flow and 3D inertial waves in rotating turbulence.
We first establish the near-resonant conditions that select particular interaction types depending on the rotation rate (\S \ref{app:KE}).
Wave-dominated interactions due to fast waves (in the sector called H, for "homochiral-only interactions"), which conserve helicity by sign, are described in \S \ref{app:homo}.
Shear-dominated interactions due to slow waves
(in the sector called A, for "all interactions" -- both homochiral and heterochiral), which do not conserve helicity by sign, are described in \S\ref{app:shear}. 
We derive analytical expressions for the wave energy flux through scales in this system
in \S\ref{app:fluxloop}, and compute the resulting total energy transfer between the waves and the mean flow, both analytically and numerically.
Finally, we introduce a wavenumber cutoff for the QL theory in \S \ref{app:cutoff}, above which wave-wave or eddy-eddy dominate over mean-fluctuation interactions. This cutoff controls the energy balance at low rotation and the transition to a regime of eddy turbulence.

\RestoreTOC

\setcounter{tocdepth}{3} 
\tableofcontents

\section{Numerical setup}

We consider the three-dimensional Navier-Stokes equations (3DNSE) written in the rotating frame, 
\begin{align}
	\partial_t \bu + \bu \cdot \nabla \bu &=- 2\Omega \be_z \times \bu  - \nabla p + \nu \nabla^2 \bu + \nu_h \nabla^{2h} \bu +  \bforc,
	\label{eq:3DNSE}
\end{align} 
in a periodic domain $(L_x,L_y,L_z)$ of renormalized volume $\Vol = (L_x L_y L_z)/(2\pi)^3$.
Forcing $\bforc$ is such that $\langle \bforc_{\bp} (t) \bforc_{\bq} (t') \rangle = \epsilon  (\delta_{i,j} - \frac{p_i p_j}{p^2}) \chi_{\bp} \delta_{\bp+\bq} \delta(t-t')$, with correlation 
$\chi_{\bp} = \delta_{p - k_f \pm 1} /(4 \pi k_f^2 dk \Vol)  $
and $\epsilon$ the energy injection rate.
Nondimensional Reynolds and Rossby numbers at the forcing scale $l_f\equiv 2\pi/k_f$ are defined as
$Re= \epsilon^{1/3} k_f^{-4/3} /\nu$
and
$Ro = \epsilon^{1/3} k_f^{2/3} /(2\Omega)$.
We simulate Eq.~\eqref{eq:3DNSE} with the pseudospectral code GHOST \citep{mininni2011hybrid}, with resolution $(N_x,N_y,N_z)=128 (L_x,L_y,L_z) /\pi$.
The inertial range of the forward energy cascade is controlled by 8-order hyperviscosity $\nu_8= 6 \times 10^{-29}$.

Forcing injects energy isotropically between 3D ($p_z\neq 0$) and 2D ($p_z=0$) modes, such that
\begin{align}
    \epsilon = \epsilon_{\rm 3D} + \epsilon_{\rm 2D}
\end{align}
with $\epsilon_{\rm 2D} = \frac{l_f}{2L_z} \epsilon \ll \epsilon_{\rm 3D}$, and $\epsilon_{\rm 3D} =  \epsilon (1 - \frac{l_f}{2L_z})$.

We decompose the 3D field into helical modes, 
$\bu(\bx,t) =\sum_{\bp,s=\pm 1} \bh_{\bp}^s ~a_{\bp}^s (t) ~  e^{i\omega_{\bp}^s t + i\bp \cdot \bm{x}} $,
with the helical basis
\begin{equation}
\bh_{\bp}^s = \frac{\bp \times [\bp \times \be_z]   - is p \bp \times \be_z}{\sqrt{2} p p_\perp} 
= \frac{1}{\sqrt{2} p p_\perp}\begin{bmatrix}
      	p_x p_z - i s p p_y	 \\
		p_z p_y  + i  s p p_x  \\
	   - p_x^2  -  p_y^2    	
         \end{bmatrix},          
 \label{eq:helical_basis}
\end{equation}
with $\bp=(p_x,p_y,p_z)$, $p=|\bp|$ and $p_\perp^2 = p_x^2+p_y^2$
\citep{lesieur1987turbulence}.
This basis satisfies $i \bp \times \bh_{\bp}^s = s \bh_{\bp}^s $ hence diagonalizes the curl operator.
From such decomposition on the instantaneous field $\bu$, we then measure the 3D-2D energy transfer
\begin{align}
    \Pi^{s_1,s_2} = -\sum_{|\bk| < 5} \sum_{\bk\in 2D, \bp,\bq \in 3D} \langle \bu_{\bk} \cdot (\bu_{\bp}^{s_1} \times {\bomega_{\bq}^{{s_2}})} \rangle \delta_{\bk\bp\bq},
\end{align}
where $\bu_{\bp}^{s_1}$ (resp. $\bomega_{\bp}^{s_2}$),  denote the projection of $\bu$ (resp. $\bomega$) on the helical mode ($\bp$, $s_1$).
Figure 1(d) in the main text shows the energy transfer due to homochiral waves, $\Pi^{++}+\Pi^{--}$, and due to heterochiral waves, 
$\Pi^{+-}+\Pi^{-+}$,
for various values of $Ro$ and $Re$.
Note that we only decompose the energy flux with regard to the chiralities $s_1,s_2$ of the 3D modes, and do not further decompose contributions from the two chiralities of the 2D modes, as done in \citep{buzzicotti2018energy, biferale2012inverse}.
Throughout this paper, we always refer to $\bk$ as being a 2D mode, and $\bp$ and $\bq$ as 3D waves.

\section{Mean-wave kinetic theory}
\label{app:KE}

We consider a stationary mean flow 
$\bU=\langle \bu\rangle= U(y) \be_x$ consisting of two $x$-invariant jets, corresponding to the condensate in the 
rectangular
domain $L_y=2L_x$.
The average $\langle \cdot \rangle$ denotes both ensemble and vertical average.
In the following, we denote by 
\begin{align}
    \Up \equiv \sqrt{\frac{1}{L_y}\int (\partial_y U)^2 dy }
    \label{eq:Uprime}
\end{align}
the root-mean-square  shear rate, averaged over $y$.
This defines a typical time scale $1/\Up$
over which turbulent fluctuations interact with the condensate.

Interactions with the condensate dominate over eddy-eddy or wave-wave interactions:
$1/\Up \ll \tau_{nl}$ (with $\tau_{nl}$ the eddy-eddy or wave-wave turnover time), hence
$\bu\cdot \nabla \bu \ll \bu \cdot \nabla \bU $. 
%
Under this quasi-linear (QL) approximation
\citep{marston2023recent}, 
the momentum equations for the mean flow and fluctuations $\bu'$ are written as
\begin{align}
\partial_t U &= - \partial_y \uv + \nu \partial_y^2 U =0 \label{eq:mf}
\\
\partial_t \bu' + U(y) \partial_x \bu' &= - \partial_y U v \be_x - 2 \Omega \be_z \times \boldsymbol{u'} -  \nabla p' + \bforc,
\label{eq:QLNSE}
\end{align}
where $u$ and $v$ denote the $x$ and $y$ components of the fluctuating velocity $\bu'$.

We decompose the fluctuation field as 
$\bu'(\bx,t) =\sum_{\bp,s=\pm 1} \bh_{\bp}^s ~a_{\bp}^s (t) ~  e^{i\omega_{\bp}^s t + i\bp \cdot \bm{x}} $, 
where $a_{\bp}^s(t)$ is the time-varying amplitude of a helical mode $(\bp,s)$ and $\omega_{\bp}^s$ is the frequency of the corresponding inertial wave.
%
%
%
%
 We write the energy balance 
for the waves $E_{\bp}^s = \frac{1}{2}\langle a_{\bp}^s a_{\bp}^{s*} \rangle$,
as an energy increment at current time $t$, from an arbitrary $t=0$,
\begin{align}
   \frac{E_{\bp}^s(t)-E_{\bp}^s(0)}{t} &= \frac{1}{2t} \sum_{\bq, \bk, \tilde{s}}\Big[ \Vp^{s\tilde{s}}  U_{\bk}^* ~ \langle a^{s*}_{\bp} a_{\bq}^{\tilde{s}*} \rangle  (t) e^{-i \omega_{\bp\bq}^{s\tilde{s}} t} 
 \Big] \delta_{\bk\bp\bq}   +c.c+\epsilon \chi^s_{\bp}   \label{eq:e_1}
\end{align} 
where 
\begin{equation}
    \omega_{\bp\bq}^{s\tilde{s}} \equiv \omega_{\bp}^s+\omega_{\bq}^{\tilde{s}}= 2\Omega p_z\left(\frac{s }{p} - \frac{\tilde{s}}{q}\right).
\end{equation}
and with coupling coefficients $\Vp^{s,\tilde{s}} =  - i p_x ~( \bh_{\bp}^{-s}  \cdot \bh_{\bq}^{-\tilde{s}})  + i k_y ~ (h_{\bp,x}^{-s} ~ h_{\bq,y}^{-\tilde{s}})$.

We now assume that wave-mean flow interactions occur on a typical time scale of the order of $1/\Up$, so that over a much shorter time scale, of order $t\ll 1/U'$, the wave amplitude correlators can be assumed constant.
Thus, performing a partial time average up to time $t\ll 1/\Up$,
\begin{align}
    \frac{E_{\bp}^s(t)-E_{\bp}^s(0)}{t} 
    &=\frac{1}2\sum_{\bq, \bk, \tilde{s}}\Big[ \Vp^{s\tilde{s}}  U_{\bk}^* ~ \langle a^{s*}_{\bp} a_{\bq}^{\tilde{s}*} \rangle \frac{1}{t}\int_0^t e^{-i \omega_{\bp\bq}^{s\tilde{s}} t'}dt' 
 \Big] \delta_{\bk\bp\bq}  +c.c+\epsilon \chi^s_{\bp} \nonumber  \\
 &\approx 
 \frac{1}2\sum_{\bq, \bk, \tilde{s}}\frac{1 - e^{- i \omega_{\bp\bq}^{s\tilde{s}}t}}{i \omega_{\bp\bq}^{s\tilde{s}}t}\Big[ \Vp^{s\tilde{s}}  U_{\bk}^* ~ \langle a^{s*}_{\bp} a_{\bq}^{\tilde{s}*} \rangle 
 \Big] \delta_{\bk\bp\bq}  +c.c+\epsilon \chi^s_{\bp} \nonumber  \\
 & = \frac{1}2\sum_{\bq, \bk, \tilde{s}}\Delta_{\bp}^{s\tilde{s}}(t)\Big[ \Vp^{s\tilde{s}}  U_{\bk}^* ~ \langle a^{s*}_{\bp} a_{\bq}^{\tilde{s}*} \rangle 
 \Big] \delta_{\bk\bp\bq}  +c.c+\epsilon \chi^s_{\bp} + o \left(\frac{\Up}{\Omega}\right)
 \label{eq:KE_app}
\end{align}
where 
\begin{equation}
    \Delta_{\bp}^{s \tilde{s}} (t) \equiv  
    \frac{1 - e^{- i \omega_{\bp\bq}^{s\tilde{s}}t}}{i \omega_{\bp\bq}^{s\tilde{s}}t}.
\end{equation}

Here we will generally consider interactions between waves of all types, due to their coupling with the condensate mode $\bk$, assuming $ \langle a_{\bp}^s a_{\bq}^{-s} \rangle\neq 0$, and follow the same procedure as in \citep{gome2026waves}.
%
%
Note that such correlations are absent initially, as our forcing does not directly correlate waves of opposite polarity, $\langle f_{\bp}^{s} f_{\bq}^{-s} \rangle=0$).
An additional equation is therefore needed to close the system, given by the evolution of the correlation between opposite polarity modes $\langle a_{\bp}^{s*} a_{\bp}^{-s} \rangle$ 
\begin{align}
\frac{\langle a_{\bp}^{s*} a_{\bp}^{-s} \rangle (t) -
 \langle a_{\bp}^{s*} a_{\bp}^{-s} \rangle(0)}{t} & = \sum_{\bq,\bk,\tilde{s}} \Bigg[
 U_{\bk}^* \Vp^{-s,\tilde{s}} \langle a_{\bp}^s a_{\bq}^{\tilde{s}} \rangle^* \Delta_{\bp}^{-s,\tilde{s}} +
 U_{\bk} \Vp^{s\tilde{s}*} 
 \langle a_{\bp}^{-s} a_{\bq}^{\tilde{s}} \rangle 
 \Delta_{\bp}^{s\tilde{s}*}  \Bigg]   \delta_{\bk\bp\bq}
  \label{eq:C_hetero_app_1}
\end{align}
%
%

Similarly to \citep{gome2026waves}, we approximate the oscillating factor in equations \eqref{eq:KE_app} and \eqref{eq:C_hetero_app} with a simple step function:
\begin{align}
    \Delta_{\bp}^{s \tilde{s}} (t) = 
    \frac{1 - e^{- i \omega_{\bp\bq}^{s\tilde{s}}t}}{i \omega_{\bp\bq}^{s\tilde{s}}t}
    \approx \Theta \Big( 1 - \frac{|\omega_{\bp\bq}^{s\tilde{s}}|}{\Up} \Big)  \label{eq:Heaviside} 
\end{align}
where $\Theta$ denotes the Heaviside function. 
This choice captures the leading order behavior 
$ \Delta_{\bp}^{s \tilde{s}} (t) \to 1 $ when $|\omega_{\bp\bq}^{s\tilde{s}}|/U' \ll 1$ and $ \Delta_{\bp}^{s \tilde{s}} (t) \to 0 $ when $|\omega_{\bp\bq}^{s\tilde{s}}|/U' \gg 1$ but is only a rough estimate for borderline cases $|\omega_{\bp\bq}^{s\tilde{s}}|/U'\sim 1$.

Homochiral-wave interactions are then allowed if 
\begin{align}
    \omega_{\bp}^s + \omega_{\bq}^s\approx \bk\cdot \nabla_{\bp} \omega_{\bp}^s  < \Up 
    \label{eq:reso_homo}
\end{align}
is satisfied, where we have used scale separation between $k$ and $p$ $p$ and recalling that $\bq=-\bp-\bk$. In the present work this condition will be assumed to be satisfied for all considered modes $\bp$.

For cross-helicity interactions $\tilde{s}= -s$, similarly using scale separation we get $\omega_{\bp\bq}^{s,-s} \simeq 2 \omega_{\bp}^s = 4s\Omega p_z/p$. Thus, the step-function approximation \eqref{eq:Heaviside} results in a condition delimiting allowed interactions from those that are off-resonant:
%
%
\begin{align}
\Delta_{\bp}^{s,-s}(t)=\frac{1- e^{i \omega_{\bp\bq}^{s,-s}t}}{i \omega_{\bp\bq}^{s,-s} t} \approx 
\begin{cases}
    1, ~~ \omega_{\bp}^s + \omega_{\bq}^{-s} \leq  \Up \Leftrightarrow ~~\frac{4 p_z}{p} \leq \frac{U'}{\Omega} \\
    0, ~~   \omega_{\bp}^s + \omega_{\bq}^{-s} > \Up \Leftrightarrow ~~ \frac{4 p_z}{p} > \frac{U'}{\Omega}   
\end{cases}
\label{eq:hetero_filter}
\end{align}
Conditions \eqref{eq:hetero_filter} lead to the definition of two separate sectors A and H:   
 \begin{align}
    \begin{cases}
    \text{Sector H: }  2\omega_{\bp}^s > \Up \Leftrightarrow \frac{4 p_z}{p} > \frac{U'}{\Omega} 
    \text{ (\emph{homochiral} interactions)}
    \\
     \text{Sector A: }   2\omega_{\bp}^s < \Up \Leftrightarrow  
     \frac{4 p_z}{p} \leq \frac{U'}{\Omega} 
    \text{ (\emph{all} interactions),}
    \end{cases}
    \label{eq:sectors}
\end{align}
For modes $\bp$ in sector H, heterochiral-wave interactions decorrelate ($\Delta_{\bp}^{s,-s}$ averages out at long times) and only homochiral-wave interactions ($\tilde{s} = s$) are present. For modes in sector A, on the other hand, both types of interactions are allowed by the resonance condition and these modes experience the full 3D QL dynamics.
Note that the resonant condition for homochiral-wave interactions \eqref{eq:reso_homo} is less restrictive than that for heterochiral-wave interactions \eqref{eq:hetero_filter} due to scale separation. The former becomes restrictive for modes in sector H only at sufficiently high rotation rates. At large rotation ($\frac{\Up}{\Omega} < l_f/(  L_y) $), there are modes at the forcing scale for which $\Delta_{\bp}^{ss}\approx0$ so they consequently do not interact with the condensate, as discussed in~\cite{gome2026waves}, and in the QL dynamics are treated as if unforced.


Note that the small parameter in our expansion is $U'/4\Omega$ (the ratio between the wave and non-linear time scale), where we keep the leading order term $O(U'/4\Omega)$.
Indeed,
the oscillations of the waves will be felt by the system (meaning that some interactions will average out to zero) only for times much larger than $1/\max(\omega_{\bp\bq}^{s\tilde{s}})=1/(4\Omega)$, which occurs in interactions between opposite-helicity waves, $s=-\tilde{s}$. Hence, such an expansion makes sense for $1/(4\Omega) \ll t\ll 1/U'$, requiring a sufficient time-scale separation. In particular, 
the condition $U'/4\Omega  < 1$ is required for the existence of Sector H.
\subsection{The limit of scale-separation and its general consequences}
In addition to the above separation of time-scales we also consider separation of scales between the mean flow and the waves $k/p\ll1$. We now consider the lowest complex-conjugate modes $k_y= \pm 2\pi/L_y$ for the condensate $U_{\bk} = \frac{\Up}{i \sqrt{2} k_y}$, and expand the coefficients in the limit $L k_f \gg 1$:
\begin{align}
    \Vp^{ss}& =  -i p_x - s  k_y\left(\frac{p_x^2p_z}{p_\perp^2 p} +\frac{p_z}{2p}\right) -i  k_y\frac{p_x p_y}{2p^2}  + O\left(\frac{k_y^2}{p^2 }\right)
    ,  ~~~~~~~~~~~  
    \Vp^{s, -s} = 2 i k_y V_{\bp}^{s,-s}  + O\left(\frac{k_y^2}{p^2 }\right),\label{eq:Vp_expand}
    \\
    &= -  i p_x - 2 i k_y V_{\bp}^{ss}  + O\left(\frac{k_y^2}{p^2 }\right), \nonumber
\end{align}
where we have defined
\begin{align}
    V_{\bp}^{ss} \equiv -i s \frac{p_z}{p} \left(1+\frac{2p_x^2}{p_\perp^2 } \right) +  \frac{p_x p_y}{p^2}, ~~~~~~~~~~~
    V_{\bp}^{s, -s} \equiv \SG{2  h_{\bp,x}^{-s} h_{\bp,y}^{s}} =
    \frac{p_x p_y}{p^2} \Big(1 + \frac{ p_z^2}{p_{\perp}^2} \Big)
+ is \frac{p_z}{p} \Big(1 - \frac{2 p_x^2}{p_{\perp}^2} \Big).
\end{align}
When considering condition \eqref{eq:hetero_filter} for heterochiral-wave interactions, and assuming that homochiral waves resonate with the condensate ($\omega_{\bp}^s + \omega_{\bq}^s < \Up$), 
we obtain the coupled equations \SG{by expanding correlators and coefficients in $L k_f \gg 1$}:
%
%
\begin{align}
\partial_t E_{\bp}^s& =
 \frac{\Up}{\sqrt{2}} \Bigg( \Big[ p_x \partial_{p_y}   +\Re[V_{\bp}^{ss}] \Big] \Phi_{\bp\bk}^s 
-  \mathbb{1}_{\bp\in A}~ \Re ( V_{\bp}^{s,-s} \Psi_{\bp\bk}^{s,-s*})
\Bigg)  +  \epsilon \chi_{\bp}^s  \label{eq:e_hetero_app}\\
\partial_t \langle a_{\bp}^{s*} a_{\bp}^{-s} \rangle & =
2\frac{\Up }{\sqrt{2}} \Bigg(  \Big[ p_x \partial_{p_y}   +  V_{\bp}^{ss\SG{*}} 
\Big] 
(\Psi_{\bp\bk}^{s,-s*} +\Psi_{\bp\bk}^{-s,s})
~~~~ - \mathbb{1}_{\bp\in A} ~  V_{\bp}^{s,-s*} (\Phi_{\bp\bk}^{s*} + \Phi_{\bp\bk}^{-s} )
\Bigg), 
\label{eq:C_hetero_app}
\end{align}
with correlators
\begin{equation}
\begin{split}
    \Phi_{\bp\bk}^s &\equiv \frac{\langle a_{\bp}^s a_{-\bp+\bk}^s\rangle+\langle a_{\bp}^s a_{-\bp-\bk}^s\rangle}{2}  =
    \frac{ \langle a_{\bp}^s a_{-\bp+\bk}^s\rangle+c.c}2
    +O\left(\frac{k_y}p\right)
    \end{split}
\end{equation}
and
\begin{equation}
\begin{split}
    \Psi_{\bp\bk}^{s,-s}\equiv \frac{\langle a_{\bp}^s a_{-\bp+\bk}^{-s}\rangle+\langle a_{\bp}^s a_{-\bp-\bk}^{-s}\rangle }{2}
    \end{split}.
\end{equation}
%
Here $\Im(\Phi_{\bp\bk}) = O(k_y/p) $, while $\Psi_{\bp\bk} \in \mathbb{C}$ at the leading order in $k_y/p$.
\AF{Equation \eqref{eq:C_hetero_app} demonstrates that for modes in the homochiral sector, $\bp\in H$, if cross-helicity correlators are initially zero, they do not get spontaneously generated.}

To write equation \eqref{eq:e_hetero_app} in a more familiar form we recall that the wave kinetic equations \eqref{eq:e_hetero_app}-\eqref{eq:C_hetero_app} are coupled with the mean-flow equation \eqref{eq:mf}. The latter is obtained by Reynolds averaging over time $T \gg 1/\Up$ and over $z$, in a time window where the ensemble-averaged correlator $\langle a_{\bp}^s a_{\bq}^{s'} \rangle$ is stationary. In particular, there is exchange of energy between the 3D modes and the 2D flow, 
given by
\begin{align}
    \Tthree
 & = \sum_{\bp,s, \tilde{s}}  U'_{-\bk} \langle u_{\bp}^s v_{-\bp + \bk}^{\tilde{s}} \rangle +  U'_{\bk} \langle u_{\bp}^s v_{-\bp-\bk}^{\tilde{s}} \rangle    
 = \frac{\Vol \Up}{\sqrt{2}} \int {\rm d}\bp\sum_{s,\tilde{s}=\pm1 } \uv^{s\tilde{s}}_{\bp\bk}  \label{eq:T2_app}
\end{align}
Note that the $1/\sqrt{2}$ prefactor is a normalization factor due to our choice of definition for $\Up$ \eqref{eq:Uprime}.
In Eq.~\eqref{eq:T2_app} the Reynolds stress due to homochiral interactions is
\begin{align}
    \uv_{\bp\bk}^{ss} \equiv \frac{1}{2} \Big( \langle u_{\bp}^s v^s_{-\bp-\bk} \rangle +  \langle u_{\bp}^s v_{-\bp+\bk}^s \rangle \Big) + c.c
    =- \frac{p_x p_y}{p^2} ~ \Phi_{\bp \bk}^s ~ + O\left(\frac{k_y}{p}\right)
    \label{eq:uv_homo_approx}
\end{align}
using scale separation (see also~\cite{gome2026waves}). 

The contribution of heterochiral interactions to the Reynolds stress is given by
\begin{align}
    \uv_{\bp\bk}^{s,-s} & \equiv \frac{1}{2}\langle u_{\bp}^s v_{-\bp+ \bk}^{-s} + u_{\bp}^s v_{-\bp-\bk}^{-s} \rangle  + c.c
    \nonumber\\ 
    &=  \frac{1}{2T} \int_0^T ~dt~ \Big(  h_{\bp,x}^{-s} h_{-\bp+\bk,y}^s \langle a_{\bp}^{s} a_{-\bp + \bk}^{-s}
    \rangle e^{ i (\omega_{\bp}^s + \omega_{-\bp+\bk}^{-s})  t}       + (\bk\to-\bk) \Big) + c.c \nonumber \\
    &\simeq   \AF{\mathbb{1}_{\bp\in A}}\Re(V_{\bp}^{s,-s} \Psi_{\bp\bk}^{s,-s}) 
    +O\left(\frac{k_y}{p}\right)
    \label{eq:hetero_corr}
\end{align}
%
%
%
This heterochiral correlator contributes to the total Reynolds stress in Eq.~\eqref{eq:T2_app}, in addition to the contribution of homochiral waves via Eq.~\eqref{eq:uv_homo_approx}.

Now we can recognize in the RHS of Eq.~\eqref{eq:e_hetero_app} the heterochiral Reynolds-stress correlator \eqref{eq:hetero_corr}, as expected from the conservation of energy in the system including the mean flow. 
The energy balance Eq.~\eqref{eq:e_hetero_app} is therefore of the form
\begin{align}
    \partial_t E_{\bp}^s + \partial_{p_y}   \Pi^s_{\bp} & = 
 - \frac{\Up}{\sqrt{2}} \left(\uv_{\bp\bk}^{ss} +
 \uv_{\bp\bk}^{s,-s} \right)
       +   \epsilon \chi_{\bp}^s
       \label{eq:Econs_2}
\end{align}
\SG{with the energy flux $\Pi_{\bp}^s = -\frac{U' p_x}{\sqrt{2}} \Phi_{\bp\bk}^s$.}
This energy balance is valid in both Sectors A and H (with $\uv_{\bp\bk}^{s,-s} = 0 $ in the latter). 
Note that in steady state, Eq.~\eqref{eq:Econs_2} reflects the balance between the wave energy flux and the transfer to the mean-flow:
\begin{align}
    \partial_{p_y}   \Pi^s_{\bp} &    
 =- \frac{\Up}{\sqrt{2}} \uv_{\bp\bk}       +   \epsilon \chi_{\bp}^s,
\end{align}
with the total Reynolds stress correlator $\uv_{\bp\bk} = \uv_{\bp\bk}^{ss} + \uv_{\bp\bk}^{s,-s}$. Away from the forcing scale ($\chi_{\bp}^s = 0$), waves either lose energy to the 2D flow (the wave energy flux $\Pi_{\bk}^s$ decreases with $p_y$ and the transfer to the 2D flow $\frac{\Up}{\sqrt{2}} \uv_{\bp\bk} > 0$), or gain energy from the 2D flow.

Now focusing on helicity, multiplying \eqref{eq:Econs_2} by $s p$ leads to the spectral helicity balance
\begin{align}
\partial_t (s p E_{\bp}^s )+ 
    \partial_{p_y} ( s p \Pi^s) = 
    -  \mathbb{1}_{\bp\in A} s p ~  \frac{\Up}{\sqrt{2}} \uv_{\bp\bk}^{s,-s}
       +  s p \epsilon \chi_{\bp}^s. \label{eq:Hcons_hetero_app}
\end{align}
The first term on the RHS of Eq.~\eqref{eq:Hcons_hetero_app} corresponds to the exchange of helicity between modes of different helicity sign, 
mediated by the condensate.
\AF{Note that 
\begin{equation}
    \uv_{\bp\bk}^{s,-s}=\uv_{\bp\bk}^{-s,s}
\end{equation}
 at the leading order in $O(k_y/p)$, since the Reynolds stress \eqref{eq:hetero_corr} is symmetric when taking $\bp \to -\bp$ (which is the same as exchanging $s\to -s$ in \eqref{eq:hetero_corr} in the scale-separated limit $\bk \to $).
 This implies}
the waves conserve their total helicity, independently of the type of condensate-wave interaction (homochiral or heterochiral) in the scale-separated limit.
In particular, a finite energy flux (and positive as is natural due to shearing) at arbitrary large scales $p$, $\lim_{p\to \infty}\Pi^s>0$ becomes possible without breaking total helicity conservation.
Furthermore, the helicity flux $s p \Pi^s$ must increase with $p$ and be of sign $s$ in such a case. 
From Eq.~\eqref{eq:Hcons_hetero_app}, this then implies that $\frac{\Up}{\sqrt{2}} \uv_{\bp\bk}^{s,-s}<0$ for large enough $p$, which in turn implies that the cross-helicity Reynolds stress acts on the mean flow to extract energy from it. Therefore, in the presence of cross-helicity interactions, modes of positive and negative helicity extract energy from the mean flow (symmetrically in our case), to compensate for the increase of helicity of a particular sign due to the $p_y$-energy flux.

\AF{We are interested in the transfer of energy between the small scale fluctuations and the condensate. Due to total energy conservation, this transfer can be determined from the spectral energy flux to large $p_y$, captured by equations \eqref{eq:e_hetero_app} and \eqref{eq:C_hetero_app} at steady state. We will see below that the $p_y$ spectral line can be divided into regions with modes belonging to the $H$ sector and those belonging to the $A$ sector. It turns out that these equations can be solved in each sector separately, each with the corresponding basis that helps diagonalize the system. We will discuss these solutions in detail the following two sections. How the two are matched at a common boundary is discussed in the main text.} 

\section{Homochiral-wave interactions}
\label{app:homo}

Here we focus on Sector H, where heterochiral interactions are off-resonant, hence $\langle a_{\bp}^s a_{-\bp+\bk}^{-s} \rangle = 0$.
each sign of the helicity of the modes is conserved in this case, and \eqref{eq:e_hetero_app} reduces to a continuity equation. In steady state, it reads
in \citep{gome2026waves}. 
%
\begin{equation}
    \partial_{p_y}\left(-sp  \frac{U'p_x}{\sqrt2}\Phi^s_{\bp\bk}\right)\equiv\partial_{p_y}\left(\Pi_{H^s}\right)=sp \epsilon \chi_p^s
    \label{eq:Econs}
\end{equation}
which is solved by:
\begin{align}
    \Phi_{\bp\bk}^s &= - \frac{  \sqrt{2} \epsilon}{  \Up p_x} 
    \int_{-\infty}^{p_y} \frac{q}{p}  ~\chi^s_{\bq}  ~dq_y  ~~ \text{ for } \Up p_x < 0    ; \nonumber \\
     \Phi_{\bp\bk}^s &= - \frac{ \sqrt{2} \epsilon}{\Up p_x} 
    \int_{\infty}^{p_y} \frac{q}{p}  ~\chi^s_{\bq}  ~dq_y ~~
    \text{ for } \Up p_x > 0   .
    \label{eq:sol_stat}
\end{align}
where we have used different boundary conditions depending on the sign of $U' p_x$:
\begin{align}
  \Pi_{H^s}(p_y\to -\infty)&=0 ~~ \text{ for } ~~ \Up p_x < 0 \\
  \Pi_{H^s}(p_y\to \infty)&=0 ~~ \text{ for } ~~ \Up p_x > 0 
  \label{eq:BC}
\end{align}
implying that for $\Up p_x < 0$ ($\Up p_x > 0$) the flux of helicity is from the forcing scale to $p_y\to \infty$ ($p_y\to -\infty$), i.e. to positive (negative) $p_y$. 

Solution \eqref{eq:sol_stat} is simplified when $\chi_{\bp}^s = \chi_{\bp}^{-s} = \delta(p-k_f) /(8 \pi k_f^2 \Vol)$ and leads to the small-scale energy flux:
\begin{align}
 \Pi^s_{\bp} \equiv \frac{-\Up p_x}{\sqrt{2}}\Phi_{\bp\bk}^s 
 =
 \left\{
    \begin{aligned}
     &~ 0 ~~ \text{ if }  \sigma p_y  < -\pyf \\
&   \frac{\sigma \epsilon^s_{p_x,p_z}}{2}  \frac{k_f}{p }   
 ~~ \text{ if }  - \pyf <  \sigma p_y < \pyf \\
& \sigma \epsilon^s_{p_x,p_z}    \frac{k_f}{p }  
~~ \text{ if } \sigma  p_y  > \pyf
\end{aligned}
  \right.
  \label{eq:sol_homo}
\end{align}
with $\sigma \equiv \sgn (-U' p_x)$ the relative sign of the shear and
\begin{equation}
\begin{split}
   \epsilon^s_{p_x,p_z} &\equiv  \int_{-\infty}^{\infty} \chi^s_{\bq}~ \mathbb{1}_{\omega_{\bq,-\bq-\bk}^{ss} < U'}~ dq_y = \frac{1}{8\pi k_f^2 \Vol}\int_{-\infty}^{\infty} \delta\left(q-k_f\right)  \mathbb{1}_{\omega_{\bq,-\bq-\bk}^{ss} < U'}
   ~dq_y\\&=\frac{1}{8\pi k_f^2 \Vol}\frac{2k_f}{p_y} 
   =\frac{  \epsilon  }{ 4 \pi k_f\Vol    \sqrt{k_f^2 - p_x^2 - p_z^2 } } 
   \end{split}
\end{equation}
the energy injection rate in each $(p_x,p_z)$ line in sector $s$. 
Here we have assumed that rotation is not too large, $U'/\Omega > l_f/L_y$, so that all homochiral waves are coupled to the condensate by near resonances, $\omega_{\bp,-\bp-\bk}^{ss} < U'$. 

For $U'/\Omega < l_/L_y$, 3D waves start to decouple from the 2D flow as $\Delta_{\bp}^{ss} \delta_{p- k_f} \approx 0 \Leftrightarrow \omega_{\bp\bq}^{ss} \delta_{p-k_f} < \Up $ at the forcing scale, which implies that $\Delta_{\bp}^{s,-s}\approx 0$ for all $\bp$ in the isolated mean-wave system. 
Such modes only interact with other 3D waves, and are then treated as if they are unforced in the QL system. Then, $\Phi_{\bp\bk}^s=0$ and the corresponding Reynolds stress is zero.
The decoupling of 2D-3D interactions as $\Omega \to \infty$ is discussed in \citep{gome2026waves}.
%

Solution \eqref{eq:sol_homo} reflects the conservation of helicity within each polar sector: $p \Pi^s = k_f \epsilon_{p_x,p_z}^s$ with the energy flux $ \Pi^s = - U' p_x \Phi_{\bp\bk}^s$. This shows that the conservation of helicity by sign imposes a decaying energy flux to small scales, $\Pi^s \sim (k_f/p) \epsilon_{p_x,p_z}^s$.
The sign of $\Pi^s$ depends on the sign of $-U' p_x p_y$, as follows from the choice of boundary conditions \eqref{eq:BC}: it is positive, e.g.\, when $p_y >0$ and $\Up p_x <0$, corresponding to the shearing of the modes by the mean flow.

\begin{figure}
    \centering
\includegraphics[width=0.5\columnwidth]{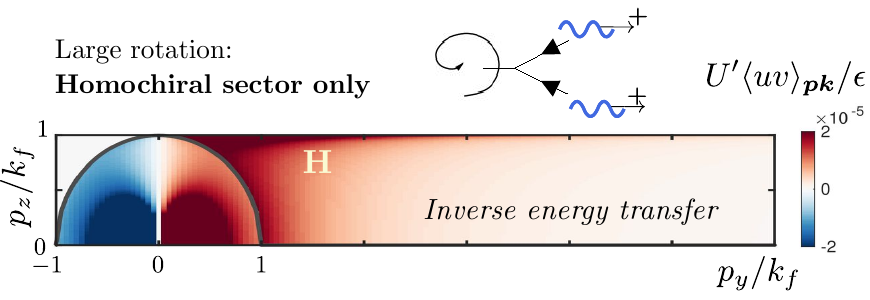}
    \caption{Reynolds stress in the spectral plane ($p_y,p_z,p_x=8\pi/L_x$), computed when interactions are homochiral (Sector H). This corresponds to the high-rotation regime.
    Advection by the shear flow generates a forward flux of wave energy, but which decays as $1/p$: energy is therefore is transferred to the mean flow ($\Up \uv_{\bp\bk} > 0$). }
    \label{fig:uv_H}
\end{figure}

For completeness we also write the corresponding Reynolds transfer (the local 2D-3D spectral energy transfer) 
\begin{align}
    \frac{\Up}{\sqrt{2}} \uv_{\bp\bk} &=
    - \frac{\Up}{\sqrt{2}} \frac{p_x p_y}{p^2} ( \Phi_{\bp\bk}^+ +  \Phi_{\bp\bk}^-)  \\
    &= \left\{
    \begin{aligned}
     &~ 0 ~~ \text{ if }  \sigma p_y  < -\pyf \\
&   \frac{\sigma \epsilon_{p_x,p_z}}{2}  \frac{k_f p_y}{p^3 }   
 ~~ \text{ if }  - \pyf <  \sigma p_y < \pyf \\
& \sigma \epsilon_{p_x,p_z}    \frac{k_f p_y}{p^3 }  
~~ \text{ if } \sigma  p_y  > \pyf
\end{aligned}
  \right.
  \label{eq:uv_homo}
\end{align}
with 
\begin{align}
\epsilon_{p_x,p_z} &= \epsilon_{p_x,p_z}^+ + \epsilon_{p_x,p_z}^- 
= \frac{\epsilon }{2\pi \sqrt{k_f^2 - p_x^2 -p_z^2} k_f \Vol}
\end{align}
the total energy injected in the $(p_x,p_z)$ layer.
This solution is shown in Fig.~\ref{fig:uv_H} in the spectral plane $(p_y,p_z)$.
Importantly, $ \frac{\Up}{\sqrt{2}} \uv_{\bp\bk} $ is positive when $\sigma p_y >0$: energy goes from the waves to the 2D condensate.
Note that this solution is restricted to homochiral near-resonances, $\omega_{\bp\bq}^{ss} < U'$ at the forcing scale $p=k_f$.

\newpage

\section{Heterochiral interactions: governing equations}
\label{app:shear}

System \eqref{eq:e_hetero_app}-\eqref{eq:C_hetero_app} when all modes $\bp$ are in Sector A describes \emph{rotation-free} interactions between 3D modes, excited at scale $l_f$, and an imposed large-scale mean flow $\bU (y)$.
This is due to the frequencies of 3D modes in this sector being too slow, so that the oscillating factor $\Delta_{\bp}^{+s,-s}\approx1$ and the interactions are not restricted. Then, rotation has no effect on the QL dynamics at leading order.
For these modes, the same dynamical system would be obtained if $\Omega$ were put to $0$ in Eq.~\eqref{eq:QLNSE}.

However, the helical basis does not diagonalize the stationary system of correlators $\Phi_{\bp}^s, \Psi_{\bp}^{+-}$. Instead, the system can be triangularized, and subsequently solved analytically, when formulating the fluid equations in terms of primary variables, as is classically done in mean-fluctuation descriptions of shear flows \citep{schmid2012stability, farrell1993optimal}.
In this section we will present the solution for the energy flux and Reynolds stress for this non-rotating, directly forced case. This solution will be useful for the treatment of modes in Sector A, which also exhibit such dynamics and can either be directly forced, or receive their energy through a flux in $\bp$ space, coming from modes in the homochiral sector.
%

We consider a non-rotating system of mean flow $\bU=U(y) \be_x$ with a \emph{given} mean shear rate $\Up$, and 3D perturbations $[u,v,w](x,y,z)$ such that $\bu\cdot\nabla\bu \ll \bu \cdot \nabla \bU$, which obeys the Rayleigh equations
\begin{align}
\partial_t \begin{bmatrix}
          \nabla^2 v \\
          \eta
           \end{bmatrix}
          =
          \begin{bmatrix}
           - (U \nabla^2 - \partial_y^2 U)\partial_x & 0 \\
           - \Up \partial_z  & - U \partial_x  
           \end{bmatrix}          
           \begin{bmatrix}
           v \\
          \eta
           \end{bmatrix}
           + 
           \begin{bmatrix}
			\nabla^2 f_y\\
			(\nabla\times \bforc)_y
           \end{bmatrix}
           \label{eq:OS}
  \end{align}
with $\eta \equiv (\nabla\times \bu)_y$ the $y$-component of vorticity (see e.g \citep{farrell1993optimal}).
Restricting Eq.~\eqref{eq:OS} to our sinusoidal mean flow
$U(y) = \Up  L_y/(\sqrt{2} \pi)  \sin( 2 \pi y/L_y), U_{\bk} = \frac{U'}{i \sqrt{2} k_y}$ and using scale separation $L_y k_f \gg 1$ (consequently, $\partial_y^2 U \ll U \nabla^2$), we obtain the simplified system in Fourier space:
\begin{align}
\partial_t ( p^2 v_{\bp})
  & =  \frac{ p_x \Up}{\sqrt{2} k_y} \Bigg[ |-\bp-\bk|^2 v_{-\bp-\bk}^* - |- \bp + \bk|^2 v_{-\bp+\bk}^* \Bigg] 
  + p^2 f_{\bp,y} \\
 \partial_t \eta_{\bp}
 &= \frac{\Up}{\sqrt{2}} \Bigg[
 \frac{p_x}{k_y}
 (\eta_{-\bp -\bk}^*  - 
 \eta_{-\bp +\bk}^* ) 
 -i p_z ( v_{-\bp-\bk}^* 
 +
 v_{-\bp + \bk}^* )\Bigg] +  g_{\bp,y} \label{eq:OS_lowk}
  \end{align}
with $k_y = 2\pi/L_y$ and $g_{\bp,y} = i p_z f_{\bp,x} - i p_x f_{\bp,z}$ and where we have used that $q_x=-p_x, q_y=-p_y \pm k_y, q_z=-p_z$.


We now write the equation for the ensemble-averaged cross-flow energy
$\langle |v_{\bp}|^2 \rangle/2$, velocity-vorticity correlation $\langle \eta_{\bp} v_{-\bp}\rangle$
and cross-flow enstrophy
$\langle |\eta_{\bp}|^2 \rangle$ at $O(k_y/p)$:
\begin{align}
    \partial_t \frac{\langle |v_{\bp}|^2 \rangle}{2}
    &= \frac{\Up}{\sqrt{2}} \Big[ p_x \partial_{p_y}
    + \frac{4 p_x p_y}{p^2} \Big]
    \Re (\vv_{\bp\bk})
    +  \epsilon \chi_{\bp,y}      \label{eq:v2} \\
    \partial_t \frac{\langle \eta_{\bp} v_{-\bp} \rangle}{2} 
    &=\frac{\Up}{\sqrt{2}} \Bigg( \Big[ p_x \partial_{p_y} 
    + 2 \frac{p_x p_y}{p^2} \Big] \etav_{\bp\bk}
       - i p_z \vv_{\bp\bk} \Bigg) 
    \label{eq:etav_dyn}\\
    \partial_t 
    \frac{\langle |\eta_{\bp}|^2 \rangle}{2}
    &= \frac{\Up}{\sqrt{2}} \Big( p_x \partial_{p_y}  \Re(  \etaeta_{\bp\bk})
    - 2 p_z 
    \Im ( \etav_{\bp\bk}) \Big) + \epsilon p^2 \chi_{\bp,y}
    \label{eq:eta2}
\end{align}
where we have defined
%
\begin{align}
    \vv_{\bp\bk} \equiv \frac{1}{2}\Big(\langle v_{\bp} v_{-\bp + \bk} \rangle + \langle v_{\bp} v_{-\bp - \bk} \rangle\Big),~~~~~
    \etav_{\bp\bk} \equiv  \frac{1}{2} \Big( \langle \eta_{\bp} v_{-\bp + \bk} \rangle + \langle \eta_{\bp} v_{-\bp - \bk} \rangle \Big), ~~~~~
    \etaeta_{\bp\bk} \equiv \frac{1}{2} \Big(\langle \eta_{\bp} \eta_{-\bp + \bk} \rangle + \langle \eta_{\bp} \eta_{-\bp - \bk} \rangle\Big)
    \label{eq:correlators}
\end{align}
and 
used that $\vv_{\bp\bk}^*= \vv_{\bp\bk} $ since $\langle v_{\bp} v_{-\bp + \bk} \rangle=\langle v_{-\bp} v_{\bp + \bk} \rangle + O(k_y/p)$, as well as
\begin{align}
   \langle a_{\bp} b_{-\bp - \bk} \rangle &=\langle b_{-\bp} a_{\bp - \bk} \rangle + O(k_y/p),\\
  k_y\partial_{p_y} \langle ab\rangle_{\bp\bk} &=
\langle a_{\bp+\bk} b_{-\bp} \rangle - \langle a_{\bp} b_{-\bp+\bk} \rangle  +  O(k_y/p) =\langle a_{\bp} b_{-\bp-\bk} \rangle
-\langle a_{\bp-\bk} b_{-\bp}\rangle+ O(k_y/p), 
\end{align}
for the pair $(a,b)$ either $(v,v)$, $(\eta,v)$ or $(\eta,\eta)$. For the forcing terms in Eqs.~\eqref{eq:correlators}, we have further used  that $\langle v_{-\bp} g_{\bp,y}\rangle = 0$ in Eq.~\eqref{eq:etav_dyn}, and have introduced the $y-$component correlator 
\begin{equation}
   \chi_{\bp,y} = (1- p_y^2/p^2) \chi_{\bp}/2 
\end{equation}
and \SG{written the $y$-enstrophy injection rate as $\langle g_{\bp,y}  g_{\bp, y}^* \rangle = (p_x^2 + p_z^2 ) \chi_{\bp} =\AF{2} p^2 \chi_{\bp,y}$}.
In steady state, Eqs.~\eqref{eq:v2}-\eqref{eq:eta2} lead to
\begin{subequations} \label{eq:corr_etav}
\begin{align}
     \partial_{p_y} \Big(p^4 
     \vv_{\bp\bk} 
     \Big)
    &= - p^4 \frac{\sqrt{2} \epsilon \chi_{\bp,y} }{ p_x \Up} \label{eq:vv} \\
     \partial_{p_y} \Big(p^2 
     \etav_{\bp\bk}  \Big)
    &= i p^2 \frac{p_z}{p_x} \vv_{\bp\bk}  \label{eq:etav} \\
     \partial_{p_y} \etaeta_{\bp\bk} 
    &=  2 \frac{p_z}{p_x}  \Im \etav_{\bp\bk}
    - \frac{\sqrt{2}\epsilon p^2 \chi_{\bp,y}}{ p_x \Up} \label{eq:etaeta} ,
\end{align}
\end{subequations}
after simple algebraic manipulations, and noting that $\Im(\vv_{\bp\bk}) $ and $\Im (\etaeta_{\bp\bk} ) $ are $ O(k_y/p)$. 

Note that, due to incompressibility, $u_{\bp}$ and $w_{\bp}$ are connected to $v_{\bp}$ and $\eta_{\bp}$ via
\begin{align}
u_{\bp} = - i \frac{p_z}{p_x^2 + p_z^2} \eta_{\bp} - \frac{p_x p_y}{p_x^2 + p_z^2} v_{\bp}, ~~~~~~~~~~
w_{\bp} = i \frac{p_x}{p_x^2 + p_z^2} \eta_{\bp} - \frac{p_y p_z}{p_x^2 + p_z^2} v_{\bp},
\end{align}
from which we can express the Reynolds stress $\uv_{\bp\bk}$ as a function of correlators $\vv_{\bp\bk}$ and $\etav_{\bp\bk}$,
\begin{align}
   \uv_{\bp\bk}
  =  - \frac{ \SG{2} p_x p_y }{p_x^2 + p_z^2} \vv_{\bp\bk}  +    \frac{\SG{2} p_z}{p_x^2 + p_z^2}  \Im \etav_{\bp\bk},
  \label{eq:uv_eta_v}
\end{align}
\AF{The total energy is given by}
\begin{equation}
    E_{\bp}= \frac{p^2}{(p_x^2+ p_z^2)}\frac{ |v_{\bp}|^2 }{\AF{2}}  + \frac{1}{(p_x^2+ p_z^2)}\frac{|\eta_{\bp}|^2 }{\AF{2}} 
\end{equation}\SG{and the energy flux is:}
\begin{align}
    &\Pi_{\bp}
    = -\frac{p_x \Up }{\sqrt{2}(p_x^2+ p_z^2)} \Big(p^2 \vv_{\bp\bk} + \etaeta_{\bp\bk} \Big).
\end{align}
as can be obtained from the
balance of kinetic energy $E_{\bp}$ (obtained from $p^2 \times \eqref{eq:vv} + \eqref{eq:etaeta}$):
\begin{align}
    \partial_t E_{\bp} + \partial_{p_y} \Pi_{\bp}
    = 
    \underbrace{
    \frac{U'}{\sqrt{2}}
    \left[  \frac{2p_x p_y }{p_x^2+p_z^2} \vv_{\bp\bk}  - \frac{2 p_z}{p_x^2+p_z^2}  \Im \etav_{\bp\bk}     \right]}_{ = - \frac{U'}{\sqrt{2}} \uv_{\bp\bk} }
   + \epsilon \chi_{\bp},
 \label{eq:energy_etav}
\end{align}
with $\uv_{\bp\bk} $ the Reynolds-stress transfer to the condensate in \eqref{eq:uv_eta_v}. In statistical steady state, Eq.~\eqref{eq:vv}, \eqref{eq:etav} and \eqref{eq:energy_etav} can be consecutively integrated over $p_y$, with the relevant spectral boundary conditions for the correlators $\vv_{\bp,\bk}$ and $\etav_{\bp,\bk}$ and the energy flux $\Pi_{\bp}$.

Primitive variables $(v_{\bp},\eta_{\bp})$ are associated to 
to the helical variables $(a_{\bp}^+, a_{\bp}^-)$ using basis \eqref{eq:helical_basis}:
\begin{align}
    v_{\bp} = v_{\bp}^+ + v_{\bp}^-, ~~~~
    \eta_{\bp} = i p v_{\bp}^+  - ip  v_{\bp}^-, ~~~~ \text{with} ~~ v_{\bp}^s = a_{\bp}^s h_{\bp,y}^s
\end{align}

For the correlators, this corresponds to applying the transformation
\begin{align}
\vv_{\bp\bk}
&= \frac{p_0^2}{2p^2} (\Phi_{\bp\bk}^+ + \Phi_{\bp\bk}^-)  ~~~+
\Re(Z) (\Psi_{\bp\bk}^{+-} + \Psi_{\bp\bk}^{-+} ) + i \Im(Z)  (\Psi_{\bp\bk}^{+-} - \Psi_{\bp\bk}^{-+})  \label{eq:transform_v2}\\
\etav_{\bp\bk} &=  p \frac{p_0^2}{2 p^2}  (\Phi_{\bp\bk}^+ - \Phi_{\bp\bk}^-) 
+ p (Z \Psi_{\bp\bk}^{+-} - Z^*
 \Psi_{\bp\bk}^{-+} )\\
\etaeta_{\bp\bk}
& = \frac{p_0^2}{2} (\Phi_{\bp\bk}^+ + \Phi_{\bp\bk}^-)
~~~ -  p^2 \Re(Z) (\Psi_{\bp\bk}^{+-} + \Psi_{\bp\bk}^{-+}) 
- i p^2 \Im(Z) 
(\Psi_{\bp\bk}^{+-} - \Psi_{\bp\bk}^{-+}) \label{eq:transform_eta2}
\end{align} 
to system \eqref{eq:e_hetero_app}-\eqref{eq:C_hetero_app}, with $p_0^2 = p_x^2 + p_z^2$ and $Z= h_{\bp,y}^s h_{-\bp,y}^{-s} = \frac{p_z^2 p_y^2 - p_x^2 p^2}{2 p^2 p_\perp^2}  + i \frac{2 p_x p_y p_z}{2 p_\perp^2 p} $.
With zero helicity injection, $\Psi_{\bp\bk}^{+-} = \Psi_{\bp\bk}^{-+ *}$ and $\Phi_{\bp\bk}^+ = \Phi_{\bp\bk}^-$, and the transformation is simplified.
Equations \eqref{eq:v2}-\eqref{eq:eta2} can then be directly obtained from Eqs.~\eqref{eq:e_hetero_app}-\eqref{eq:C_hetero_app} through this transformation (we do not detail the computation here).

Note that in sector H, where cross-chirality correlations are absent: $\Psi_{\bp\bk}^{+-}=0$, and if both helicity signs are excited isotropically ($\Phi_{\bp\bk}^+ = \Phi_{\bp\bk}^-$), the correlation between velocity and vorticity vanishes: $\etav_{\bp\bk}=\langle v^+\eta^+\rangle_{\bp\bk}+\langle v^-\eta^-\rangle_{\bp\bk} \underset{k/p \ll 1}{\approx} i p \langle v^+v^+\rangle_{\bp\bk}-ip \langle v^-v^-\rangle_{\bp\bk}=0$. 
%
\SG{Furthermore, velocity and vorticity correlators are enslaved in Sector H:
\begin{align}
    \vv_{\bp\bk} = \frac{p_0^2}{p^2} \Phi_{\bp\bk}^s = \frac{\etaeta_{\bp\bk}}{p^2}.
\end{align}
}

\section{Spectral flux across sectors}
\label{app:fluxloop}

We now combine the results for the two sectors above, where the value of 
\begin{align}
    S\equiv \frac{\Up}{\Omega}
\end{align}
dictates which heterochiral-wave interactions with the condensate are possible. Note that the rotation rate also enters in the resonant condition for the homochiral interactions, Eq.~\eqref{eq:reso_homo}. However, in what follows, we consider the regime where all homochiral-wave interactions contribute and do not start to decouple due to stringent resonances, corresponding to $S \gtrsim l_f/L_y$. Regime $S < l_f/L_y$ and the origin of this bound are analyzed in \citep{gome2026waves}

%

\begin{figure}[t]
    \centering
    \includegraphics[width=0.8\textwidth]{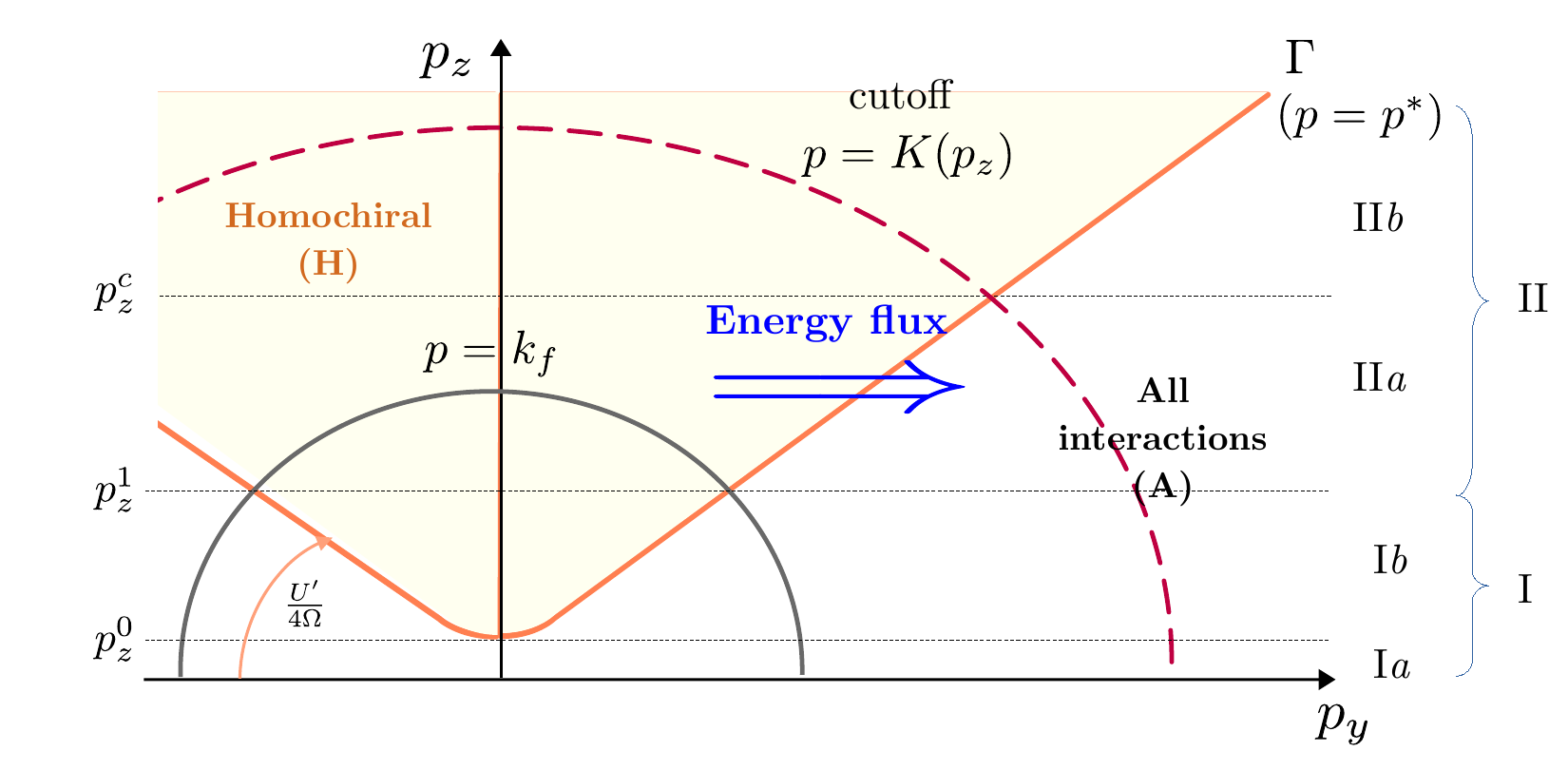}
    \caption{Schema of the different layers in the low-rotation \emph{mixed state}: modes in sector H (yellow zone) are only homochiral and solve Eq.~\eqref{eq:Econs}; modes in sector A (white zone) sustain both homochiral and heterochiral interactions and solve System~\eqref{eq:corr_etav}.
    The shear flow advects the energy from low to high $p_y$, hence from sector H to sector A.
    Cutoff curve $p=K$ is represented as a dashed red line (whose shape is arbitrary).
    Different cases emerge depending on the position of forcing and cutoff in each sector.
    Modes in layers Ia are all in sector A. In layers Ib, forcing occurs in A but some modes below the forcing shell are in H.
    In layers II (a and b), forcing occurs in H.
    Case IIa has $\KO> p^*$; 
    Case IIb has $\KO<p^*$.
    %
%
    }
    \label{fig:schema_computation}
\end{figure}

For fixed $(p_x,p_z)$, due to the resonant condition when $\tilde{s}=-s$ \eqref{eq:hetero_filter}, heterochiral interactions in Sector A emerge for $|p_y| > p_y^*$, with
\begin{align}
 p^*(p_z) &=  \frac{4 |p_z|}{S}=\frac{4 |p_z| \Omega}{U'},  ~~~~~~~~~~ p_y^*(p_x,p_z) = \sqrt{p^{*2} - p_x^2 - p_z^2},  \qquad \text{if} \qquad {p^*}^2\geq p_x^2+p_z^2  
 \label{eq:Gamma}
\end{align}
The surface $\Gamma=(p_x,p_y^*,p_z)$ delimiting sectors $H$ and $A$ is represented in Fig.~\ref{fig:schema_computation} (orange line) in the $(p_y,p_z)$ plane:
at fixed $p_z$, modes $p_y < p_y^*$ are in Sector $H$, while modes $p_y \geq p_y^*$ are in Sector $A$.

Modes in Sector H have only homochiral interactions and follow the homochiral energy balance 
\eqref{eq:Econs}. They generate a decaying small-scale energy flux \eqref{eq:sol_homo}.
Meanwhile, modes in Sector $A$ sustain both types of interactions and obey system \eqref{eq:corr_etav}. At the meeting point between the two sectors at $p_y^*$ we apply boundary conditions imposed by the continuity of the energy flux and the same-helicity correlator more generally, as well as the assumption that cross-helicity correlators vanish in sector H. 
%
%
For simplicity in this section we restrict calculations to $\sigma= - {\rm sgn} (\Up p_x )= 1$ (i.e. $p_x<0$) and multiply the end result by two (the direction of the $p_y$ integration being reversed for $\sigma = -1$ but the result of the calculations being identical).

An important intersection point of the delimiting surface $\Gamma$ is given by 
\begin{align}
 p_z^1 \equiv \frac{k_f S}{4}   : \qquad p^*(p_z)=k_f,
\end{align}
the $p_z$-value at which $p_y^*$ meets the forcing shell $p= k_f$.
Depending on the position of $p_z$ with regard to $p_z^1$, the forced modes either lie in sector $H$ or in sector $A$ (see Fig.~\ref{fig:schema_computation}), leading to two distinct cases:
\begin{itemize}
    \item[--] Layer II, $p_z>p^1_z$: Sector H is forced. The energy flux emerging from Sector H is transferred to Sector A.
    \item[--] Layer I, $p_z<p^1_z$: Sector A is forced. Energy is then transferred to sector H with increased $p_y$ before going back to sector A, or, for the lowest $p_z$, stays in sector A for all $p_y$.
\end{itemize}

Our goal here is to derive the energy transfer from the fluctuations to the mean flow, and its leading-order behavior with variable $S=U'/\Omega$ and domain geometry.
Due to the conservation of energy, this total transfer can be computed either by integrating the Reynolds stress transfer $\Up \uv_{\bp\bk}$, or by computing the difference between the total energy flux to small scales and the energy injection rate per mode $\epsilon_{p_x,p_z}$:
\begin{align}
\frac{\Up}{\sqrt{2}} \uv_{p_x,p_z} &\equiv \int dp_y \frac{\Up}{\sqrt{2}} \uv_{\bp\bk} \\ \nonumber
&= \epsilon_{p_x,p_z} - \Pi_{p_x,p_y\to \infty,p_z},
\end{align}
Thus, it is sufficient to compute $\Pi (p_x,p_y\to \infty,p_z)$ to obtain the total transfer of energy to the mean flow from a mode with given $p_x,p_z$.

The exact computation of the spectral energy flux $\Pi_{\bp}$ relies on integrating Eqs.~\eqref{eq:vv}, \eqref{eq:etav} and \eqref{eq:energy_etav} over $p_y$ in steady state, with appropriate spectral boundary conditions corresponding in Layers I or II.
%
%
In \S \ref{sec:layII}, we will show that the $\bp-$integrated energy flux in Layers II produces a leading-order forward energy extraction from the mean flow, of order $O  \left(  \frac{S}{4} \frac{L_x}{l_f}\right)$.
%
In contrast, 
Layers I concern smaller values of $p_z$, hence exhibit smaller shear tilting than Layers II, $p_z/p_x$ being at most $\sim p_z^1/p_x \sim \frac{S}{4} \frac{L_x}{l_f}$, and are a factor $S/4$ less numerous than in Layers I. Hence, the energy extracted in Layers I is expected to be highly suppressed, roughly of order $(S/4)^3 L_x/l_f$. If sub-leading corrections are kept, the energy injection from the $p_z\to 0$ modes into the mean flow, occurring because they behave as (quasi) 2D modes, should be accounted for.
%
%
We derive the exact Reynolds stress in Layers I in \S\ref{sec:layI}, and estimate their net contribution to the energy transfer, confirming that they are indeed subleading.
We will also compute further subleading terms in Layers II, in part due to discretization errors originating from our approximation of sums as integrals in the theory. These terms are included in the main text when comparing with DNS.

\subsection{Layers II}
\label{sec:layII}

\SG{Here we derive the energy contribution from Layers II ($|p_z| > k_f S/4$), which give the leading-order energy transfer.}

Layers II are forced in Sector $H$, at $p_y = \pm \pyf$ ($\pyf\equiv\sqrt{k_f^2 - p_x^2 - p_z^2}$). Modes $ k_f < p < p^* $ therefore obey the homochiral-only solution Eq.~\eqref{eq:sol_homo}.
As a result of single-sign-helicity conservation, the energy injected in each $(p_x,p_z)$ line, $\epsilon_{p_x,p_z}= \frac{\epsilon }{2\pi\Vol \pyf k_f}$, is transferred to sector A at wavenumber $p^* = \frac{4 |p_z|}{S}$, with conserved helicity flux:
\begin{align}
    \Pi^* \equiv \Pi (p^*) =  \sigma \epsilon_{p_x,p_z} \frac{k_f}{ p^*}=-\frac{U'p_x}{\sqrt{2}}(\Phi_{\bp^*\bk}^++\Phi_{\bp^*\bk}^-).
    \label{eq:BC2}
\end{align}
%
The other correlators at the boundary $p^*$ are determined by the absence of heterochiral correlations in sector H: 
\begin{align}
    &\vv_{\bp\bk}(p^*) = -\frac{\sqrt{2}}{U' p_x} \frac{p_0^2}{2 p^{*2}} \Pi^*, 
    ~~~~ \etav_{\bp\bk}(p^*) = 0, 
    \label{eq:BC1}
\end{align}

The steady-state energy flux in sector A then follows from solving successively Eqs.~\eqref{eq:vv}, \eqref{eq:etav} and \eqref{eq:energy_etav}, with boundary conditions \eqref{eq:BC1} and \eqref{eq:BC2} at $p_y=p_y^*$:
\begin{align}
 \vv_{\bp\bk} 
&= -\frac{\Pi^*}{\sqrt{2} \Up p_x } \frac{p_0^2 p^{*2}}{p^4} \\
\etav_{\bp\bk}
&=
i \frac{p_z}{p_x} \frac{1}{p^2} \int_{p_y^*}^{p_y} q^2   \vv_{\bq\bk} ~ dq_y  = - i \frac{\Pi^* ~ p_0 p^{*2}}{ \sqrt{2} \Up p_x   } 
    \frac{p_z}{p_x} \frac{1}{p^{2}} 
\Big[ 
\arctan \Big( \frac{p_y}{p_0}\Big) -
\arctan \Big( \frac{p_y^*}{p_0}\Big) 
\Big],
\label{eq:vv_eta_flux}
\end{align}
and
%
\begin{align}
    \Pi_{\bp} = \frac{ \Pi^*}{2} \Bigg[ 1 + 
    \frac{p^{*2}}{p^2} +  \frac{p^{*2}}{p_0^2} \frac{p_z^2}{p_x^2}
    \Big( \arctan \Big( \frac{p_y}{p_0}\Big) -
    \arctan \Big( \frac{p_y^*}{p_0}\Big) \Big)^2
    \Bigg].
    \label{eq:sol_II}
\end{align}
%
The $\bp$-dependent Reynolds transfer associated to Layers II is:
\begin{align}
 \frac{\Up}{\sqrt{2}}\uv_{\bp\bk}
  &=\sqrt{2}\Up \left(- \frac{p_x p_y }{p_0^2} \vv_{\bp\bk}  + \frac{p_z}{p_0^2}  \Im \etav_{\bp\bk} \right)  \\
  &=
  \left\{
  \begin{aligned}
  & k_f \epsilon_{p_x,p_z} \frac{p_y}{2p^3}, ~~~~~
   -\pyf < p_y < \pyf \\
 & k_f \epsilon_{p_x,p_z}  \frac{p_y}{p^3}, ~~~~~
   \pyf < p_y < p_y^*  \\
&    k_f \epsilon_{p_x,p_z}
  \Bigg[ \frac{p^* p_y}{p^4} 
  - \frac{p_z^2}{p_x^2} \frac{p^*}{p_0} \frac{1}{p^2}
  \Big(   \arctan \Big(\frac{p_y}{p_0} \Big)  
    - \arctan \Big(\frac{p_y^*}{p_0}   
  \Big) \Big)
  \Bigg], ~~~~~ p_y > p_y^*  
    \end{aligned}
\right.  
 \label{eq:uv_py_2}
\end{align}

With zero helicity injection, and if all homochiral-wave interactions are resonant ($U'/\Omega > l_f/L_y$),  the total energy balance between the small-scale flux $\Pi_{\bp}$ \eqref{eq:sol_II} and the transfer to the mean flow from layers II is therefore:
\begin{align}
\frac{\Up}{\sqrt{2}} \uv_{p_x,p_z}^{\rm II} \equiv \int dp_y \frac{\Up}{\sqrt{2}} \uv_{\bp\bk}^{\rm II} 
&= \epsilon_{p_x,p_z} - \Pi_{p_x,K_y,p_z}
= \underbrace{\epsilon_{p_x,p_z} - \Pi^* }_{ \text{energy input from H}} +  \underbrace{ \Pi^*  -\Pi_{p_x,K_y,p_z} }_{ \text{energy extraction from A}}
\label{eq:uvII_flux}
\end{align}
with $K_y$ a wavenumber cutoff related to the breaking of the QL approximation, to be discussed below. 
In the limit $K_y \to \infty$, we therefore obtain the \emph{exact} energy transfer due to the $(p_x,p_z)$ layer:
\begin{align}
 \Aboxed{
\frac{\Up}{\sqrt{2}} \uv_{p_x,p_z}^{\rm II} 
&= \epsilon_{p_x,p_z} \Bigg[
  1
   - \frac{k_f}{2p^*} 
   - \frac{p_z^2}{p_x^2}  \frac{k_f p^*}{2p_0^2}
   \Bigg(
  \frac{\pi}{2}
   - \arctan \Big( \frac{p_y^*}{p_0} \Big) 
   \Bigg)^2
 \Bigg], } ~~
 p^* = \frac{4 |p_z|}{S}, ~ p_y^*=(p^{*2}-p_0^2)^{\frac{1}{2}}, p_0^2= p_x^2 + p_z^2 
 \label{eq:uvII_2}
\end{align}
with
$ \epsilon_{p_x,p_z} = \frac{\epsilon }{2\pi \sqrt{k_f^2 - p_x^2 -p_z^2} k_f \Vol} $ the injection rate in the $(p_x,p_z)$ line.

The total energy transfer to the condensate follows from
\begin{align}
   \frac{\Tthree( S)}{\epsilon_{\rm 3D}} = \frac{1}{\epsilon_{\rm 3D}} \sum_{p_x,p_z} \frac{U'}{\sqrt{2} } \uv_{p_x,p_z}^{\rm II} 
    \equiv \mathcal{I}_{\rm II} 
    + \mathcal{D}_{\rm II}^{(2)}  
    + \mathcal{D}_{\rm II}^{(1)},
\end{align}
with $\mathcal{I}_{\rm II}$, $ \mathcal{D}_{\rm II}^{(2)}$ and $ \mathcal{D}_{\rm II}^{(1)}$ denoting, respectively, the integral of the first, second and last term in \eqref{eq:uvII_2}.
$\epsilon_{\rm 3D}$ denotes the energy injection rate into 3D modes $p_z\neq 0$: 
\begin{align}
    \epsilon_{\rm 3D} = \epsilon \left( 1- \frac{l_f}{2L_z} \right).
\end{align}

Introducing rescaled variables
\begin{align}
    a \equiv \frac{S}{4}, ~~~
    x= \frac{p_x}{k_f}, ~~~
    z= \frac{p_z}{k_f},\\
    \nu_x = \frac{2\pi}{L_x},~~~
    \nu_z = \frac{2\pi}{L_z},
\end{align}
the energy flux in layers II is summed over
\begin{align}
 &   -\sqrt{1 - a^2 } \leq x \leq  ~ -\nu_x/k_f, ~~~
a  \leq  z <  \sqrt{1 - x^2}.
\end{align}
Here we restrict the computation to sectors $p_x<0, ~ p_z>0$ ($x<0,z>0$) and premultiply the result by 4 to account for the three other $(p_x,p_z)$ sectors.
the total energy transfer is written in the continuous limit as:
%
%
%
%
\begin{align}
   \frac{\Tthree}{\epsilon_{\rm 3D}}
& = \int \dd x \dd z \frac{2}{\pi\sqrt{1-x^2-z^2}}\left[1-\frac{a}{2z}
 -\frac{z^3}{2a\,x^2(x^2+z^2)}
 \arcsin^2\!\left(a\frac{\sqrt{x^2+z^2}}{z}\right)\right].
 \label{eq:exact}
\end{align}

\subsubsection{Energy extraction}

We first consider the 
last term in \eqref{eq:uvII_2}, corresponding to energy extraction due to vortex tilting: 
\begin{align}
    \mathcal{D}_{\rm II}^{(1)}=-\frac{1}\pi \int_0^{\sqrt{1-a^2}}\frac{K(x,a)}{x^2}\dd x
\end{align}
with
\begin{align}
  K(x,a)
 = \frac{1}{a}\int_a^{\sqrt{1-x^2}} \dd z
 \frac{z^3}{(x^2+z^2)\sqrt{1-x^2-z^2}}
 \arcsin^2\!\left(a\frac{\sqrt{x^2+z^2}}{z}\right).
 \label{eq:Kza}
\end{align} 
This term is singular at $x=0$ ($p_x=0$) since $K(0,a)\sim O(1)$. As a result, the discrete sum over $p_x$ cannot be approximated with a Riemann integral. To treat this singularity first rewrite $K(x,a)$ as:
\begin{equation}
K(x,a)=K(0,a)+[K(x,a)-K(0,a)],
 \label{eq:Ksplit}
\end{equation}
with 
\begin{align}
 K(0,a)
 &= \frac{1}{a}(\arcsin a)^2\int_a^1\frac{z ~ \dd z}{\sqrt{1-z^2}}
 \nonumber\\
 &=\frac{1}{a}\sqrt{1-a^2}\, (\arcsin a)^2 \approx a+O(a^3).
 \label{eq:K0}
\end{align}
when expanding in $a\ll1$. Note that we are considering a continuous approximation in $p_y$, taking the leading order in an expansion in $l_f/L_y$, but are allowing $p_x,p_z$ to be discrete. For consistency we should also be keeping only the leading order in $l_f/L_z,l_f/L_x$ in the following. This gives
\begin{align}
    \mathcal{D}_{\rm II}^{(1)}=-\frac{1}\pi \int_0^{\sqrt{1-a^2}}\frac{K(0,a)}{x^2}\dd x -\frac{1}\pi \int_0^{\sqrt{1-a^2}}\frac{K(x,a)-K(0,a)}{x^2}\dd x 
\end{align}
where the last term in the RHS above is no longer singular since $K(x,a)-K(0,a)\sim O(x^2)$ for $x\sim 0$, and can therefore be evaluated using integration over $x$. The first term in the RHS is the leading order, singular contribution which requires using summation.  

\paragraph{\underline{Contribution from $p_x\approx 0$}:}
Using the decomposition above we now write the sum explicitly for the singular contribution, dominated by $x\approx 0$:
\begin{align}
    \mathcal{D}_{\rm II}^{(1)}
    &= - \frac{1}{\pi} \frac{k_f}{\nu_x} \sum_{n=1}^{ \lfloor (k_f/\nu_x) \sqrt{1-a^2} \rfloor} \frac{1}{n^2} 
    K \left(0,a \right)
\end{align}
where $x=\nu_x/k_f \cdot n$ and $n\in \mathbb{Z}$
while the $z$ ($p_z$) sum entering $K$ was approximated by its Riemann integral in the continuum limit $l_f/L_z \ll 1$.\\

%

Using that
$\sum_{1}^N \frac{1}{n^2} \underset{N \gg 1}{\sim} \frac{\pi^2}{6} - \frac{1}{N} + O(\frac{1}{N^2})$:
\begin{align}
    \frac{k_f}{\nu_x} \sum_{n=1}^{ \lfloor (k_f/\nu_x) \sqrt{1-a^2} \rfloor} \frac{1}{n^2}\approx \frac{k_f}{\nu_x}\frac{\pi^2}{6}-\frac{1}{\sqrt{1-a^2}}\approx \frac{L_x}{l_f}\frac{\pi^2}{6}-1
\end{align}
when expanding in $a \ll 1$.
The energy transfer to small scales resulting from this singular contribution is therefore
\begin{align}
    \Aboxed{\mathcal{D}^{(1)}_{\rm II} = - a \left( \frac{\pi}{6} \frac{L_x}{l_f} - \frac{1}{\pi} \right)}
    \label{eq:Dlead_2}
\end{align}

\paragraph{\underline{$O(a)$ corrections to $\mathcal{D}_{\rm II}^{(1)}$}:}
The regular part results in $O(a)$ corrections to Eq.~\eqref{eq:Dlead_2}, which can now be estimated with the Riemann integral
\begin{align}
\mathcal{D}^{(1)}_{\rm II} 
&=  - a \left( \frac{\pi}{6} \frac{L_x}{l_f} - \frac{1}{\pi} \right) - 
\frac{1}{\pi }
 \int_0^{\sqrt{1-a^2}}
 \frac{K(x,a)-K(0,a)}{x^2}\,\dd x.
 \label{eq:DregK}
\end{align}
We are interested in the leading order in $a$, so we can expand the $\arcsin(\cdot)$ in $K(x,a)$:
\begin{equation}
 K(x,a)\approx \int_a^{\sqrt{1-x^2}}\frac{a z \dd z}{\sqrt{1-x^2-z^2}}   = a \sqrt{1-a^2-x^2}+O(a^2),
\end{equation}
Such an expansion is only possible when $a\sqrt{x^2+z^2}/z\ll 1$. However the region where $z\sim O(a)$ while $x\sim O(1)$, where this approximation breaks, will contribute at most at order $a^2$ (e.g. due to the $z^3$ factor) which is beyond the approximation needed here.

Using that 
\begin{equation}
    \int \dd x \frac{\sqrt{A^2-x^2}}{x^2} = -\frac{\sqrt{A^2-x^2}}{x}-\arcsin{\frac{x}{A}},
\end{equation}
we obtain that
\begin{align}
\mathcal{D}^{(1)}_{\rm II} 
&=  -  a \left( \frac{\pi}{6} \frac{L_x}{l_f} - \frac{1}{\pi} \right) - \frac{ a}{\pi }
 \left(1-\frac\pi2\right)
 +O( a^2) \\
 \Aboxed{
 &= -   a \left( \frac{\pi}{6} \frac{L_x}{l_f} - \frac{1}{2} \right) +O( a^2).}
\end{align}

\paragraph{\underline{Discretization corrections}}
The integral in Eq.~\eqref{eq:K0} should be treated with care when considered as an approximation to a sum over $z$, since the weight at the last point $z=1$ diverges (this is due to the continuous in $p_y$ treatment of energy injection). This means the sampling points need to be defined so as to evaluate the sum one grid point away from the boundary, and that discretization errors then become important (though still subleading). 
%
The latter can be estimated via the Euler-Maclaurin formula at the end point $p_z=k_f- \nu_z$ (with $p_x=0$), corresponding to $z=1- \nu_z/k_f$:
\begin{align}
    K(0,a) &= \frac{\arcsin^2 a}{a}
    \left[ \sqrt{1-a^2} + \frac{\nu_z}{2 k_f} \frac{1 - \nu_z/ k_f }{\sqrt{1- \left( 1 - \nu_z/k_f \right)^2}} + O \left(\frac{\nu_z}{k_f} a \right) \right] \\
    &=\frac{\arcsin^2 a}{a}
    \left[ \sqrt{1-a^2} + \frac{1}{2}\sqrt{\frac{\nu_z}{2k_f}}
    \right] \simeq
    a 
    \left[ 1 + \frac{1}{2}\sqrt{\frac{l_f}{2L_z}}
    \right]
\end{align}
This results in a subleading contribution in $a \frac{L_x}{l_f}\sqrt{\frac{l_f}{2L_z}}$ to $\mathcal{D}_{\rm II}^{(1)}$, non negligible for our DNS parameters:
\begin{align}
\mathcal{D}^{(1)}_{\rm II} 
&= -   a \left( \frac{\pi}{6} \frac{L_x}{l_f} - \frac{1}{2}  + \frac{\pi}{12} \frac{L_x}{l_f}\sqrt{\frac{l_f}{2L_z}} 
\right)
+O \left( a^2 \right) + O \left(  a^2 \frac{L_x}{l_f} \frac{l_f}{L_z} \right).
\end{align}

However, to keep our approximation consistent we must only retain the leading order corrections in $l_f/L_x, l_f/L_z$, resulting in energy extraction equal to

\begin{align}
\mathcal{D}^{(1)}_{\rm II} 
&= -   a  \frac{\pi}{6} \frac{L_x}{l_f}
\end{align}

\subsubsection{Energy input}
Here we consider the energy input coming from the homochiral sector:
\begin{align}
   \frac{\Tthree}{\epsilon_{\rm 3D}}
& = \int \dd x \dd z \frac{2}{\pi\sqrt{1-x^2-z^2}}\left[1-\frac{a}{2z}
 \right].
\end{align}
We see that at leading order in $a$ and $l_f/L_x$ only the first term will contribute, resulting in
\begin{equation}
   \mathcal{I}_{\rm II}= \int \dd x \dd z \frac{2}{\pi\sqrt{1-x^2-z^2}}=\frac{2}{\pi}\int_a^1 dz\int_0^{\sqrt{1-z^2}}\frac{1}{\sqrt{1-x^2-z^2}}=\int_a^1 dz=1-a \approx 1
\end{equation}
where we have taken the continuum limit throughout and have used that $\int dx  \frac{1}{\sqrt{b^2 - x^2}} = \arctan \Big( \frac{x}{\sqrt{b^2 - x^2}}\Big)$.


%

\paragraph{\underline{Discretization and sub-leading corrections}}
In a discrete setting, we need to take into account the difference between $a$ and the closest grid point in $z$, introducing
\begin{align}
    m^* (a) \equiv \left\lfloor \frac{a L_z}{l_f} \right\rfloor,
\end{align}
the row number separating layer I from layer II (corresponding to $p_z^1= \nu_z m^*$, $z=\nu_z/k_f m^*$). Furthermore, modes $p_x=0$ must be removed from the sum, corresponding to a subtraction of the energy injection rate into the ring $p_x=0$, $\epsilon l_f/(2L_x)$, which do not interact with the mean flow. Therefore,
\begin{align}
    \mathcal{I}_{\rm II} &=
     \sum_{m^*(S) +1 }^{\lfloor k_f/\nu_z \rfloor} \frac{\nu_z}{k_f}  \SG{-\frac{l_f}{2L_x} }
    =  1- \SG{\frac{l_f}{2L_x} }- \frac{l_f}{L_z} m^* (a) 
\end{align}
 Note that this neglects subleading terms in $a l_f/L_x$ corresponding to the subtraction of $p_x=0$ modes.

Finally, we consider the remaining term which is of order $a$, the second term in the flux in \eqref{eq:uvII_2}, which we now sum first in $p_z$ and second in $p_x$:
\begin{align}
    \mathcal{D}_{\rm II}^{(2)}
    &=- \frac{2 \nu_x\nu_z}{\pi k_f} 
    \sum_{p_z= \nu_z (m^* +1) }^{\sqrt{k_f^2-\nu_x^2}}
    \sum_{p_x = \nu_x }^{\sqrt{k_f^2 - p_z^2}}
    \frac{1}{\sqrt{k_f^2 - p_x^2 - p_z^2}} \frac{a}{2} \frac{k_f}{p_z}
\end{align}
Replacing the $p_x$ sum by its integral, and using that $\int \frac{1}{\sqrt{1- x^2 - b^2}}dx = \arctan(x/\sqrt{1-b^2})$, we obtain:
\begin{align}
    \Aboxed{
    \mathcal{D}_{\rm II}^{(2)}
    &= - \frac{ a}{2}
    H_{m^* +1} + O \left( a \frac{l_f}{L_x} \right),}
    ~~~ \text{  with  }
    H_{m^* +1} = \sum_{m= m^*+1}^{\lfloor \frac{k_f}{\nu_z} \rfloor} \frac{1}{m}
\end{align}
the harmonic sum starting from $m^*(a) +1$.

\subsubsection{Total transfer in layer II}
%
At leading order in $L_x /l_f$ and $a$ we obtain 
\begin{align}
   \frac{\Tthree^{\rm II}}{\epsilon_{\rm 3D}}  = \frac{\mathcal{I}_{\rm II} + \mathcal{D}_{\rm II}^{(1)} + \mathcal{D}_{\rm II}^{(2)}}{\epsilon_{\rm 3D}} &
   \approx
 1  - a   \frac{\pi}{6} \frac{L_x}{l_f} 
\end{align}

For later comparison with semi-discrete numerical integration we also provide the formula including subleading correction in $a,l_f/L_x, l_f/L_z$ (but not accounting for subleading corrections in $l_f/L_y$)

%
\begin{align}
   \frac{\Tthree^{\rm II}}{\epsilon_{\rm 3D}}  = \frac{\mathcal{I}_{\rm II} + \mathcal{D}_{\rm II}^{(1)} + \mathcal{D}_{\rm II}^{(2)}}{\epsilon_{\rm 3D}} &
   \approx
 1 - \frac{l_f}{2 L_x}  - a \Bigg(  \frac{\pi}{6} \frac{L_x}{l_f} -\frac{1}{2}
 + \frac{1}{2} H_{m^* + 1}
 + \frac{\pi}{12} \frac{L_x}{l_f}\sqrt{\frac{l_f}{2L_z}} 
 \Bigg) 
 \label{eq:TII_leading}, ~~~ a= \frac{U'}{4 \Omega} 
\end{align}

The energy transfer between the waves and the 2D condensate in \eqref{eq:TII_leading} decreases with decreasing rotation (i.e.\ increasing $a= \Up/(4\Omega)$), due to the growing number of modes in sector A that break conservation of $H_{\bp}^{\pm}$ and extract mean-flow energy.
This energy extraction is larger as $L_x/l_f$ increases, due to vortex-tilting preferentially energizing modes with large $p_z/p_x \sim L_x/l_f$, via the term proportional to $(p_z/p_x)^2$ in \eqref{eq:uvII_2}.

\subsection{Layers I
}
\label{sec:layI}
Layers I require a slightly different treatment of the energy flux between sectors, since the energy input lies in sector A rather than sector H. However, this only changes the boundary conditions but not the functional form of the energy extraction term. In particular, the main contribution to energy extraction comes from small $p_x$ and the largest available $p_z$, $\propto p_z^2$. Since $p_z/k_f<a\ll 1$ in layer I, this extraction term is at most of order $O(a^2 L_x/l_f)$. Additionally, these layers contain only a few modes, $l_f/L_z<p_z/k_f<a$ so its total contribution is at most $O(a)$. All in all, the contribution from layers I to the small-scale energy flux does not enter at leading order.

For later use in numerical solutions we here detail the steady state solutions in layers I and their leading order approximate forms. 
Here, two subcases emerge, depending  if $(p^*(p_z))^2>p_x^2+p_z^2$ or not (i.e. if $p_y^*>0$), determined by the position of $p_z$ compared to
\begin{align}
 p_z^{0} \equiv  \frac{|p_x|}{ (\frac{1}{a^2} -1 )^{1/2}}.
\end{align} 
\begin{itemize}
    \item[-] Layers Ib: for $p_z> p_z^0$, forcing occurs in sector A, energy is then transmitted to sector H at $p_y=-p_y^*$, and goes back to A again at $p_y=p_y^*$.
    \item[-] Layers Ia: when $p_z < p_z^0$, then $p_y^*=0$ and the corresponding $p_z$-layer lies entirely in sector A, corresponding to the layer Ia in Fig.~\ref{fig:schema_computation}.
\end{itemize}

%
%
Note that, as we did for layers II in \ref{sec:layII}, we only focus on increasing $p_y$ and $ p_x < 0$ ($\sigma= \sgn{(- p_x U')} >0$), the other relevant quadrant being obtained by taking $p_x \to - p_x$ and $p_y \to -p_y$ (recall that quadrants $p_x p_y >0$ are not energized by the shear-flow advection).

%

%

\subsubsection{Reynolds stress in Layers I}

\paragraph{Layers Ib. }

 With increasing $p_y$, energy is first injected into Sector $A$ at $-\pyf$, then flows into Sector $H$ (for $-p_y^* < p_y <  p_y^*$), before going back to Sector $A$ (for $p_y> p_y^*$) where the 3D modes are forced again (at $p_y= \pyf$). 

Modes $p_y < - p_y^*$ in Sector A receive an energy flux $\epsilon_{p_x,p_z}/2$ from the forcing shell at $-\pyf$. 
The corresponding velocity and vorticity correlators for $ - \pyf < p_y < - p_y^*$, solving Eqs.~\eqref{eq:vv}-\eqref{eq:etav}, are therefore 
\begin{align}
  \vv_{\bp\bk}
   &= \frac{\epsilon_{p_x,p_z}}{2 \sqrt{2} |p_x \Up|}\frac{p_0^2 k_f^2}{p^4}, \nonumber\\ 
  \etav_{\bp\bk}
   &= i \frac{\epsilon_{p_x,p_z}}{2\sqrt{2} |p_x \Up|} p_0 \frac{p_z}{p_x} \frac{k_f^2}{p^2} 
\Bigg[ 
\arctan \Big( \frac{p_y}{p_0}\Big) +
\arctan \Big( \frac{\pyf}{p_0}\Big) 
\Bigg], ~~~~~~~   - \pyf <  p_y < - p_y^* 
\end{align}
compare to equations \ref{eq:vv_eta_flux} with $\Pi^*=\epsilon_{p_x,p_z}/2$ and $p^*=k_f$ where we have used that $- U' p_x >0$ to write the expressions more generally 
(In the other quadrant, the order of integration will be flipped, hence is writing $|U' p_x|$ gives the more genera; expression). 

Modes $-p_y^* < p_y < p_y^*$ are in Sector H, satisfying the helicity-by-sign conserving dynamics. In addition, we need to supply the boundary conditions when going from sector A to sector H for the correlators $\Phi_{\bp\bk}^s(\bp^*)$ and $\Psi_{\bp\bk}^{+-}(\bp^*)$. We assume the latter average out to zero in sector H, falling sharply to 0 at $p=p^*$ from their value in sector A. For the former we use continuity of the energy flux and write  $\Phi_{\bp\bk}^s(\bp^*)= \frac{p^{*2}}{p_0^2}\vv_{\bp\bk} (p_y^*)$ from Eq.~\eqref{eq:transform_v2}. This gives the solution 
\begin{align}
    \Phi_{\bp\bk}^s = \frac{\epsilon_{p_x,p_z}^s}{2 \sqrt{2} |p_x \Up  | }  \frac{k_f^2}{p^* p},
    \label{eq:flux_H_I}
\end{align}
for $-p_y^* < p_y < p_y^*$.
%
Note that Eq.~\eqref{eq:flux_H_I} exhibits $p_y \to -p_y$ symmetry (inherited from Eq.~\eqref{eq:Econs}), so the value of the correlator $\vv_{\bp\bk}$ at the transition point from sector H back to sector A at $p_y^*$ equals that at the entry point at $-p_y^*$. Note that the correlator  $\etav_{\bp\bk}(p_y^*)=0$ on the other hand. .
%


For $p_y > p_y^*$, modes are again in Sector A, and since the equation for the $\vv_{\bp\bk}$ correlator is independent of $\etav_{\bp\bk},\etaeta_{\bp\bk} $, it does not feel the passage through sector H and follows Eq.~\eqref{eq:corr_etav} with forcing at $p_y=\pyf$:
\begin{align}
&\vv_{\bp\bk} =
\vv_{\bp\bk}  (-p_y^*) \frac{p^{*4}}{p^4}
  - \frac{\sqrt{2} \epsilon}{2 \Up p_x p^4} \int_{p_y^*}^{p_y} q^4 (1 - \frac{q_y^2}{q^2} ) \chi_{\bq} dq_y  =
\left\{
    \begin{aligned}
    &\frac{\epsilon_{p_x,p_z}}{2\sqrt{2} |\Up p_x|} \frac{p_0^2 k_f^2}{p^4}, ~~~ p_y^* < p_y < \pyf \\
    &\frac{\epsilon_{p_x,p_z}}{ \sqrt{2} |\Up p_x|} \frac{p_0^2 k_f^2}{p^4}, ~~~ p_y > \pyf
    \end{aligned}
\right.
\end{align}

while the correlator $\etav_{\bp\bk}$ is then given by 
\begin{align}
&\etav_{\bp\bk} = 
    \frac{i}{p^2} \frac{p_z}{p_x} \int_{p_y^*}^{p_y}  q^2 \langle vv \rangle_{\bq\bk} dq_y  
=
\left\{
    \begin{aligned}
    &  i \frac{\epsilon_{p_x,p_z} ~  p_0 }{2 \sqrt{2} |\Up p_x |} \frac{p_z}{p_x} \frac{k_f^2}{p^2} \Bigg[
    \arctan (\frac{p_y}{p_0})
    - \arctan (\frac{p_y^*}{p_0} ) 
   \Bigg],  \nonumber\\
   & ~~~~~~~~~~~~~~~~~~~~~~~~~~~~~~~~~~~~~~~~~~
    ~~~~ p_y^* < p_y < \pyf \\
    &i \frac{\epsilon_{p_x,p_z} p_0}{ \sqrt{2}|\Up p_x |\pyf} \frac{p_z}{p_x} 
    \frac{k_f^2}{p^2} \Bigg[ \arctan(\frac{p_y}{p_0}) 
-   \frac{1}{2} \arctan(\frac{p_y^*}{p_0}) 
- \frac{1}{2} \arctan(\frac{\pyf}{p_0}) 
     \Bigg], \nonumber \\
     &~~~~~~~~~~~~~~~~~~~~~~~~~~~~~~~~~~~~~~~~~~~~~~  p_y > \pyf
    \end{aligned}
\right.
\end{align}

Using Eq.~\eqref{eq:uv_eta_v}, this leads to the Reynolds stress in layers Ib:
\begin{align}
&\uv_{\bp\bk} =\left\{
    \begin{aligned}
    &
    \frac{\epsilon_{p_x,p_z} k_f^2}{ \sqrt{2} U'  } \Bigg[
\frac{ p_y}{p^4}
-\frac{p_z^2}{p_x^2} \frac{1}{p_0}  \frac{1}{p^2}
\Big( \arctan \Big( \frac{p_y}{p_0} \Big) 
+ \arctan \Big( \frac{\pyf}{p_0} \Big) \Big) 
\Bigg],  ~~~~~
- \pyf < p_y < -p_y^* \\
&\frac{\epsilon_{p_x,p_z} k_f^2}{ \sqrt{2}  U'  } \Bigg[
\frac{p_y}{p^4} - \frac{p_z^2}{p_x^2} \frac{1}{p_0} \frac{1}{p^2} 
\Big(
 \arctan (\frac{p_y}{p_0}) 
   - \arctan (\frac{p_y^*}{p_0} )
 \Big) \Bigg] , ~~~ p_y^* < p_y < \pyf  \\
&\frac{\sqrt{2}  \epsilon_{p_x,p_z} k_f^2}{  U'  }\Bigg[
\frac{p_y}{p^4} - \frac{p_z^2}{p_x^2} \frac{1}{p_0} \frac{1}{p^2} 
\Big(
 \arctan (\frac{p_y}{p_0})
  - \frac{1}{2}\arctan (\frac{p_y^*}{p_0} )
  - \frac{1}{2} \arctan (\frac{\pyf}{p_0} )
 \Big)
 \Bigg], ~~~~~~~ p_y > \pyf
 \label{eq:uv_py_I}
\end{aligned}
\right.
\end{align}
In Fig.~2a in the main text, we visualize $\Up \uv_{\bp\bk}$ in the $(p_y,p_z)$ plane by showing Eq.~\eqref{eq:uv_py_I} in layers I, completing the Reynolds stress in layers II written in Eq.~\eqref{eq:uv_py_2}. In the figure, we fix $p_x= 8 \pi/L_x$ and $S = 1$.

From Eq.~\eqref{eq:uv_py_I}, the $p_y-$integrated Reynolds transfer up to $p_y \to \infty$ is
%
\SG{
\begin{align}
     & \frac{\Up}{\sqrt{2}}\uv_{p_x,p_z}^{\rm I}
 =
  \frac{\epsilon_{p_x,p_z}}{2 } \Bigg[1  
 -   \frac{p_z^2}{p_x^2} \frac{k_f^2}{2 p_0^2} 
 \Bigg[
 \Big(
 \frac{\pi}{2}
 -\arctan (\frac{p_y^*}{p_0})  \Big)^2 
 + \Big(
 \frac{\pi}{2}
 -\arctan (\frac{p_y^f}{p_0})  \Big)^2 
+ \Big(  \arctan (\frac{\pyf}{p_0}) 
 -\arctan (\frac{p_y^*}{p_0})  \Big)^2
 \Bigg] \Bigg]
 \label{eq:uvIb_exact}
\end{align}
}
with $p_y^* = \sqrt{ (|p_z|/a)^2 - p_0^2 }$.

Note here the prefactor $1/2$, which governs the injection of energy for $p_z\to0$:
quasi-2D layers approximately conserve horizontal enstrophy (driving an inverse transfer of horizontal energy $|| u_{\perp}||^2$) and also exhibit a forward cascade of the vertical component $w$, which produces small-scale energy (but does not extract energy from the mean flow). Energy is therefore equally distributed in the two dynamical variables, $w^2$ and $|| u_{\perp}||^2$.
Note, finally, that the Reynolds stress is continuous between layers I and layers II (compare Eq.~\eqref{eq:uvIb_exact} to Eq.~\eqref{eq:uvII_2} at $p_z=p_z^1$, where $p^* = k_f$. This will be reflected in the total energy transfer being continuous in variable $a$.

\paragraph{Layers Ia (Purely non-rotating shear flow).}

Here, sector H is not crossed and modes lie entirely in sector A:
the layers are fully unconstrained by rotation, corresponding to a non-rotating shear flow.
The $\vv$ correlator has the same expression as above (with energy flux coming from the forcing),
\begin{align}
&\vv_{\bp\bk} =
  - \frac{\sqrt{2} \epsilon}{2 \Up p_x p^4} \int_{-\pyf}^{p_y} q^4 (1 - \frac{q_y^2}{q^2} ) \chi_{\bq} dq_y  =
\left\{
    \begin{aligned}
    &\frac{\epsilon_{p_x,p_z}}{2\sqrt{2} |\Up p_x|} \frac{p_0^2 k_f^2}{p^4}, ~~~ -\pyf < p_y < \pyf \\
    &\frac{\epsilon_{p_x,p_z}}{ \sqrt{2} |\Up p_x|} \frac{p_0^2 k_f^2}{p^4}, ~~~ p_y > \pyf
    \end{aligned}
\right.
\end{align}
and generates cross-flow vorticity, feeding the $\etav$ correlation (which does take a different form):
\begin{align}
&\etav_{\bp\bk} = 
    \frac{i}{p^2} \frac{p_z}{p_x} \int_{-\pyf}^{p_y}  q^2 \langle vv \rangle_{\bq\bk} dq_y  
=
\left\{
    \begin{aligned}
    &  i \frac{\epsilon_{p_x,p_z} ~  p_0 }{2 \sqrt{2} |\Up p_x |} \frac{p_z}{p_x} \frac{k_f^2}{p^2} \Bigg[
    \arctan (\frac{p_y}{p_0})
    + \arctan (\frac{\pyf}{p_0} ) 
   \Bigg],  \nonumber\\
   & ~~~~~~~~~~~~~~~~~~~~~~~~~~~~~~~~~~~~~~~~~~
    ~~~~ -\pyf < p_y < \pyf \\
    &i \frac{\epsilon_{p_x,p_z} p_0}{ \sqrt{2}|\Up p_x |} \frac{p_z}{p_x} 
    \frac{k_f^2}{p^2} \arctan(\frac{p_y}{p_0}), \nonumber ~~~~~~~  p_y > \pyf
    \end{aligned}
\right.
\end{align}
This results in the total Reynolds stress:
\begin{align}
     & \frac{\Up}{\sqrt{2}}\uv_{p_x,p_z}^{\rm I}
 = \sqrt{2} U'
 \int_{-\pyf}^\infty  \left(- \frac{p_x p_y }{p_0^2} \vv_{\bp\bk}  + \frac{p_z}{p_0^2}  \Im \etav_{\bp\bk} \right)
 =
  \frac{\epsilon_{p_x,p_z}}{2 } \Bigg[1  
 -   \frac{p_z^2}{p_x^2} \frac{k_f^2}{ p_0^2} 
 \Bigg[ \Big(
 \frac{\pi}{2} \Big)^2
 + \arctan^2 \left(\frac{p_y^f}{p_0}\right) 
 \Bigg] \Bigg]
 \label{eq:uvIa_exact}
\end{align}


%

\begin{figure}[t]
    \centering
    \includegraphics[width=0.9\columnwidth]{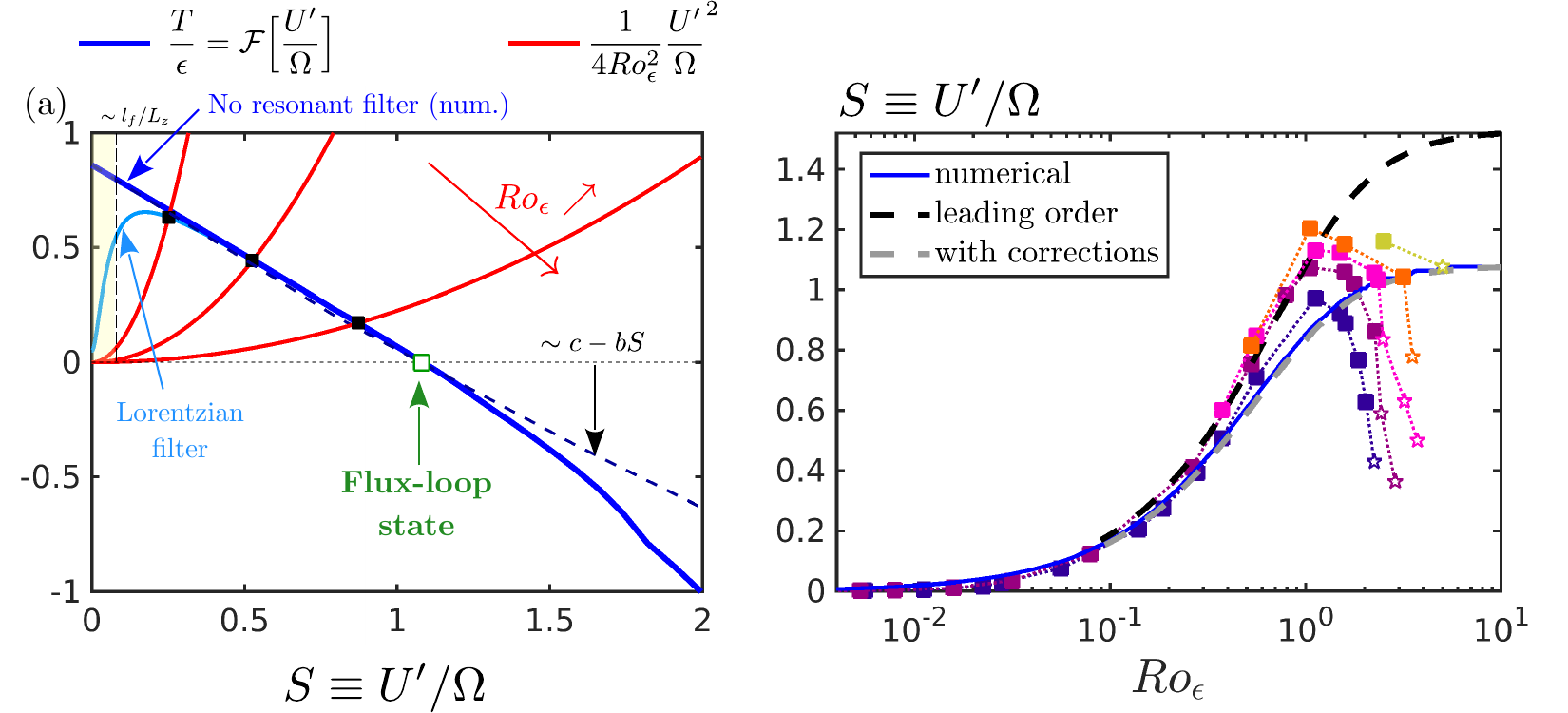} ~
    \caption{(a) Total energy transfer to the 2D condensate as a function of $S\equiv\Up/\Omega$.
    Integration is carried out with all homochiral waves contributing (\AF{blue}), and with a Lorentzian filter accounting for their progressive decoupling as $U'/\Omega \to 0 $ \eqref{eq:Fkf_Lorentzian} (\AF{light-}blue), both without cutoff ($K\to \infty$). Dashed line: leading order behavior \eqref{eq:T2_leading_corr}.
    Red line: mean-flow dissipation $\nu \Up^2 = \frac{S^2}{4\Roe^2}$ for various values of $\Roe$. The solution to the mean-wave system is found at the intersection between the blue and the red line (black squares).
    The flux-loop state $\Tthree=0$,  obtained when $\Roe \to \infty$, is marked as a green square.
    The yellow region $S< l_f/L_z$ indicates the range where interactions are mostly homochiral.
    (b) Closed solution $U'/\Omega= f(\Roe)$ without cutoff and comparison with DNS data points.
    Solid blue line: numerical estimate of the solution with the Reynolds stress in \eqref{eq:F}.
    Dashed \AF{black} line: leading-order analytical solution \eqref{eq:Sstar}.
    Dashed \AF{grey} line: analytical solution including subleading corrections (\eqref{eq:Sstar} with $b,c$ in \eqref{eq:b_c}).
    }
    \label{fig:F} 
\end{figure}

\subsubsection{Energy input from Layers I}

We focus here on the terms injecting energy into layers Ia and Ib, corresponding to the left-most terms in \eqref{eq:uvIb_exact} and \eqref{eq:uvIa_exact}, each $p_z$ layer contributing at rate $\epsilon_{p_x,p_z}/2= \epsilon/(\pi \sqrt{k_f^2 - p_x^2 - p_z^2} k_f \Vol)$ to the condensate.
The total energy input to the condensate from Layers I (rescaled by $\epsilon_{\rm 3D}$) therefore reads
\begin{align}
    \mathcal{I}_{\rm I} &= \frac{1}{\pi} 
    \int_{\nu_z}^{\frac{S k_f}{4}} \frac{dp_z}{k_f} \int_{\nu_x}^{\sqrt{k_f^2 - p_z^2}} dp_x  \frac{1}{\sqrt{k_f^2 - p_x^2 - p_z^2}}
    = \frac{1}{2} \left(a- \frac{l_f}{L_z} \right)
\end{align}
in the continuous limit $L_z \gg l_f, L_x \gg l_f$.
However, in a discrete setting, only the $p_x$ sum can be interpreted as an integral:
\begin{align}
     \mathcal{I}_{\rm I} &= \frac{ 1}{\pi} 
    \sum_{\nu_z}^{m^* \nu_z} \frac{\nu_z}{k_f} \int_{\nu_x}^{\sqrt{k_f^2 - p_z^2}} dp_x  \frac{1}{\sqrt{k_f^2 - p_z^2 - p_x^2}}\\
    &\simeq 
    \frac{1}{2} \sum_{1 }^{m^*(a)} \frac{\nu_z}{k_f} = \frac{1}{2} \frac{l_f}{L_z}  m^*(a) 
\end{align}
at the leading order in $a$, the exclusion of $p_x=0$ modes resulting in subleading $a l_f/L_x$ corrections.

Note that the effective injection rate in all layers (I and II), $\mathcal{I} \equiv \mathcal{I}=\mathcal{I}_{\rm I} + \mathcal{I}_{\rm II} + D_{\rm II}^{(2)} $ can be formulated as a continuous function of $a$ only:
\begin{align}
    \mathcal{I}  = 1 - \frac{l_f}{2L_x}- \frac{l_f}{2L_z} k - \frac{1}{2}a H_{k+1}
    &= 1  - \frac{l_f}{2L_x} - \frac{1}{2} \sum_{n=1}^{\lfloor k_f / \nu_z \rfloor} \min \left(\frac{l_f}{L_z}, \frac{a}{n} \right),
\end{align}
where the dependence on $m^*$ has disappeared. $\mathcal{I}$ is continuous in $a$ (in particular at the discrete points $a=n l_f/L_z$), but not differentiable at these points.
This continuity is inherited from the continuity of the Reynolds stress $\uv_{\bp\bk}$ at the boundary $p=p^*$ between layers I and layers II.

Note finally that 2D modes can be added separately. They result in a separate energy injection into the condensate of $l_f/(4L_z)$: half of the energy injected into 2D modes ($\epsilon_{\rm 2D} = l_f /(2L_z) \epsilon$), being transferred to small scales.

\subsubsection{Summary of the analytical energy flux}

The energy transfer to the 2D condensate finally reads:
\begin{align}
 \Aboxed{ \Tthree & =
\epsilon_{\rm 3D}
 \Bigg[ 1 - \frac{l_f}{2L_x}
 - a \Bigg( {\frac{\pi}{6}}\frac{L_x}{ l_f}   -  \frac{1}{2} +  {\frac{\pi}{12}}\frac{L_x}{l_f} \sqrt{\frac{l_f}{2 L_z}} 
 \Bigg)
 -
 \frac{1}{2} \sum_{n=1}^{\lfloor k_f/\nu_z \rfloor} \min \left( \frac{l_f}{L_z}, \frac{a}{n} \right)
 \Bigg],
 \label{eq:T2_leading_corr} }
\end{align}
including discretization corrections in $l_f/L_x$ and $l_f/L_z$.
The contribution of $p_z=0$ modes, $\Ttwotwo = \frac{\epsilon_{\rm 2D}}{2} = \epsilon l_f/(4L_z) $, must be added to this term to obtain the net energy transfer to the condensate.

The function \eqref{eq:T2_leading_corr} with our parameter choice ($L_x/l_f = L_z/l_f=5$) is shown in Figure \ref{fig:F}(a) as a dashed blue line.
$\Tthree$ reaches zero in the range $ l_f/L_z =0.2 <a< 2 l_f/L_z= 0.4$, hence the last term is
$- \frac{1}{2} (l_f/L_z) - \frac{1}{2} a H_{2}$
with $H_2= \sum_{n=2}^{5} 1/n \simeq 1.2833$. 
In this range, we therefore obtain the linear expression for the total energy transfer from the fluctuations to the condensate:
\begin{align}
   \Aboxed{ \frac{T}{\epsilon} =
   \frac{\Tthree+ \Ttwotwo}{\epsilon} 
   &=
   \frac{\Tthree}{\epsilon} + \frac{l_f}{4 L_z} =c - b \frac{U'}{\Omega}, }\\
    \text{with} ~~
    b= \frac{1}{4} \left( \frac{\pi}{6 } \frac{L_x}{l_f} - \frac{1}{2} + \frac{\pi}{12} \frac{L_x}{l_f} \sqrt{\frac{l_f}{2 L_z}} + \frac{H_2}{2} \right) \frac{\epsilon_{\rm 3D}}{\epsilon}, ~~~~
    & c=\left(1 - \frac{l_f}{2L_x} - \frac{l_f}{2L_z} \right)\left(1- \frac{l_f}{2L_z} \right) + \frac{l_f}{4 L_z} \simeq 1- \frac{l_f}{2L_x} - \frac{3 l_f}{4L_z}
    \label{eq:b_c}
\end{align}

%
%
With this estimate of the energy transfer, we obtain a flux-loop state $\Tthree = 0 $ at $S = c/b \simeq 1.07$.
We will compare this leading-order estimate with a numerical summation of the discrete Reynolds transfer \eqref{eq:F} in \S \ref{app:num_int}.


\subsection{Numerical estimate of the exact Reynolds stress and validation}
\label{app:num_int}

We now integrate numerically the exact Reynolds stress in both layers II (Eq.~\eqref{eq:uvII_2}) and layers I (Eq.~\eqref{eq:uvIb_exact} and \eqref{eq:uvIa_exact}) over a discrete set of wavenumbers $p_z = 2\pi  m_z/L_z$ and $p_x= 2\pi m_x/L_x$, where $m_z,m_x \in \mathbb{Z}$.
This amounts to making a semi-continuous approximation only along $p_y$, which is still treated as a continuous variable.
%
%
Note that we use the value of $\epsilon_{p_x,p_z}$
corresponding to the same discrete forcing as in the DNS, such that $\sum_{p_y} \epsilon_{p_x,p_z}= \epsilon$.

Figure \ref{fig:F}(a) shows the total energy transfer $\Tthree[S]$, without homochiral-wave filter (light blue), resulting from summing Eqs.~\eqref{eq:uvII_2}, \eqref{eq:uvIb_exact} and \eqref{eq:uvIa_exact} over discrete $p_x,p_z$ and with the parameters corresponding to our DNS ($L_x/l_f=L_z/l_f= 5$).
%
%
%
%
%
The asymptotic scaling \eqref{eq:T2_leading_corr}, including relevant corrections, is shown as a blue dashed line in Fig.~\ref{fig:F}(a).
It reproduces the dominant behavior of the numerical integration of $\Tthree(S)$ (light blue line).
%
%
%
We have also verified that \eqref{eq:T2_leading_corr} was realized with larger values of $L_x/l_f$ and $L_z/l_f$, which we do not show here.

At small values of $S$ homochiral interactions become constrained and average out. To take this effect into account we introduce a filter $F_{k_f}$ acting on the homochiral sector, and show the corresponding transfer $\Tthree(S)$ in Fig.~\ref{fig:F}(a) as a solid \AF{light-}blue line. 
We chose a Lorentzian form for the homochiral-wave filter,
\begin{align}
     F_{\rm Lor}
     = \frac{1}{ 1+ \left( \frac{1}{S}  \frac{4\pi p_z \sqrt{k_f^2 - p_x^2 - p_z^2}}{ L_y k_f^3} \right)^2 }.
    \label{eq:Fkf_Lorentzian}
\end{align}
premultiplying the Reynolds stresses \eqref{eq:uvII_2}, \eqref{eq:uvIb_exact} and \eqref{eq:uvIa_exact}.

When including this filter, the theory captures the decoupling between the waves and the 2D flow occurring at large rotation ($S<l_f/L_y$), when resonances are more stringent.
It leads to the decay of $\Tthree$ when $S\to0$.
With such a smooth filter, the function $\Tthree$ exhibits a smooth transition from the decoupling to the heterochiral regime, around $S\simeq 0.2$.
A sharper filter for such resonant interactions, like the Heaviside function \eqref{eq:Heaviside}, results in a continuous but sharp transition between the two regimes.

 \subsection{Closed analytical solution for the mean flow}


%
To obtain a solution for the shear rate $U'$ we use the energy balance 
 \begin{align}
     \nu \Up^2 = T,
 \end{align}
 and the expression for the energy transfer
 \begin{align}
     \frac{T}{\epsilon} [S]=c-bS
 \end{align}
 where $S=U'/\Omega$, e.g. when including subleading correction (though not fully consistent with the $l_f/L_y$ discretization errors) it appears in \eqref{eq:T2_leading_corr} (and the coefficients $c,b$ have a slight $S$ dependence then),
 valid in the regime where homochiral-wave interactions have not yet decoupled ($S> l_f/L_y$).
In non-dimensional form we get
 \begin{align}
     \frac{1}{4 \Roe^2} S^2 = \frac{\Tthree}{\epsilon} [S] + \frac{l_f}{4 L_z},
     \label{eq:mf_app}
 \end{align}
with non-dimensional parameters $\Roe = \frac{1}{2\Omega}\sqrt{\frac{\epsilon}{\nu}}$ and $S = \Up/\Omega$.
With the transfer $\Tthree$ written explicitly in \eqref{eq:T2_leading_corr}, 
Eq.~\eqref{eq:mf_app} is of the form
\begin{align}
    \frac{1}{4 \Roe^2} S^2 = c   - b S
    \label{eq:closure}
\end{align}
with $b$ and $c$ written in \eqref{eq:b_c} for the range $\frac{4 l_f}{L_z} < S < \frac{8 l_f}{L_z}$, hence $ 0.8 < S< 1.6$.
The solution to Eq.~\eqref{eq:closure} is: 
\begin{align}
      S= 2
     b \Roe^2 
     \Bigg(-1 + \sqrt{1 + \frac{c}{b^2 \Roe^2 } }
     \Bigg),
     \label{eq:Sstar}
\end{align}
If we use the leading order estimates $c=1$ and $b= \frac{\pi}{24} \frac{L_x}{l_f} $ the solution reads
\begin{align}
      U'= \Omega
     \frac{\pi}{12}\frac{L_x}{l_f} \Roe^2 
     \Bigg(-1 + \sqrt{1 + \left(\frac{24}{\pi}\frac{l_f}{L_x}\right)^2 \frac{1}{\Roe^2 } }
     \Bigg),
     \label{eq:Sstar}
\end{align}
In Fig.~\ref{fig:F}(b), we show as a dashed grey line the leading-order solution in $L_x/l_f$, with $c=1$ and $b= \frac{\pi}{24} \frac{L_x}{l_f} $ and as a dashed black line the solution with subleading corrections, with $b$ and $c$ given in \eqref{eq:b_c}. 

\section{Cutting-off wave-wave and eddy-eddy interactions}
\label{app:cutoff}

\subsection{Choice of cutoff}

We now discuss the introduction of a wavenumber cutoff $K$, which, as we will see, provides a validity range for both quasi-linear hypothesis and wave kinetics.
We first introduce various relevant timescales:
\begin{align}
    &\tau_\Omega = \frac{p}{\Omega p_z}, ~~\text{twice the wave period} \label{eq:tau_Omega} \\
    &\tau_{nl}^{\rm eddy} =
    \epsilon^{-1/3} p^{-2/3}   \text{, the eddy-eddy interaction time} \label{eq:tnl_eddy}\\
    &\tau_{nl}^{\rm wave} = \Omega^{1/2} \epsilon^{-1/2} p_z^{1/2} p_\perp^{-3/2} \text{, the wave-wave interaction time} \label{eq:tnl_wave}
\end{align}
The eddy turnover time $\tau_{nl}^{\rm eddy}$ is determined from the Kolmogorov scaling, while the wave-wave interaction time scale $\tau_{nl}^{\rm wave} $ is predicted from wave-turbulent theory \citep{galtier2003weak}.

These time scales lead to the definition of various $p_z$-dependent wavenumbers:
\begin{align}  
 k_\Omega &=
            \Omega^{3/5} \epsilon^{-1/5} p_z^{3/5}
            = (2 Ro)^{\AF{-}3/5} (p_z/k_f)^{3/5}k_f
            , \label{eq:kOmega} ~~~\text{the crossover from waves to eddies}\\
k_U^{\rm wave}  &= 
              (\Omega p_z)^{1/3} {\Up}^{2/3} \epsilon^{-1/3}= (2Ro)^{-1} (p_z/k_f)^{1/3} S^{2/3}k_f,
              ~~~\text{the crossover from condensate- to wave-dominated interactions} 
              \label{eq:kUwt} \\    
 k_U^{\rm eddy} &=
           {\Up}^{3/2} \epsilon^{\AF{-1/2}}= S^{3/2} (2Ro)^{-3/2}k_f, ~~~\text{the crossover from condensate- to eddy-dominated interactions.}  
           \label{eq:kUeddy}
\end{align}
These various wavenumbers are explained in the following.

Wavenumber $k_\Omega$ in \eqref{eq:kOmega} denotes the crossover wavenumber delimiting wave-like ($p<k_\Omega \Longleftrightarrow \tau_\Omega < \tau_{nl}^{\rm eddy}$) from eddy-like ($p>k_\Omega \Longleftrightarrow  \tau_{nl}^{\rm eddy}< \tau_\Omega $) behaviors, called the Zeman wavenumber.
Here we define by ``waves" rotation-dominated modes, as opposed to ``eddies".

Nonlinear interactions between fluctuation modes occur over a nonlinear time scale $\tau_{nl}$, whose estimation depends on whether mode $\bp$ is wave-dominated ($p< k_\Omega$) or eddy-dominated ($p>k_\Omega$).
When modes are wave-dominated, 
they interact with other waves over a slow nonlinear timescale $\tau_{nl}^{\rm wave} $ in \eqref{eq:tnl_wave}, which is $\gg \tau_\Omega$.
In contrast, when modes are eddy-dominated, they interact with each other over the Kolmogorov time scale $\tau_{nl}^{\rm eddy}$ in \eqref{eq:tnl_eddy}.

The quasi-linear assumption is valid when mean-fluctuation interactions dominate over nonlinear interactions within fluctuations, i.e when $1/\Up < \tau_{nl}$.
Depending on whether modes are eddy or wave-dominated, this leads to two definitions for the crossover wavenumber $k_U$, $k_U^{\rm wave}$ in \eqref{eq:kUwt} and $k_U^{\rm eddy}$ in \eqref{eq:kUeddy}, above which interactions with the mean flow are negligible.

%
%

%

\begin{figure}[t]
\includegraphics[width=\columnwidth]{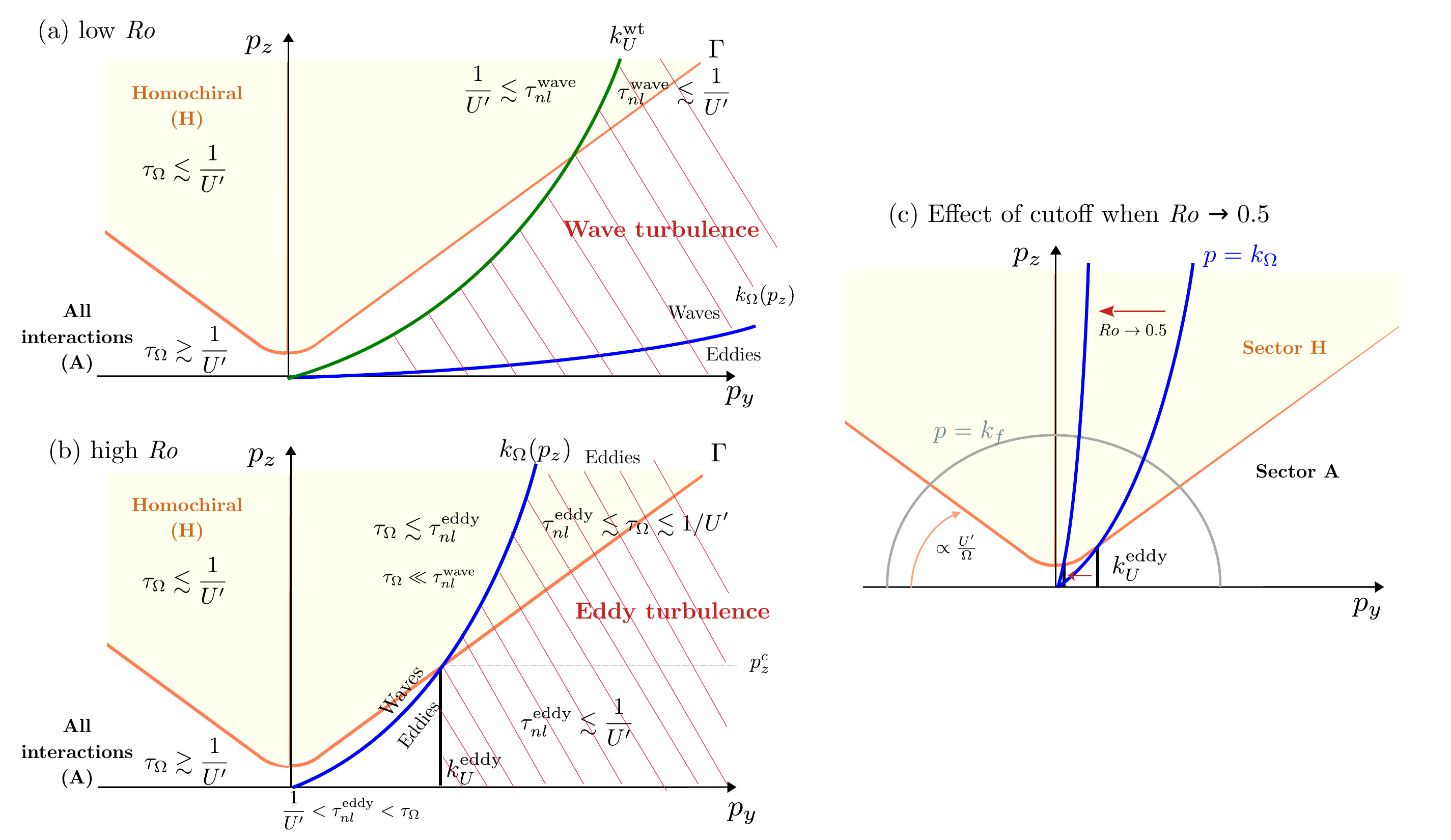}
\caption{Hierarchy of time scales and choice of cutoffs at (a) high and (b) low rotation. 
Sector H ($\tau_\Omega <1/\Up$) is colored in yellow, and Sector A ($\tau_\Omega \geq 1/\Up$) is in white.
Here we fix the value of $U'/\Omega$, hence the boundary between sectors H and A.
The Zeman scale $k_\Omega$ \eqref{eq:kOmega} (blue curve) delimits wave-like from eddy-like behaviors.
Waves dominate at low $Ro$ (a), and wave-wave interactions dominate $p>k_U^{\rm wave} (p_z)$ \eqref{eq:kUwt} (green line).
Eddies dominate at large $Ro$ (b), and eddy-eddy interactions dominate for $p>k_U^{\rm eddy}$ \eqref{eq:kUeddy} in sector A and $p> k_\Omega(p_z)$ in sector H (blue lines).
Modes above these cutoffs do not interact with the condensate and are not considered in the QL theory.
Note that $k_U^{\rm wave} \ll k_U^{\rm eddy}$ in (a), while $k_U^{\rm eddy} \ll k_U^{\rm wave}$ in (b).
(c) Evolution of cutoff $K(p_z)$ at low rotation, when $Ro \to 0.5$.
At fixed $S=\Up/\Omega$, fewer waves in Sector H are excited as $Ro\to 0.5$.     When $Ro=0.5$ and $k_\Omega (p_z=k_f) = k_f$, no homochiral waves are excited and the energy input to the condensate vanishes.
}
   \label{fig:schema_cutoff}
\end{figure}

Wave-dominated modes ($p<k_\Omega$), if $\tau_{nl}^{\rm wave} < 1/U'$, are outside the QL framework. 
Therefore, modes above $k_U^{\rm wave}$ \eqref{eq:kUwt} 
must be cutoff from the QL computation.
Waves obeying the QL assumption, i.e such that $p< k_U^{\rm wave}$, can lie either in Sector H or in Sector A, depending if $\tau_\Omega$ is above or below $1/U'$.
When rotation is sufficiently large, most modes behave as waves, and only the cutoff $k_U^{\rm wave}$ needs to be considered.
This high-rotation case is illustrated in Fig.~\ref{fig:schema_cutoff}a.

Eddy-dominated modes ($p>k_\Omega$), which appear at low rotation, obey a different time-scale hierarchy, illustrated in Fig.~\ref{fig:schema_cutoff}b.
When eddies $(\tau_{nl}^{\rm eddy} < \tau_\Omega)$ dominantly interact with the mean flow $(1/\Up < \tau_{nl}^{\rm eddy}$), they are such that 
$1/\Up < \tau_\Omega$, hence, from condition \eqref{eq:sectors}, they necessarily belong to the heterochiral Sector A.
This is a consequence of the fact that eddies \emph{cannot be homochiral-only when interacting with the condensate}, as expected from them being unaffected by rotation hence not subject to resonant constraints.
%
%
As a consequence, the homochiral Sector H only concerns wave-like modes and not eddies and must cutoff at $k_\Omega$. 
In contrast, Sector A can sustain both waves and eddies: the former being cutoff at $k_U^{\rm wave}$, 
the latter at $k_U^{\rm eddy}$.
Modes in Sector A are therefore necessarily cutoff at $\min(k_U^{\rm eddy}, k_U^{\rm wt})$ to obey the QL condition.
%

All in all, we define the cutoff wavenumber as
\begin{align}
   & K (p_z, \Omega, \Up) = 
    \begin{cases}
              k_U^{\rm wave} 
     ~~~
     \text{ if } 
     k_U^{\rm wave} < k_\Omega  \text{ and }  k_U^{\rm wave} < k_U^{\rm eddy} \vspace{1em}\\
     \begin{cases}
            k_\Omega 
            \text{ if }
            \tau_\Omega < \frac{1}{\Up} ~~ (\text{Sector H})\\
            \min( k_U^{\rm eddy},  k_U^{\rm wave})
            \text{ if }
            \tau_\Omega \geq \frac{1}{\Up} ~~(\text{Sector A})
            \end{cases} 
    \end{cases}
    \label{eq:cutoff}
\end{align}
to cover both wave-dominated and eddy-dominated cases.

The dominant effect of this cutoff appears at low rotation, where both $k_\Omega$ and $k_U^{\rm eddy} \ll k_U^{\rm waves}$.
Here, Sector H is cutoff at $k_\Omega$ and Sector A at $k_U^{\rm eddy}$.
The presence of a cutoff in Sector H at $k_\Omega$ strongly affects the energy input to the condensate from Sector H.
As $Ro \to 0.5$, $k_\Omega(p_z=k_f) \to k_f$, and the contribution from sector H progressively vanishes, as illustrated in Fig.~\ref{fig:schema_cutoff}(c).
This causes the condensate to vanish at $Ro \simeq 0.5$ to a state of pure eddy turbulence.
%

\subsection{Reynolds stress with cutoff}

In the following, we derive the Reynolds stress with the inclusion of a wavenumber cutoff $K$, above which modes do not interact with the condensate due to the dominance of other interactions (eddy-eddy or wave-wave) breaking the QL assumption.
We first introduce 
\begin{align}
p_z^{c} = a \KO
\end{align}
the value of $p_z$ at which the cutoff-curve $K(p_z)$ intersects the A-H boundary $\Gamma$ (see Fig.~\ref{fig:schema_computation}).
Introducing a cutoff at $K$ therefore leads to two further distinctions in layers II:
\begin{itemize}
      \item[--] Layers {\rm II}a -- ($p_z\AF{<} p_z^c$)
         The energy flux emerging from Sector H is transferred to Sector A, and is eventually cutoff. This case is solved by \eqref{eq:sol_II}, cut-off at $K$.
        \item[--] Layers {\rm II}b --  ($p_z \AF{>} p_z^c$) 
        The energy flux emerging from Sector H is cutoff before it can enter sector A ($K < p^*$).
        This case is solved by Eq.~\eqref{eq:sol_homo}, cut-off at $K$.
\end{itemize}

The introduction of cutoff $K$ then simply follows from evaluating the flux $\Pi_{\bp}$ at $p_y=K_y \equiv \sqrt{K^2 - p_x^2 - p_z^2}$ in Eq.~\eqref{eq:uvII_flux} (for layers IIa), and integrating similarly Eq.~\eqref{eq:uv_py_2} up to $p_y=K_y$. This leads to the following total energy transfer, written as a function of the discrete $(p_x,p_z)$ variables:
\begin{align}
    &\frac{\Tthree}{\epsilon_{\rm 3D}} = \frac{4\Up }{\sqrt{2}\epsilon_{\rm 3D}} 
    \sum_{p_x=-k_f}^{-\frac{2\pi}{L_x}} 
    \Bigg(\sum_{p_z=\nu_z}^{ \lfloor a k_f \rfloor } \uv_{p_x,p_z}^{\rm I}
    + \sum_{ p_z= \lfloor a k_f \rfloor + 1}^{\sqrt{k_f^2 - p_x^2}}  \uv_{p_x,p_z}^{\rm II} 
    \Bigg)
    , \label{eq:F}\\
 &\uv_{p_x,p_z}^{\rm I}
 = \left\{
    \begin{aligned}
    &
 =
  \frac{\sqrt{2}\epsilon_{p_x,p_z}}{2 U' } \Bigg[  \Big( 1 - \frac{k_f^2}{\KO^2}
 \Big) 
 -   \frac{p_z^2}{p_x^2} \frac{k_f^2}{ p_0^2} 
 \Bigg[ \arctan^2 (\frac{K_y}{p_0})
 + \arctan^2 (\frac{p_y^f}{p_0}) 
 \Bigg] \Bigg], ~~~ p_z < \frac{|p_x|}{\sqrt{1/a^2- 1}}\\
&~~~~~~~~~~~~~~~~~~~~~~~~~~~~~~~~~~~~~~~~~~~~~~~~~~~~~~~~~~~~~~~~~~~~~~~~~~~~~~~~~~~~~~~~~~~~~~~~~~~  ~ [{\rm Layers~ I\AF{a}}, K = \min(k_U^{\rm wave},k_U^{\rm eddy})]   \nonumber\\
 & \frac{\sqrt{2} \epsilon_{p_x,p_z}}{2  \Up} \Bigg[
 \Big( 1 - \frac{k_f^2}{\KO^2}
 \Big) 
 -   \frac{p_z^2}{2p_x^2} \frac{k_f^2}{p_0^2} 
 \Bigg[
 \Big(
 \arctan (\frac{\KyO}{p_0}) 
 -\arctan (\frac{p_y^*}{p_0})  \Big)^2 
 + \Big(  \arctan (\frac{K_y}{p_0}) 
 -\arctan (\frac{\pyf}{p_0})  \Big)^2 \nonumber\\
 & ~~~~~~~~~~~~~~~~~~ ~~~~~~ 
+ \Big(  \arctan (\frac{\pyf}{p_0}) 
 -\arctan (\frac{p_y^*}{p_0})  \Big)^2
 \Bigg]
 \Bigg], ~~~ p_z > \frac{|p_x|}{\sqrt{1/a^2- 1}}  ~~~~    [{\rm Layers~ Ib}, K = \min(k_U^{\rm wave},k_U^{\rm eddy})] 
   \end{aligned}
   \right.\\
  &\uv_{p_x,p_z}^{\rm II} 
  = \left\{
    \begin{aligned}
& \frac{\sqrt{2}\epsilon_{p_x,p_z}}{\Up} \Bigg[
   1  - \frac{k_f}{2 p^*} \Bigg(
   1 + \frac{ p^{*2}}{\KO^2}
  + \frac{p_z^2}{p_x^2}  \frac{p^{*2} }{p_0^2}
  \Bigg(
   \arctan \Big( \frac{\KyO}{p_0} \Big) 
   - \arctan \Big( \frac{p_y^*}{p_0} \Big) 
   \Bigg)^2
   \Bigg)
 \Bigg] ~ ~ \text{ if }
      a k_f < p_z < a K \nonumber\\
   &~~~~~~~~~~~~~~~~~~ ~~~~~~ ~~~~~~ ~~~~~~ ~~~~~~ ~~~~~~ ~~~~~~ ~~~~~~ ~~~~~~ ~~~~~~ ~~~~~~ ~~~~~~ ~~~~~~ [{\rm Layers~ IIa}, K = \min(k_U^{\rm wave},k_U^{\rm eddy})], \\
 &\frac{\sqrt{2}\epsilon_{p_x,p_z}}{\Up} \Bigg[
   1- \frac{k_f}{\KO} \Bigg]
   ~~~~~~~ \text{ if }
     p_z \geq a K ~~ [{\rm Layers ~ IIb}, K= \min( k_\Omega, k_U^{\rm wave})]
   \end{aligned}
   \right.\label{eq:uv_cutoff} \\
 & ~~~~~~ \text{ with } ~~ 
 p^* =  |p_z|/a,  ~~~~
  p_y^* = (  p^{*2} - p_0^2 )^{1/2}, 
  ~~~p_0^2=p_x^2+ p_z^2 ,~~~~~  \KyO =  (\KO^2 - p_0^2)^{1/2} , ~~~ a \equiv\frac{\Up}{4\Omega}
\end{align}    

Note that we treat 2D modes $p_z=0$ slightly differently than the 3D modes: instead of using $k_U^{\rm eddy}$ as a cutoff for $\Ttwotwo= l_f/(4L_z) \epsilon$, we put their contribution to zero when $S$ is such that $k_\Omega (k_f) = (2 Ro)^{-3/5} k_f < 20 k_f$. This way, the contribution of 2D modes does not survive when $a \to \infty$, corresponding to the observation from DNS.
%

%

%



%




\newpage

\subsection{Numerical QL results}
\label{app:num}

We now discuss numerical solutions of the quasi-linear Reynolds stress \eqref{eq:F}, and their dependence on the system parameters, in particular through cutoff $K$.
The condensate amplitude $\Up/\Omega$ solves in steady state the mean-flow energy balance
\begin{align}
    &\frac{1}{4 \Roe^2 } \frac{\Up^2}{\Omega^2}  = \frac{\Tthree}{\epsilon} \Big[ \frac{\Up}{\Omega}, Ro \Big] + \Ttwotwo
    \label{eq:QL_formal_app}
\end{align}
with $\Tthree\Big[ \frac{\Up}{\Omega}, Ro\Big]$ written in Eq.~\eqref{eq:F}, and with the inclusion of the energy input from 2D modes, $\Ttwotwo$. The additional $Ro$ dependence arises through cutoff $K$ defined in \eqref{eq:cutoff}.

In order to compare the QL theory with our DNS of rotating 3DNSE, we numerically integrate the Reynolds stress in Eq.~\eqref{eq:F} over a discrete set of wavenumbers $p_z = 2\pi  m_z/L_z$ and $p_x= 2\pi m_x/L_x$, where $m_z,m_x \in \mathbb{Z}$.
This amounts to making a semi-continuous approximation only along $p_y$, which is still treated as a continuous variable.
%
%
Note that we use the value of $\epsilon_{p_x,p_z}$
corresponding to the same discrete forcing as in the DNS, such that $\sum_{p_y} \epsilon_{p_x,p_z}= \epsilon$.

\begin{figure*}[t]
\includegraphics[width=0.8\textwidth]{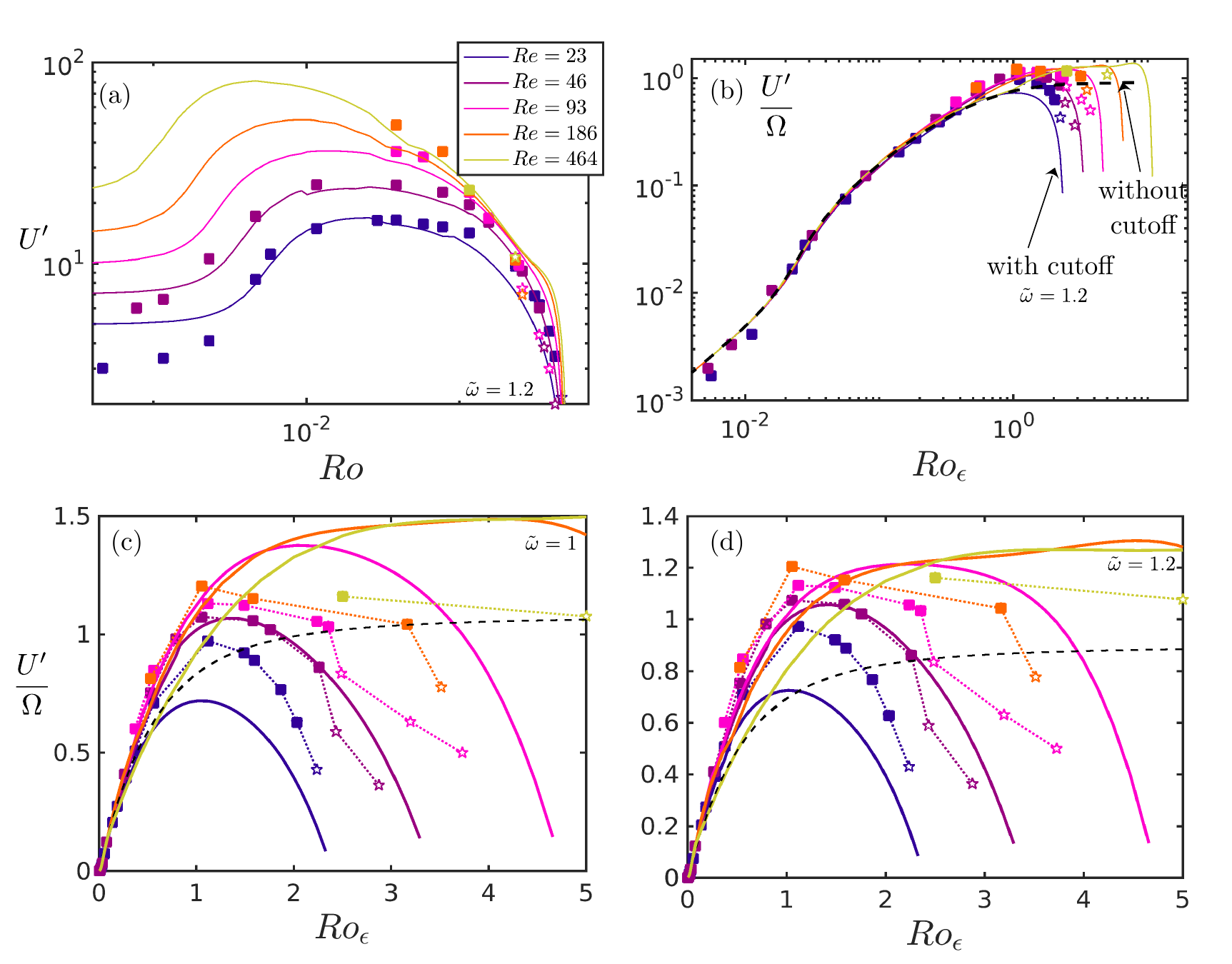} 
\caption{Numerical solution of the quasi-linear system \eqref{eq:QL_formal_app}, with (solid lines) and without (dashed lines) cutoff $K$ \eqref{eq:cutoff}.
(a) $\Up$ as a function of $Ro$, showing the collapse on $Ro$ at large $Ro$.
(b) $\Up/\Omega$ as a function of $\Roe = Ro Re^{1/2}$, 
showing the collapse on $\Roe$ at low $\Roe$.
All QL results in (a,b) are obtained
with $\tilde{\omega}=1.2$.
(c,d) Effect of $\tilde{\omega}$ on solution $\Up/\Omega$, visualized in linear scale:
(c) $\tilde{\omega}=1$, (d) $\tilde{\omega}=1.2$, which better fits the DNS result in the presence of a cutoff.
}
\label{fig:QL}
\end{figure*}

Once $\Tthree [\frac{U'}{\Omega}, Ro]$ is computed from \eqref{eq:F}, we solve \eqref{eq:QL_formal_app} by finding the intersection between $S^2/(4\Roe^2)$ and $\Tthree/\epsilon [S, Ro]$ (where $S= \Up/\Omega$) for all values of $Ro$ and $Re$, see Fig.~\ref{fig:F}. 
The intersection is found by a numerical root-finding procedure.
This results in a function
\begin{align}
    \frac{\Up}{\Omega} = \mathcal{F}[ \Roe, Ro],
    \label{eq:sol_QL_formal}
\end{align}
which we show in Fig.~\ref{fig:QL}(a) as a function of $Ro$, and Fig.~\ref{fig:QL}(b) as a function of $\Roe = Ro \times Re^{1/2}$.
The solution is shown for various values of $Re$.

When integrating numerically the Reynolds stress, we consider the Lorentzian filter for the homochiral-wave resonances \eqref{eq:Fkf_Lorentzian}.
It yields a smooth transition between the homochiral regime and the mixed-interaction regime around $\Roe \simeq 0.1$ (see Fig.~\ref{fig:QL}).
At intermediate $\Roe$, this near-resonant treatment is more accurate than simply removing off-resonant modes $\omega_{\bp\bq}^{ss}>U'$, as it captures the contribution of small near-resonances.
However, it fails in the limit $\Roe\to 0$, as explained in \citep{gome2026waves}.

We furthermore introduce a parameter $\tilde{\omega}$ setting the time-scale ratio over which heterochiral interactions occur: we use the near-resonant condition
\begin{align}
    \omega_{\bp}^{+} + \omega_{\bq}^{-} < \tilde{\omega} \Up
\end{align}
in place of \eqref{eq:sectors}. This is equivalent to changing $S$ to $\tilde{\omega} S$ in \eqref{eq:F}.
Parameter $\tilde{\omega}$, which must be $\sim 1$, provides an adaptable broadening of the resonant condition. It corresponds to approximating the oscillating factor $\Delta(t)^{s,-s}$ \eqref{eq:Heaviside} with a parametrized Heaviside function $\Theta( 1 - | \omega_{\bp\bq}^{+-}|/(\tilde{\omega} \Up) $.
%
In the results established so far, and in the solution shown in Fig.~4 of the main text, $\tilde{\omega}$ is set to 1.
In general, $\tilde{\omega}$ can be used as a tunable parameter of the numerical integration which can better fit the DNS data points.
In Figs.~\ref{fig:QL}(a,b), we use $\tilde{\omega}=1.2$. Solutions with $\tilde{\omega}=1$ and $\tilde{\omega}=1.2$ are compared in Fig.~\ref{fig:QL}(c) and (d), respectively.

Due to the flux-loop behavior at low rotation, the high-$Re$ solution of $\Up/\Omega$ in \eqref{eq:sol_QL_formal} changes its functional behavior as $Ro$ is increased.
It deviates from a single dependence on $\Roe\equiv Ro Re^{1/2}$ at low $Ro$ to 
a single dependence in $Ro$ at high $Ro$. This is visible in Fig~\ref{fig:QL}(a).
Cutoff $K(p_z)$ in \eqref{eq:cutoff} controls the number of modes interacting with the condensate, hence its energy input, controlled by the value of $Ro$.
Without cutoff, $\Up/\Omega$ tends to a $Ro-$independent value when $\Roe \to \infty$, hence $\Up \propto Ro^{-1}$. 
With cutoff, whose effect is particularly important at high $Ro$ where $K(p_z) \simeq k_\Omega(p_z)$, the number of waves shrinks when $Ro\to 0.5$, as $k_\Omega(p_z=k_f)\simeq k_f$. This wave shortage causes the condensate amplitude $\Up/\Omega$ to decay to 0 at $Ro \simeq 0.5$, as seen in Fig.~\ref{fig:QL}(a).

The introduction of a cutoff $K$ has direct consequences on the dependence of $\Up/\Omega$ on $Re$. This is illustrated in Figs.~\ref{fig:QL}(b-d), where the integration is shown without cutoff ($K\to\infty$, dashed lines) and with cutoff (finite $K$, colors). 
Without cutoff, $\Up/\Omega$ only depends on $\Roe$ and saturates as $\Roe \to \infty$.
With cutoff, $\Up/\Omega$ decays at a finite  $\Roe^* \simeq 0.5/Re^{1/2}$.
%
When $Re$ is increased, the decay of $\Up/\Omega$ at $\Roe^*$ is shifted to even larger $\Roe^*$  when $Re\to \infty$. 
Therefore, the solution at large $\Roe$ never collapses to the single $\Roe$-dependence obtained in the non-cutoff case, where $\Up/\Omega \underset{\Roe\to\infty}{\to} \rm{cste}$.
The cutoff solution therefore captures the behavior of $U'/\Omega$ in the DNS when $\Roe \sim 1$, at least qualitatively.

Note that introducing a cutoff generates an overshoot of the solution at high $Re$ and $\Roe \sim 1$. See the colored lines in Fig.~\ref{fig:F}(c)) compared to the non-cutoff case (dashed black line in Fig.~\ref{fig:F}(c)). 
This overshoot is associated to a non-monotonous trend with $Re$:
at fixed $\Roe$, $\Up/\Omega$ first increases and then decreases with $Re$, hence the value at finite $Re$ overshoots that at $Re\to\infty$, which gets closer to the non cutoff case.
This behavior is due to a non-monotonous dependence of the energy transfer $\Tthree$ \eqref{eq:F} on cutoff wavenumber $K(p_z)$ at fixed $\frac{\Up}{\Omega}$, which is inherent to our computation. 
When increasing cutoff $K$, the number of modes in Sector H first increases and so does $\Tthree$, until all Sector H contributes. 
However, increasing $K$ further results in larger contribution of the heterochiral sector A.
This causes $\Tthree$ to decrease with $K$ as more and more modes take energy away from the mean flow.
This behavior is independent of the specific choice of the function $K(p_z)$. 
The overshoot from the non-cutoff case visible 
around $\Roe \simeq 1$ is due to the heterochiral sector A being cutoff, which increases the energy transfer to the condensate and its amplitude.

\begin{figure*}[t]
\includegraphics[width=0.6\textwidth]{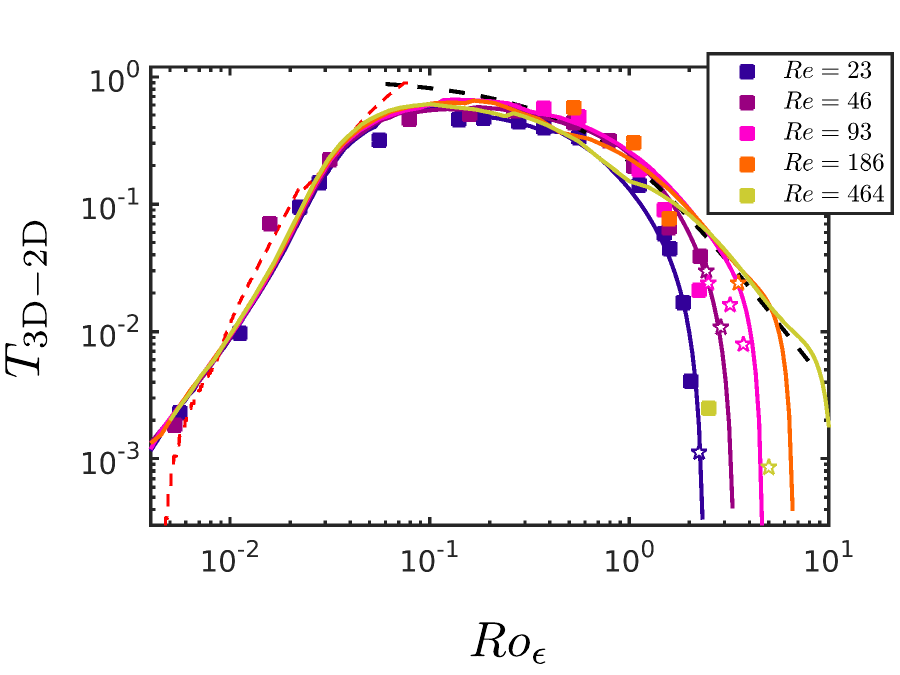}
\caption{Energy transfer to the 2D flow, $\Tthree/\epsilon$ as a function of $\Roe$.
Theoretical lines show the numerical QL solution with $\tilde{\omega}=1.2$.
The black dashed line shows the analytical solution without cutoff.
}
\label{fig:QL2}
\end{figure*}

The QL computation involving cutoff $K$ is compared to the DNS data points in Figs.~\ref{fig:QL}(c-d).
The agreement is quantitatively good up to $\Roe \simeq 0.5$.
At larger $\Roe$, 
the behavior is sensitive to the cutoff and $\Up/\Omega$ decays due to wave shortage, and reaches zero at $Ro\simeq 0.5$.
Our QL results qualitatively reproduce the decay of $\Up/\Omega$ at fixed $Re$, and the transition to strong turbulence seen in the DNS.
However, quantitative agreement between DNS and QL theory is delicate to obtain for all $Re$: first, our theory neglects finite-size effects in $L_y$ that might affect the value of $U'/\Omega$.
%
%
Second, the overshoot due to the introduction of a cutoff in the theory, and the resulting non monotonicity in $Re$, cannot be confirmed with our DNS, whose $Re-$range is limited.

\end{document}